\documentclass[a4paper,11pt]{article}
\pdfoutput=1

\usepackage{jcappub}
\usepackage[T1]{fontenc}
\usepackage{amsmath,amssymb,graphicx}
\usepackage{soul}
\usepackage[utf8]{inputenc}
\usepackage{dsfont}
\usepackage{dcolumn}
\usepackage{subfigure}
\usepackage{tabularx}
\usepackage{array}
\usepackage[normalem]{ulem}
\newcolumntype{L}{>{\centering\arraybackslash}m{2.6cm}}
\usepackage[section, below, above]{placeins}
\usepackage{afterpage}
\usepackage{mathtools}
\usepackage{bbold}

\newcommand{\be}{\begin{equation}}
\newcommand{\ee}{\end{equation}}
\newcommand{\bea}{\begin{eqnarray}}
\newcommand{\eea}{\end{eqnarray}}

\newcommand{\rDelta}{\overline{r}_{_{\hspace{-0.06cm} \Delta}}}

\title{\boldmath Properties of oscillons in hilltop potentials: energies, shapes, and lifetimes}

\author{Stefan Antusch, Francesco Cefal\`{a}, Francisco Torrent\'{i}}
\affiliation{Department of Physics, University of Basel, Klingelbergstr.\ 82, CH-4056 Basel, Switzerland.}

\emailAdd{stefan.antusch@unibas.ch}
\emailAdd{f.cefala@unibas.ch}
\emailAdd{f.torrenti@unibas.ch}

\abstract{Oscillons are spatially localised strong fluctuations of a scalar field. They can e.g.\ form after inflation when the scalar field potential is shallower than quadratic away from the minimum. Although oscillons are not protected by topology, they can be remarkably stable and have a significant impact on the (p)reheating phase. In this work we investigate the properties of oscillons in hilltop-shaped potentials, in particular the typical energies, shapes and lifetimes. In the first part of the paper, we simulate oscillon creation and stabilization with (3+1)-dimensional classical lattice simulations, and extract the typical energies, radii and amplitudes of the oscillons. In the second part we approximate the oscillons as spherically symmetric, and simulate single oscillons until their decay. We find that typical oscillons live up to about $10^4$-$10^5$ field oscillations, with the individual lifetime of the oscillons depending mainly on the initial shape of the oscillon and the power-law coefficient characterising the particular hilltop model. We also observe a breathing mode in the oscillon radii and amplitudes, and find that stronger breathing implies shorter lifetimes. }

\begin{document}
\maketitle
\flushbottom

\section{Introduction}
\label{sec:introduction}

The dynamics of scalar fields in the early universe often features spatially localised oscillatory field configurations with large amplitude, called oscillons (see \cite{Bogolyubsky:1976nx,Segur:1987mg,Gleiser:1993pt,Copeland:1995fq,Honda:2001xg,Copeland:2002ku,Adib:2002ff,Broadhead:2005hn,Farhi:2005rz,Fodor:2006zs,Graham:2006vy,Farhi:2007wj,Gleiser:2007te,Fodor:2008es,Amin:2010jq,Amin:2010dc,Gleiser:2011xj,Amin:2011hj,Amin:2013ika,Achilleos:2013zpa,Antusch:2015nla,Bond:2015zfa,Antusch:2015ziz,Lozanov:2017hjm,Hong:2017ooe,Cotner:2018vug} for a partial list of references). It has been shown that oscillons can form when the potential for the scalar field is shallower than quadratic \cite{Amin:2011hj} in some region around the minimum of the potential. Although oscillons are not protected by a topological invariant, they can be remarkably long lived, and are sometimes referred to as quasi-solitons.  In previous studies, the lifetime of oscillons has been investigated e.g.\ in \cite{Graham:2006xs,Hindmarsh:2006ur,Saffin:2006yk,Gleiser:2008ty,Gleiser:2009ys,Hertzberg:2010yz,Salmi:2012ta,Andersen:2012wg,Saffin:2014yka,Gleiser:2014ipa,Mukaida:2016hwd,Ikeda:2017qev,Gleiser:2018kbq,Ibe:2019vyo,Gleiser:2019rvw,Olle:2019kbo,Muia:2019coe}. It has been emphasised that the lifetime is strongly dependent on the form of the potential and the initial oscillon shape. In some papers it has been conjectured that, under certain conditions and  for specific initial configurations, they might even live until today (see e.g.\ \cite{Ikeda:2017qev,Gleiser:2019rvw}).
 It has also been discussed that oscillons can be a strong source of gravitational waves (GWs) during their production and oscillation stage as long as they are sufficiently asymmetric \cite{Zhou:2013tsa,Antusch:2016con,Liu:2017hua,Antusch:2017flz,Antusch:2017vga,Amin:2018xfe,Kitajima:2018zco,Liu:2018rrt,Lozanov:2019ylm,Sang:2019ndv}. The GW signal is instead suppressed when the oscillons get symmetric \cite{Zhou:2013tsa,Antusch:2016con,Amin:2018xfe}. 

Oscillons can form efficiently in scalar field potentials of the hilltop type \cite{Antusch:2015nla,Antusch:2015ziz,Antusch:2016con,Antusch:2017flz}, as they appear in various particle physics models of the early universe, e.g.\ in \textit{hilltop inflation} \cite{Linde:1981mu,Izawa:1996dv,Izawa:1997df,Senoguz:2004ky,Boubekeur:2005zm,Kohri:2007gq,Antusch:2013eca,Antusch:2014qqa}, in the phase transition after K\"ahler-driven \textit{tribrid inflation} \cite{Antusch:2012jc}, in \textit{flavon inflation} \cite{Antusch:2008gw}, or for string theory moduli (cf.\ \cite{Antusch:2017flz}). One of the main production mechanisms is \textit{tachyonic oscillations} \cite{Desroche:2005yt,Brax:2010ai} but they can also be produced via parametric self-resonance of the scalar field \cite{Kofman:1994rk,Kofman:1997yn}. While the production phase and initial stage of the oscillon dynamics can be studied well with 3-dimensional lattice simulations, it is typically not possible to continue these simulations until the oscillons decay, given the high resolution needed for reliably simulating the oscillons and limited numerical capabilities. It is therefore desirable to obtain estimates for the oscillon lifetime from dedicated studies making simplifying assumptions. 

In this paper we present a detailed study of some oscillon properties in hilltop-shaped potentials, using different kinds of numerical techniques. In particular, we will focus on a particular class of \textit{hilltop inflation} models, and study the energies, shapes, and lifetimes of oscillons in these models. We will divide our work in two parts. In the first part we perform classical lattice simulations of the post-inflationary dynamics of the inflaton field in 3 spatial dimensions, and extract the oscillon shapes and energies from the lattice, well after the oscillons have stabilized. In the second part we investigate the lifetime of oscillons in hilltop potentials,  using numerical simulations of single classical oscillons in the approximation of spherical symmetry. We investigate the dependence of the oscillon lifetime on the parameters of the hilltop potentials, in particular on the \textit{steepness} of the potential.

The structure of this paper is as follows. In Section \ref{sec:model} we describe some relevant properties of the inflationary hilltop models that we consider in this work, and review oscillon formation after inflation. In Section \ref{sec:oscillon-shapes} we present results from the (3+1)-dimensional lattice simulations, and extract the shapes of the oscillons. In Section \ref{sec:spherical-symmetries} we present results from the spherically symmetric simulations of single oscillons, and provide estimates for the oscillon lifetime in hilltop models. Finally, in Section \ref{sec:discussion} we summarize and discuss our results.

\section{Hilltop models}\label{sec:model}
\begin{figure}
\centering
\includegraphics[width=0.5\textwidth]{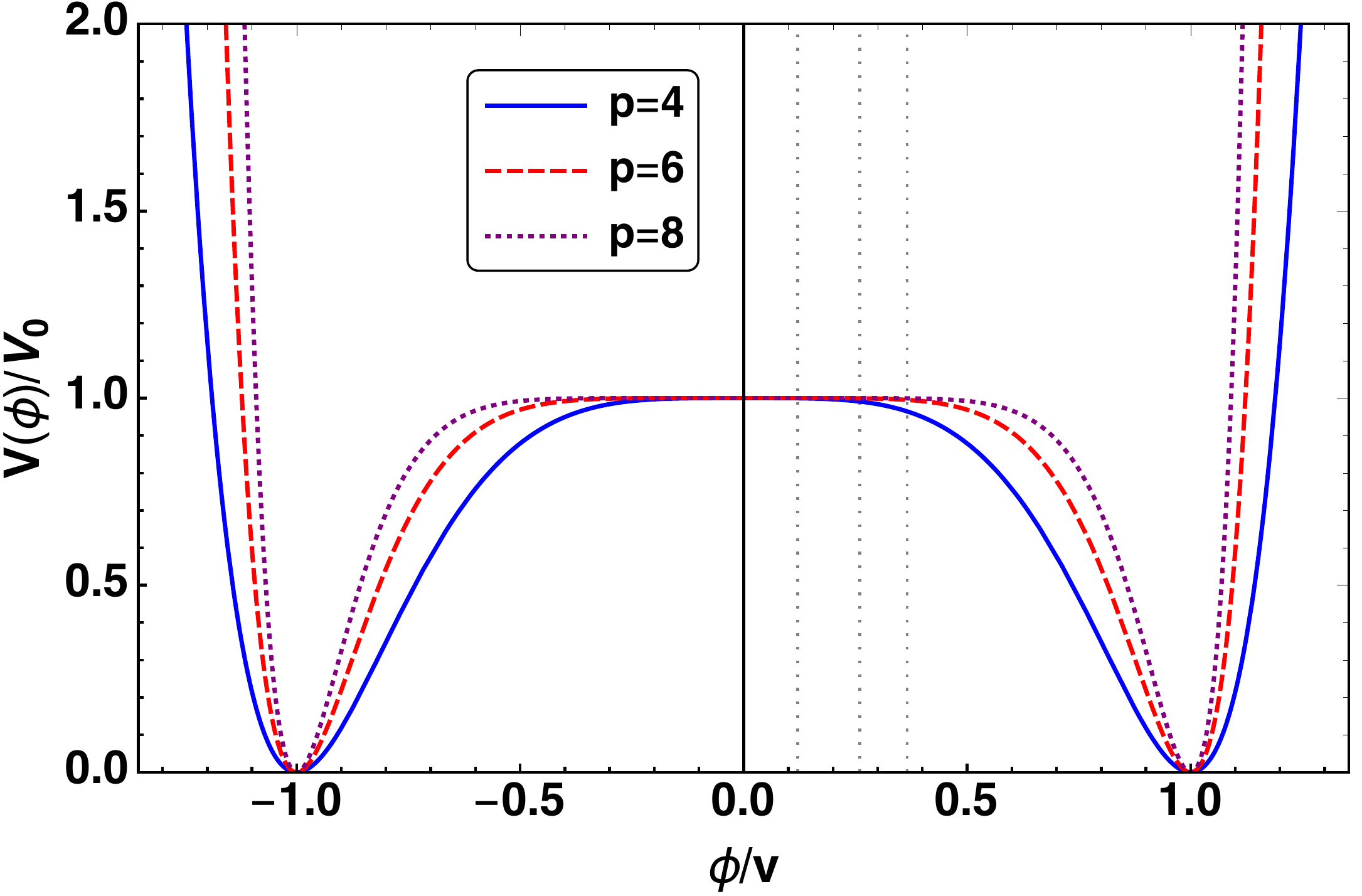}
\caption{We show the hilltop potential [Eq.~(\ref{eq:scalar_potential})] as a function of $\phi$, for $p=4,6,8$. We also indicate with vertical dashed lines the field amplitudes $\phi (t_i)$ at which the oscillatory behaviour of the inflaton starts, as defined in the bulk text.}
\label{fig:hilltop-potential}
\end{figure}

As a specific example, we consider a real scalar field $\phi$ with the following \textit{hilltop}-shaped potential
\be V(\phi)\,=\,V_0\,\left(1-\frac{\phi^p}{v^p}\right)^2 \ , \label{eq:scalar_potential} \ee
where $v$ and $V_0$ have dimensions of energy and $\text{(energy)}^4$ respectively, and $p$ is a positive even integer with $p \geq 2$. We show the shape of this potential as a function for $\phi$  in Fig.~\ref{fig:hilltop-potential}, for $p=4,6,8$.  We observe that the potential features a plateau around $\phi \approx 0$, as well as two minima at $\phi = \pm v$. The power-law coefficient $p$ governs the flatness of the plateau, as well as the steepness of the potential towards large field values, with larger $p$ corresponding to a more pronounced plateau and a steeper large field potential around the minimum.

Let us assume in this work that $\phi$ is an inflaton field. In this case, inflation takes place when the inflaton field $\phi$ is placed close to the ``top of the hill'' at $\phi \simeq 0$. During inflation, the field slowly rolls down the potential towards any of the two minima at $ \phi = \pm v$. Let us define the time $t_i$ of end of inflation when the slow-roll parameter becomes $\epsilon=1$: this time signals approximately the onset of the oscillatory regime of the inflaton homogeneous mode, and it is when we impose the initial conditions of our (3+1)-dimensional lattice simulations in Section \ref{sec:oscillon-shapes}. In Table \ref{table:hilltop-parameters} we provide the model parameters that we consider in this work, as well as the initial values for the field amplitude and its derivative at time $t=t_i$. The value of $V_0$ is obtained by fitting the amplitude of the CMB anisotropies, while we have chosen $v = 0.01 m_{p}$ as an example. The values of $\phi (t_i)$ and $\dot{\phi} (t_i)$ are obtained by solving numerically the field and scale factor equations, with initial conditions deep in slow-roll at $\phi \simeq 0$, and assuming that the inflaton decays towards the positive-amplitude vacuum.

While the scalar field rolls towards its minimum in the region of the scalar potential where $V''(\phi(t)) < 0$, fluctuations of the inflaton field grow, and there is an initial stage of tachyonic preheating. As the field rolls through this tachyonic region, all the modes $\delta \phi_k$ with $k^2/a^2 + \partial^2 V/\partial\phi^2<0$ grow exponentially. In hilltop models, tachyonic preheating is generically followed by another, even more efficient mechanism for the growth of perturbations: tachyonic oscillations. The frequency of the field oscillations around the minimum of the potential is governed by the mass $m_\phi$ of the scalar field at the minimum,
\be m_{\phi}^2 \equiv \frac{\partial^2 V}{\partial \phi^2} \Big\rvert_{\phi = v} =  \frac{2 p^2 V_0}{v^2} \ . \label{eq:inf-mass} \ee
Tachyonic oscillations happen when the oscillating field crosses periodically the inflection point $\partial^2V/\partial\phi^2=0$, entering the region where $V''(\phi(t)) < 0$. For each oscillation, the field perturbations grow in the tachyonic region as the field accelerates toward the minimum, and they decrease as the field decelerates toward the plateau (i.e.\ when it moves uphill). This behaviour, combined with the expansion of the Universe and the periodic crossing of the inflection point, leads to a net growth of the field perturbations peaked at a characteristic wavenumber $k_p \lesssim m_\phi $, closely related to the frequency of the oscillations of the homogeneous mode \cite{Brax:2010ai,Antusch:2015nla}. 

\begin{table}
\begin{center}
    \begin{tabular}{| c || c | c | c |  c |}
    \hline
    $p$ & $v\,({\rm GeV})$ & $V_0\,({\rm GeV}^4)$ & $\phi (t_i)\,({\rm GeV})$ & $\dot{\phi} (t_i)\,({\rm GeV}^2)$ \\ \hline
    4 & $2.44 \cdot 10^{16}$ & $7.0 \cdot 10^{51}$     & $2.9\cdot 10^{15}$  & $2.4 \cdot 10^{24}$\\ \hline
    6 & $2.44 \cdot 10^{16}$ & $3.5 \cdot 10^{54}$  & $6.3 \cdot 10^{15}$ & $2.8 \cdot 10^{24}$ \\ \hline
    8 & $2.44 \cdot 10^{16}$ & $1.8 \cdot 10^{55}$  & $8.9 \cdot 10^{15}$ & $2.9 \cdot 10^{24}$ \\ \hline
    \end{tabular}
\end{center} \label{table:hilltop-parameters}  
\caption{Numerical values for the parameters of the hilltop potential, used in the lattice simulations of Sections \ref{sec:oscillon-shapes} and \ref{sec:spherical-symmetries}. $\phi (t_i)$ and $\dot{\phi} (t_i)$ are the values of the field amplitude and derivative for each case at time $t = t_i$, derived as explained in the bulk text.}
\end{table}

It has been shown that these fluctuations can grow so strongly that the scalar field can, at some localised regions, overshoot temporarily towards the ``wrong vacuum'', creating periodically expanding and collapsing bubbles \cite{Antusch:2015nla}. Although this early phase of collapsing bubbles typically  lasts only a few oscillations, it can efficiently trigger the formation of oscillons in hilltop models. The aim of this work is to understand better the properties of such oscillons. In Section \ref{sec:oscillon-shapes} we will focus on the oscillon energies and shapes, while in Section \ref{sec:spherical-symmetries} we give a first estimate on their typical lifetimes. With this aim, we have carried out two different kinds of numerical simulations of the tachyonic oscillatory regime in hilltop models, which cover different regimes of the oscillon dynamics:

\begin{itemize}
\item \textbf{(3+1)-dimensional lattice simulations (Section \ref{sec:oscillon-shapes}):} We present results from a set of classical lattice simulations in (3+1)-dimensions, which start when the inflationary slow-roll conditions break at $t = t_{\rm i}$, and capture the first $\mathcal{O} (10^3)$ oscillations of the inflaton homogeneous mode. In these simulations, we observe the initial field perturbation growth described above, as well as the corresponding formation and stabilization of oscillons. Well after the oscillons have stabilized, we extract the oscillon shapes and energies by applying an appropriate \textit{extraction procedure}, which is described in detail in Section \ref{sec:fitting-proc}. We then fit the oscillon shapes to spherically symmetric Gaussian functions, which are parametrized by their \textit{amplitude} and \textit{radius}. This way, we are able to give estimates to the typical oscillon energies and shapes, for different power-law parameters $p$ of the hilltop potential.

\item \textbf{Spherically symmetric simulations  (Section \ref{sec:spherical-symmetries}):} Typical oscillons continuously lose energy due to the emission of small-amplitude field waves \cite{Segur:1987mg}, and they eventually decay. However, the oscillons lifetime is very large in comparison with the inflaton oscillation period, so the decay process cannot be typically observed with (3+1)-dimensional lattice simulations. Moreover, the oscillon radius in the lattice decreases with time, so there exists a lack of good enough spatial resolution at late times. For this reason, in Section \ref{sec:spherical-symmetries} we take an alternative approach: we simulate the dynamics of a \textit{single} oscillon, and make the approximation of \textit{spherical symmetry}. This way, although we neglect both possible oscillon asymmetries and oscillon-oscillon interactions, we reduce the complexity of the system, and can solve numerically the equations of motion much faster. These spherically symmetric simulations are initiated just when the (3+1)-dimensional simulations end, and the initial conditions are provided by the fitted oscillon shapes in Section \ref{sec:oscillon-shapes}. An appropriate truncation technique is implemented in order to avoid unphysical interference effects due to boundary conditions. In these simulations we are able to observe the oscillons decay, so we  provide first estimates for the oscillon lifetime, for different initial oscillon configurations and power-law coefficients $p$.
\end{itemize}

\section{Oscillon shapes from lattice simulations in 3+1 dimensions} \label{sec:oscillon-shapes}

We present in this section results from (3+1)-dimensional lattice simulations of the hilltop model. The tachyonic oscillatory regime in these models has already been studied with lattice simulations in the past  \cite{Antusch:2015nla,Antusch:2015ziz,Antusch:2016con,Antusch:2017flz}, but these simulations mainly focused on the initial oscillon formation regime. Our aim in this section is, instead, to extract the typical oscillon energies and shapes, well after the oscillons have stabilized. These will constitute the initial conditions for the spherically symmetric simulations in Section \ref{sec:spherical-symmetries}.

Let us begin by introducing the field and scale factor equations describing the post-inflationary stage of the hilltop model in a FLRW spacetime. They are
\bea
\ddot{\phi} - \frac{1}{a^2}\nabla^2\phi + 3 H(t) \dot{\phi} + \frac{\partial V}{\partial\phi}\,=\,0 \ , \label{eq:eom1}  \\
H^2 (t) \equiv \left( \frac{\dot{a}}{a} \right)^2 = \frac{1}{3 m_p^2} \left( \frac{1}{2} \dot{\phi^2} + \frac{1}{2 a^2} |\nabla \phi |^2 + V(\phi) \right) \ , \label{eq:eom2} 
\eea
where $\dot{} \equiv d / dt$, and the potential function is given in Eq.~(\ref{eq:scalar_potential}). From now on we set the scale factor initially as $a(t_{\rm i}) = 1$. It is convenient to define new dimensionless spacetime and field variables as 
\be \bar{t} \equiv m_{\phi} t \ , \hspace{0.4cm} \bar{x}^i \equiv m_{\phi} x^i \ , \hspace{0.4cm} \Phi \equiv \frac{\phi}{v} \ , \ee
where $m_{\phi}$ is the inflaton effective mass at the minimum of the potential, defined in Eq.~(\ref{eq:inf-mass}). We will refer to this new set of variables as \textit{natural variables}. This way, the negative/positive-amplitude vacua of the inflaton field correspond to field amplitudes  $\Phi = -1, 1$ respectively. The equations of motion (\ref{eq:eom1}) and (\ref{eq:eom2}) can be then rewritten as
\bea 
\Phi'' - \frac{1}{a^2}\nabla_{\bar{x}}^2 \Phi + 3 \mathcal{H} \Phi' + \frac{1}{p} \Phi^{p-1} (\Phi^p - 1 ) = 0 \ , \label{eq:eom3} \\
\mathcal{H}^2  \equiv m_{\phi}^{-2} H^2  = \frac{1}{6} \left( \frac{v}{m_p}\right)^2 \left( \Phi'^2 + \frac{1}{a^2} |\nabla_{\bar{x}} \Phi |^2 + \frac{1}{p^2} (1 - \Phi^p)^2 \right) \ , \label{eq:eom4} \eea
where we have defined $' \equiv \partial / \partial \bar{t}$ and $\nabla_{\bar{x}} \equiv \partial / \partial \bar{x}$.  Note that for the model parameters displayed in Table \ref{table:hilltop-parameters}, there is a suppressing global factor $\sim (v /m_{pl})^2 \sim 10^{-4}$ on the right hand side of Eq.~(\ref{eq:eom4}), so the initial scale factor growth rate is much smaller than the inflaton oscillation frequency, $H(t_i) \ll m_{\phi}$. Due to this, it takes $\mathcal{O} (10^3)$ oscillations of the inflaton homogeneous mode for the scale factor to grow a factor 5. Finally, let us write the local energy density of the system as
\bea \rho &\equiv & \frac{1}{2} \dot{\phi}^2 + \frac{1}{2 a^2} |\nabla \phi |^2 + V(\phi) \nonumber \\ 
&=& V_0 \left( p^2 \Phi'^2 + \frac{p^2}{a^2} | \nabla_{\bar x}\Phi |^2 + (1 - \Phi^p)^2 \right) \label{eq:local-energy} \ , \eea
where in the second line we have written the expression in terms of natural variables.

As said, we have performed lattice simulations of the post-inflationary regime of the hilltop model in 3+1 dimensions. Let us first provide some technical details about these simulations. Our discretization technique, as well as the way in which the fields have been initialized, are very similar to \textsc{LatticeEasy} \cite{Felder:2000hq}. We refer to its documentation for details. However, there are two important differences in our lattice approach. First, \textsc{LatticeEasy}  approximates the continuum Laplacian in the field \textit{eom}, with a discrete Laplacian that is second-order accurate in space, while we use instead a fourth-order accurate one \cite{DiscreteDerivatives}. Let us write the position vector of a point in the lattice as $\vec{n} = \Delta {\rm x} (n_i \hat{\imath} + n_j \hat{\jmath} + n_k \hat{k})$,  where $(n_i,n_j,n_k)$ are positive integers, $\Delta {\rm x}$ is the step length of the box, and $(\hat{\imath} ,\hat{\jmath}, \hat{k})$ are vectors of length $\Delta {\rm x}$ in the $(x,y,z)$ directions respectively. We use 
\bea 
\nabla^2 \Phi (\vec{n}) \simeq \frac{1}{12 \Delta {\rm x}^2} &\times & \left[16  (\Phi (\vec{n} + \hat{\imath}) + \Phi (\vec{n} + \hat{\jmath}) + \Phi (\vec{n} + \hat{k})+ \Phi (\vec{n} - \hat{\imath}) - \Phi  (\vec{n} - \hat{\jmath}) - \Phi (\vec{n} - \hat{k}) ) \right. \nonumber \\
&& \hspace*{-0.8cm}  \left.    - ( \Phi (\vec{n} + 2 \hat{\imath}) +  \Phi (\vec{n} + 2 \hat{\jmath})  + \Phi (\vec{n} + 2 \hat{k})  + \Phi (\vec{n} - 2 \hat{\imath})  + \Phi (\vec{n} - 2 \hat{\jmath})  + \Phi  (\vec{n} - 2 \hat{k}) )  \right. \nonumber \\
&& \hspace*{-0.8cm}  \left. - 90 \Phi (\vec{n} ) \right] + \mathcal{O} (\Delta {\rm x}^4 ) \label{eq:discrete-laplacian}\ . \eea
This way we improve the spatial resolution of the oscillons, and hence the extraction and fitting of their shapes. The second difference with respect to \textsc{LatticeEasy}, is that we set the initial fluctuations for the scalar field and its derivative according to the following expressions [$\phi_k \equiv \phi (|\vec{k}| = k)$ in momentum space], 
 \bea 
 \phi_k &=& \frac{1}{\sqrt{2}} (a_k e^{i \theta_1} + b_k e^{i \theta_2} ) \ , \label{eq:inflaton-influc} \\
 \dot{\phi}_k &=& \frac{1}{\sqrt{2}} i \omega_k (a_k e^{i \theta_1} - b_k e^{i \theta_2} )  - H(t_i) \phi_k \ , 
\label{eq:inflaton-influc2} \eea
where $\omega_k = \sqrt{(k/a)^2 + m_{\rm eff}^2}$ is the frequency of the field mode, and $m_{\rm eff}^2 \equiv \partial^2 V / \partial \phi^2 \rvert_{\phi = \phi (t_i)}$ is the initial effective mass of the inflaton. Here, $\theta_1$ and $\theta_2$ are two random phases that vary randomly at all lattice points, and $a_k$ and $b_k$ are two real numbers that vary according to the following Rayleigh distribution,
\be P(|a_k|) = \frac{2}{\langle |a_k|^2 \rangle} |a_k| e^{-|a_k|^2/\langle |a_k|^2 \rangle} \ , \hspace{0.4cm} \langle |a_k|^2 \rangle = \frac{1}{2 a^3 \omega_k} \ , \ee
and similarly for $b_k$. This is similar to the initialization procedure of   \textsc{LatticeEasy}, but with one important change: in Eqs.~(\ref{eq:inflaton-influc})-(\ref{eq:inflaton-influc2}) there are four random numbers for each lattice point, while in \textsc{LatticeEasy} there are only three, because it imposes $a_k = b_k$. This ensures that the initial spectrum of fluctuations is Gaussian \cite{Frolov:2008hy}.

We have carried out several lattice simulations for the power-law coefficients $p=4,6,8$ in the hilltop potential (\ref{eq:scalar_potential}). In Table \ref{table:hilltop-parameters} we give the model parameters and initial field homogeneous modes used in the lattice simulations, for each value of $p$. The results we present in this section have been obtained with lattices of $N^3 = 256^3$ points and infrared cutoff $p_{\rm min} \equiv 2\pi / L = 0.18 m_{\phi}$, with $L$ the length side of the box. However, we have done additional simulations with $N^3 = 512^3$ points and different box sizes, in order to check the consistency of our results. We have simulated the system until the scale factor becomes $a \simeq a_{\rm e} \equiv 5$, which corresponds to simulation times $m_{\phi} t_{\rm e} \simeq  6870, 10450,14290$ for $p=4,6,8$ respectively. It is at these times when we extract and fit the oscillon shapes, which will be used as initial conditions for the spherically symmetric simulations in Section \ref{sec:spherical-symmetries}. The subindex `e' in the quantities above means \textit{extraction}, and we denote the time $t_{\rm e}$ as \textit{extraction time}.

\subsection{Fitting procedure for oscillon shapes}\label{sec:fitting-proc}

\begin{figure}
\centering
\includegraphics[width=0.4\textwidth]{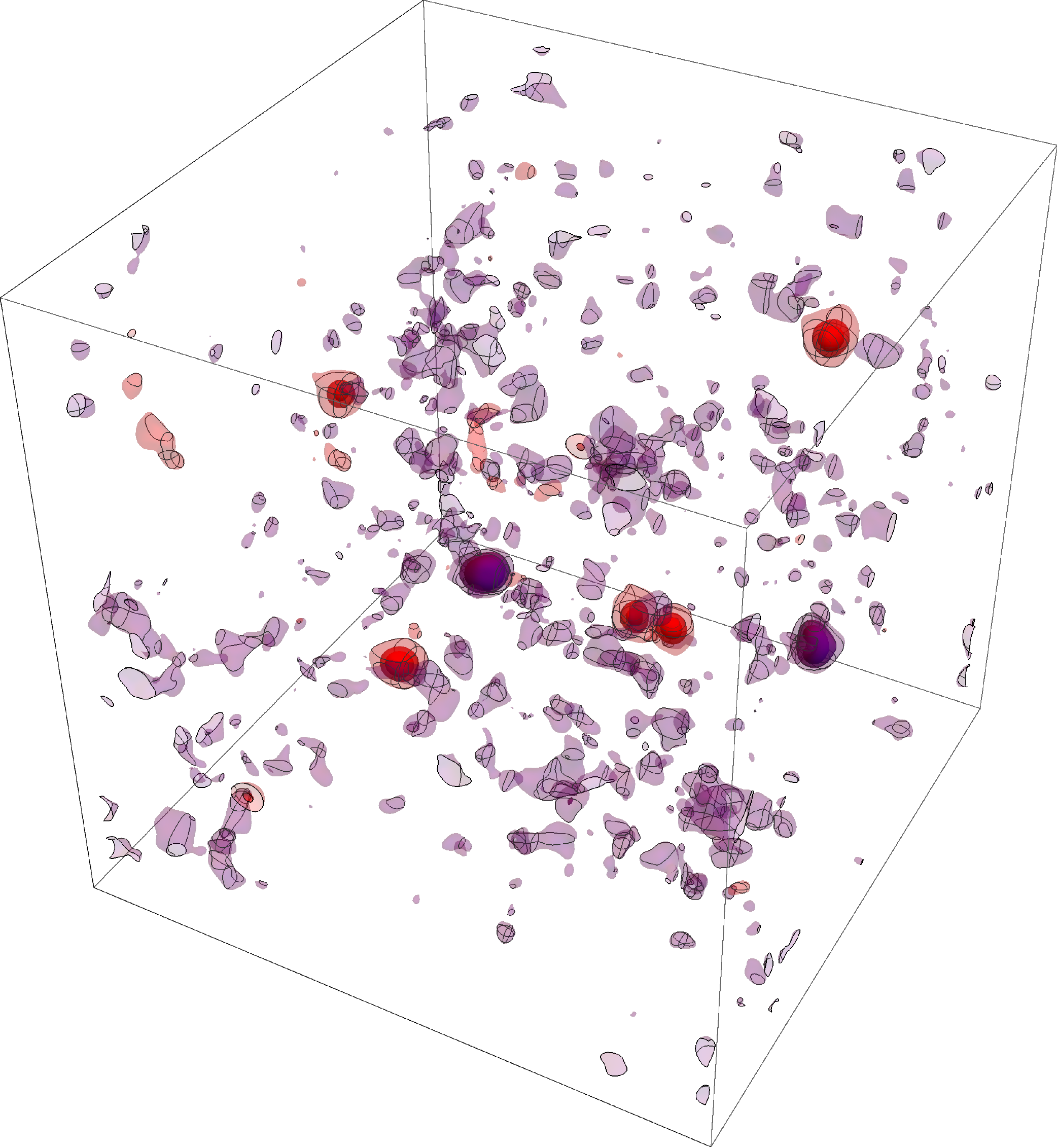}  \hspace{0.5cm}
\includegraphics[width=0.4\textwidth]{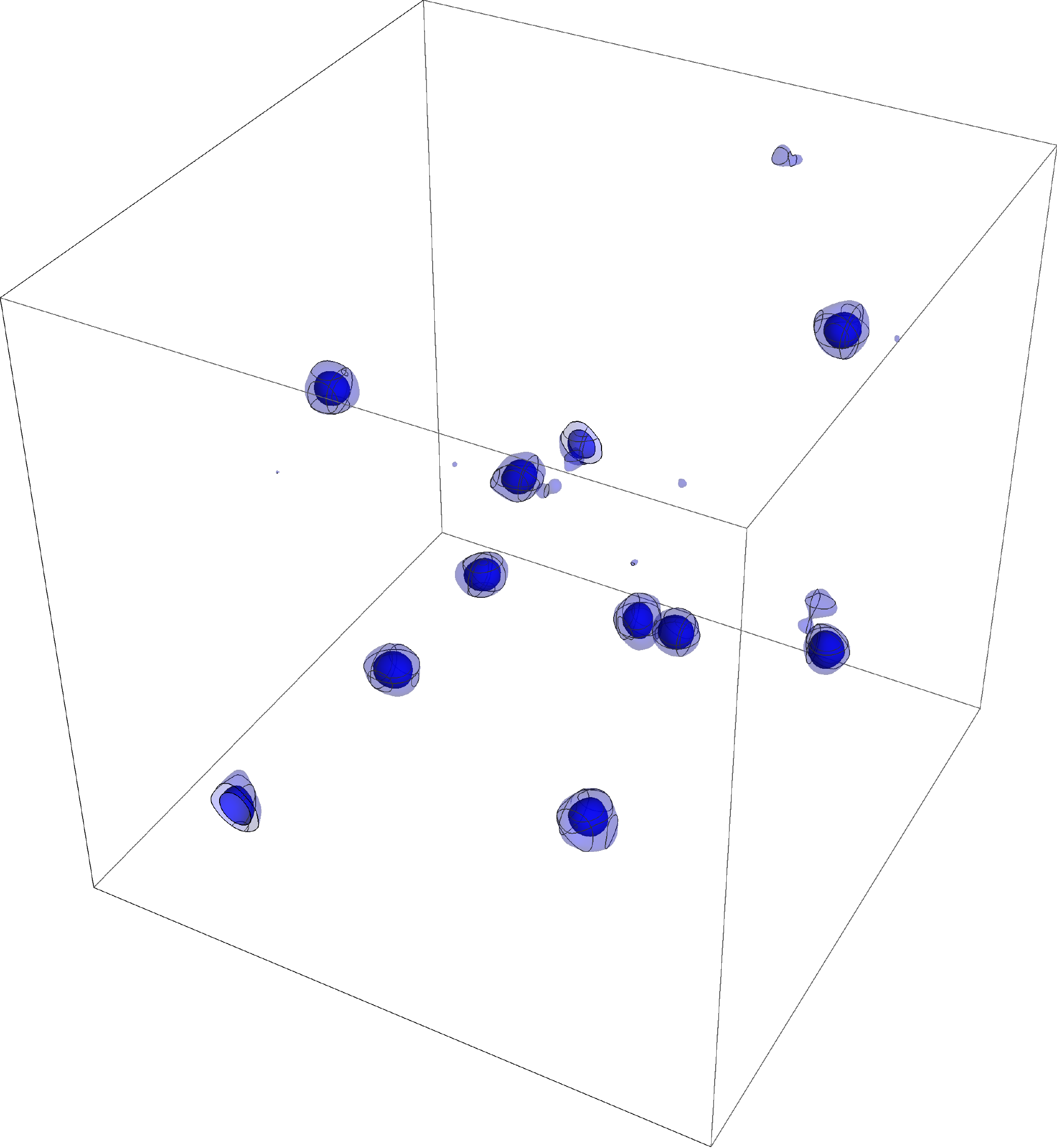}
\caption{Left panel: We show a 3-dimensional slice of the inflaton amplitude distribution, for a simulation with power-law coefficient $p=4$, and a lattice with $N^3=256^3$ points and length side $L=34.90 m_{\phi}$. The field values have been extracted at scale factor $a \simeq a_{\rm e} \equiv 5$. Red/purple surfaces indicate where the field is in the left/right part of the vacuum (i.e. $\Phi < 1$ and $\Phi > 1 $ respectively): $\Phi =$ 0.75 (dark red), 0.85 (red), 0.95 (light red), 1.03 (light purple), 1.06 (purple), and 1.09 (dark purple). Right panel: We depict the energy overdensity regions for the same lattice simulation. Dark blue surfaces indicate where the overdensity is $\Delta_{\rho} = 50$, while light blue surfaces indicate where it is $\Delta_{\rho} = 10$. According to the criteria explained in the bulk text, we observe 10 oscillons within the box, whose amplitudes and radii we parametrize.}
\label{fig:3Dslices}
\end{figure}

Let us now explain the details of our \textit{extraction} and \textit{fitting} procedure for the oscillon shapes, which we apply at time $t = t_{\rm e}$. The first step is to identify all the oscillons that are present in a lattice simulation. As we want to do many simulations, it is useful to find a generic criterium that can be systematically applied, for all the three power-law coefficients $p=4,6,8$. As known, oscillons are characterized by a local energy density much larger than the average energy present in the Universe. Hence, in our particular scenario, we have defined oscillon as the \textit{set of all adjacent points in the lattice} with local energy density $\rho$ [defined in Eq.~(\ref{eq:local-energy})] 10 times larger than the average energy density in the box, i.e.,
\be \Delta_{\rho} (\vec{n} ) \equiv \frac{ \rho (\vec{n}) }{ \langle \rho \rangle_{_{L^{3}}} } \gtrsim 10 \ , \label{eq:osc-defin}\ee 
where $\langle \dots \rangle_{L^3}$ means spatial average over the lattice. Note that, while the average energy density redshifts due to the expansion of the Universe, the oscillon energy remains approximately constant, so the oscillon energy overdensity $\Delta_{\rho}$ grows with time. The factor 10 in Eq.~(\ref{eq:osc-defin}) is just an appropriate choice to capture all oscillons within the box in all our simulations, at the time $t  = t_{\rm e}$ and for the power-law coefficients $p=4,6,8$. It is also useful to require, for at least one of the points in the oscillon, to have an energy overdensity larger than $\Delta_{\rho} > 50$. This way, we rule out regions with exceptionally large vacuum fluctuations, but too small to be considered as oscillons. With this definition, and for the lattice boxes indicated above, we typically observe $\sim 10$ oscillons in our simulations, which span approximately $\mathcal{O} (10^3)$ lattice points. As an example let us focus on Fig.~\ref{fig:3Dslices}, where we show the field amplitude and energy distributions in a lattice simulation with $p=4$, plotted when the scale factor is $a\simeq a_{e} \equiv 5$. In the left panel we observe regions where the field is either on the left hand side of the inflaton vacuum ($\Phi < 1$) or on the right hand side ($\Phi > 1$). Note that at this time, there are no regions with $\Phi < 0$. We can clearly observe several oscillons, which are in a local extremum of their oscillations at the time when the slice is plotted. Correspondingly, as seen in the right panel of Fig.~\ref{fig:3Dslices}, these regions present a very large energy overdensity $\Delta_{\rho} \gtrsim 10$, and are identified as oscillons. However, let us also remark that there are also other regions with large energy overdensity, which do not have a counterpart in the left panel: these are also oscillons, but they are in the middle of their oscillations (i.e. $\Phi \approx 1$) at the time when the slice is plotted, so they cannot be observed in an amplitude distribution.

\begin{figure}
\centering
\includegraphics[width=8cm]{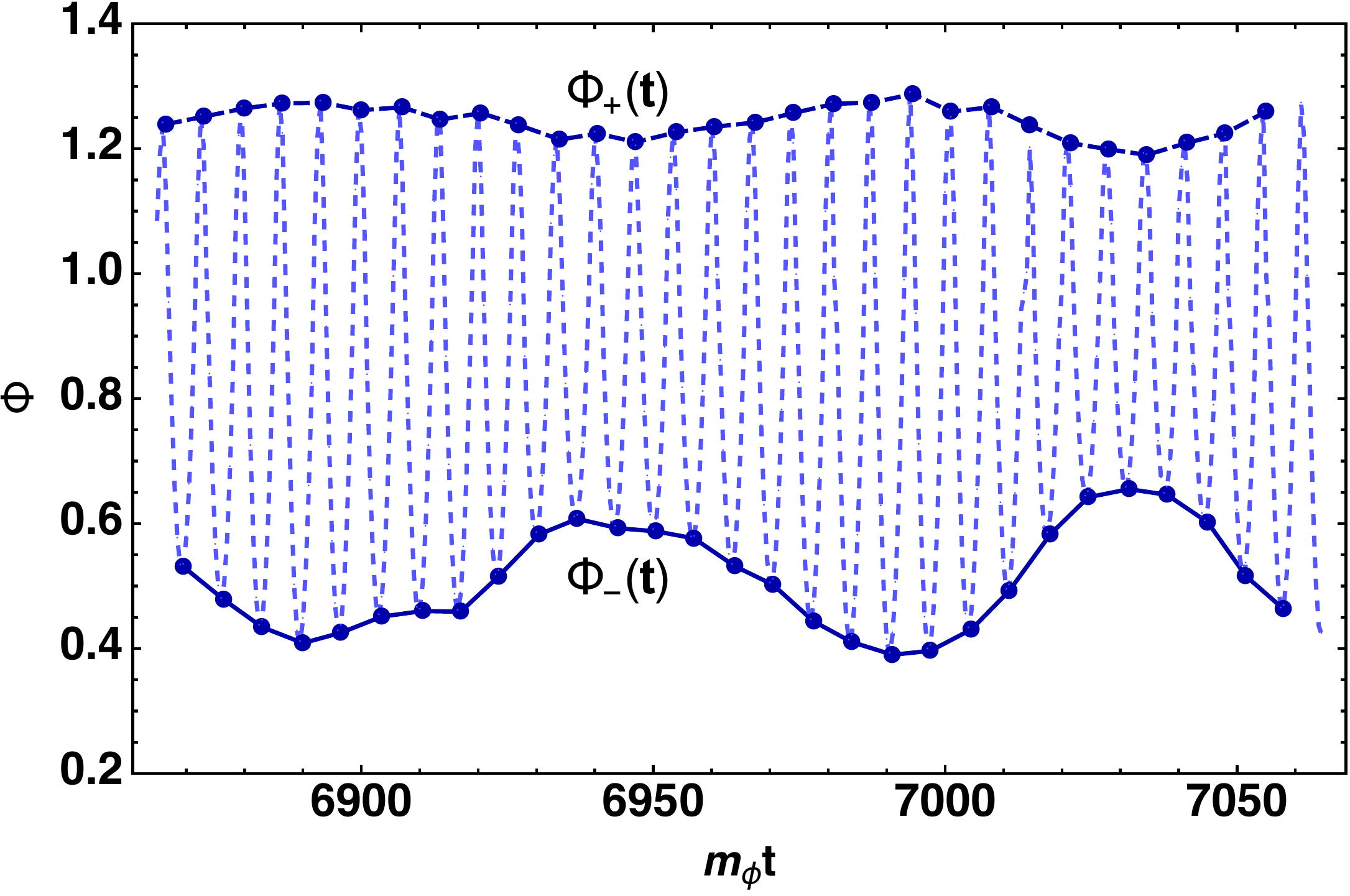}
\caption{We show the time-evolution of a single oscillon, extracted from a (3+1)-dimensional lattice simulation with $p=4$. The dashed blue line shows the field amplitude at the center of the oscillon, during several oscillations at approximately $a \simeq a_{\rm e} \equiv 5$. The thick blue lines show the upper and lower envelopes of the oscillations, $\Phi_+ (t)$ and $\Phi_- (t)$ .}
\label{fig:single-oscillon}
\end{figure}

Having identified all oscillons within the box, we want to fit their shapes. For this, let us focus on Fig.~\ref{fig:single-oscillon}, where we show the time evolution of a particular oscillon, extracted from a (3+1)-dimensional lattice simulation. The dashed line in the figure indicates the field amplitude at the center of the oscillon. It oscillates with a period $T_{\rm osc} \gtrsim 2 \pi$ in natural time units. We also plot the upper and lower envelopes of these oscillations, which we denote as $\Phi_+$ and $\Phi_-$ respectively. From this, we can define the \textit{oscillon amplitude} $A_{-}$ as
\be A_{-} (t) \equiv 1 - \Phi_{-} (t) \ .  \label{eq:amplitude_def}\ee
This quantity measures the deviation of the lower envelope with respect to the inflaton vacuum at $\Phi = 1$. The larger the amplitude of the oscillations, the larger $A_{-}$ is. Of course, we could  have also defined the oscillon amplitude in terms of the upper envelope as  $A_{+} (t) \equiv \Phi_{+} (t) - 1$. As the potential is not symmetric around the minimum, we have $A_{+} \neq A_{-}$, and in fact, $A_{-}>A_{+}$ due to the existence of a flat region of the potential in the left hand side of the vacuum. In this work we present our results for the oscillon shapes in terms of $A_-$. Note also that both $A_{-}$ and $A_{+}$ evolve with time: this is related to the existence of a \textit{breathing} mode, which we discuss in more detail in Section \ref{sec:shapes-results}.

Let us define $A_{-}^{\rm (e)} \equiv A (t_{\rm e})$ as the oscillon amplitude when the scale factor is $a \simeq a_{\rm e} \equiv 5$ [from now on, we will use the superscript ``(e)'' to indicate quantities evaluated at the extraction time $t=t_{\rm e}$]. We have been able to extract the numerical values of $A_{-}^{\rm (e)}$ for all the oscillons observed in the (3+1)-dimensional lattice simulations.  We can then fit the oscillons to the following spherically symmetric Gaussian function,
\be \Phi(\overline{r},t_{\rm e} )\,=\,1-A_{-}^{\rm (e)}\,e^{-\frac{1}{2}\left(\frac{\overline{r}}{R^{\rm (e)}}\right)^2} \ , \label{eq:Gaussian-fit} \ee
where $\bar{r}$ is the radial distance to the center of the oscillon in natural units, and $R^{(e)} \equiv R (t_{\rm e})$ is the \textit{oscillon radius} at time $t= t_{\rm e}$, obtained directly from the fit\footnote{More specifically, the fit is carried out at a time $t = t_* (\geq t_{\rm e})$, when the field at the center of the oscillon is in a local minimum of the oscillation. For example, in the case of the oscillon depicted in Fig.~\ref{fig:single-oscillon}, the extraction time is $m_{\phi} t_{\rm e} = 6861$, while the fitting time is $m_{\phi} t_* = 6869$.}. All lattice points at a radial distance $\overline{r} < 5R^{\rm (e)}$ with respect to the center of the oscillon are used for the fitting.

\begin{figure}
\centering
\includegraphics[width=0.47\textwidth]{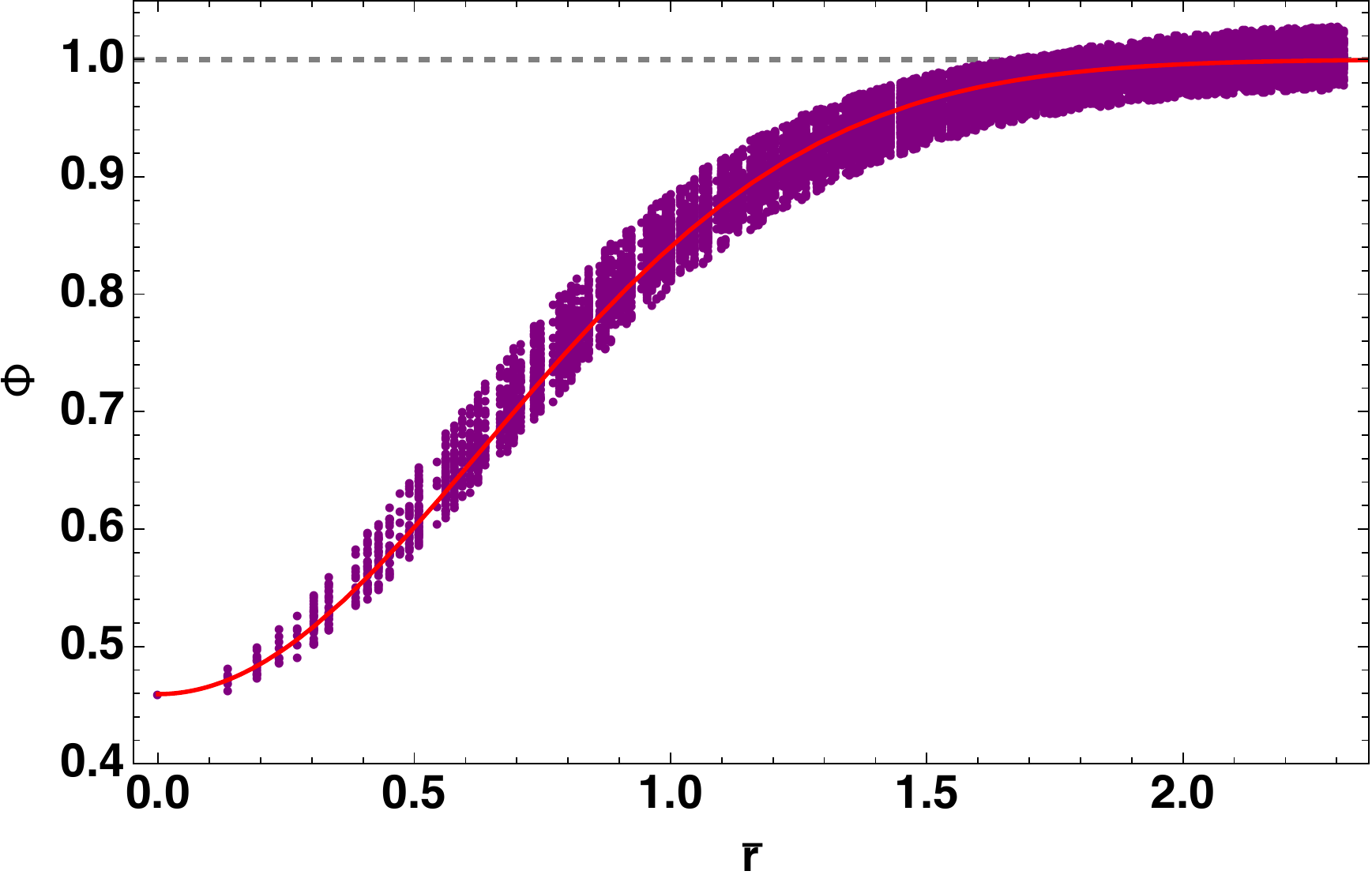} \hspace{0.3cm}
\includegraphics[width=0.47\textwidth]{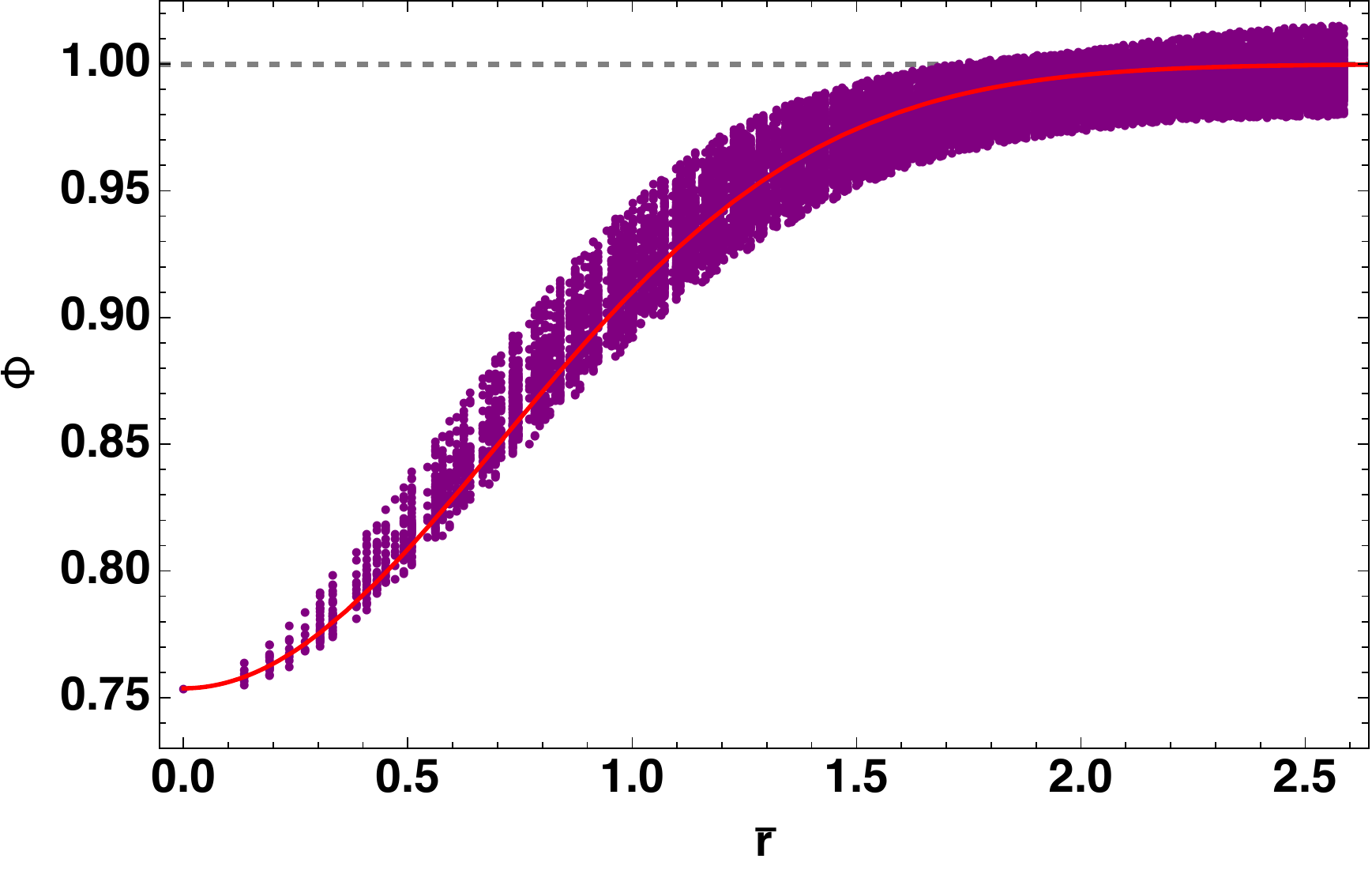} 
\caption{We show the profile of two typical oscillons for $p=4$ (left) and $p=8$ (right), extracted when the scale factor is $a \simeq a_{\rm e} \equiv 5$. Each purple dot corresponds to a particular lattice point within the oscillon: the horizontal axis shows the radial distance of that point from the center of the oscillon, and the vertical axis shows the field amplitude at that point. We have fitted the oscillons to Eq.~(\ref{eq:Gaussian-fit}), and show the fit with a red line in each of the two panels. For the oscillon in the left panel, the field value at the center of the oscillon is $\Phi_- \simeq 0.46$, so the amplitude is $A_-^{\rm (e)} \equiv 1 - \Phi_- \simeq 0.54$. On the other hand, from the fit we obtain $R^{\rm (e)} \simeq 0.64 $, so the physical radius is $R_p^{\rm (e)} \equiv a_{\rm e} R^{\rm (e)} \simeq 3.2$. For the oscillon in the right panel, we have $\Phi_- \simeq 0.75$, $A_-^{\rm (e)}\equiv1-\Phi_-  \simeq 0.25$, $R^{\rm (e)}\simeq0.7$, and $R_p^{\rm (e)} \equiv a_{\rm e} R^{\rm (e)} \simeq 3.5$. }
\label{fig:gaussian-prof}
\end{figure}

We show in Fig.~\ref{fig:gaussian-prof} two examples of this fitting procedure, for two different oscillons with $p=4$ and $p=8$. We see that the Gaussian expression (\ref{eq:Gaussian-fit}) fits remarkably well the real oscillon solution. The small difference is due mainly to two reasons: first, the real oscillons are not exactly Gaussian, and second, they are slightly asymmetric. Despite this caveat, we are able to obtain estimates for $A_{-}^{\rm (e)}$ and $R^{\rm (e)}$ for all oscillons, and to qualitatively describe the oscillon shapes with just these two quantities. We present a simple statistical analysis of these quantities in Section \ref{sec:shapes-results}.

Before proceeding to the analysis of the oscillon shapes, let us define the \textit{physical radius} of an oscillon as $R_p \equiv a R$. The radius decreases approximately as $R \sim a^{-1}$, so as we shall see in Section \ref{sec:spherical-symmetries}, $R_p$ remains quite constant during the time evolution of the oscillon. For this reason, in Section \ref{sec:shapes-results} we will quote our results for the oscillon shapes in terms of the physical radius at time $t=t_{\rm e}$, i.e. $R_p^{\rm (e)} \equiv a_e R^{\rm (e)}$.

\subsection{Results for oscillon shapes and energies}\label{sec:shapes-results}

Let us now present our results for the oscillon shapes in hilltop potentials. According to the procedure described in Section \ref{sec:fitting-proc}, we have obtained the amplitude $A_-^{\rm (e)}$ and physical radius $R_p^{(e)}$ of all oscillons observed in the (3+1)-dimensional lattice simulations, for the three power-law coefficients $p=4,6,8$. As explained above, we extract the oscillon shapes when the scale factor is $a \simeq a_e \equiv 5$, which is well after the oscillons have formed and stabilized.  In our simulations we typically observe $\sim$10 oscillons in the lattice (see Fig.~\ref{fig:3Dslices}), which is not enough to do a simple statistical analysis of the data. Therefore, in order to accumulate enough results, we have decided to run five different simulations for each individual value of $p$. The lattice and model parameters are the same for all five simulations, but the seeds that generate the initial random fluctuations of the scalar field [see Eqs.~(\ref{eq:inflaton-influc})-(\ref{eq:inflaton-influc2})] are different. As the initial field distribution is different in each simulation, so is the number and shape of the oscillons.

\begin{figure}
\centering
\includegraphics[width=0.3\textwidth]{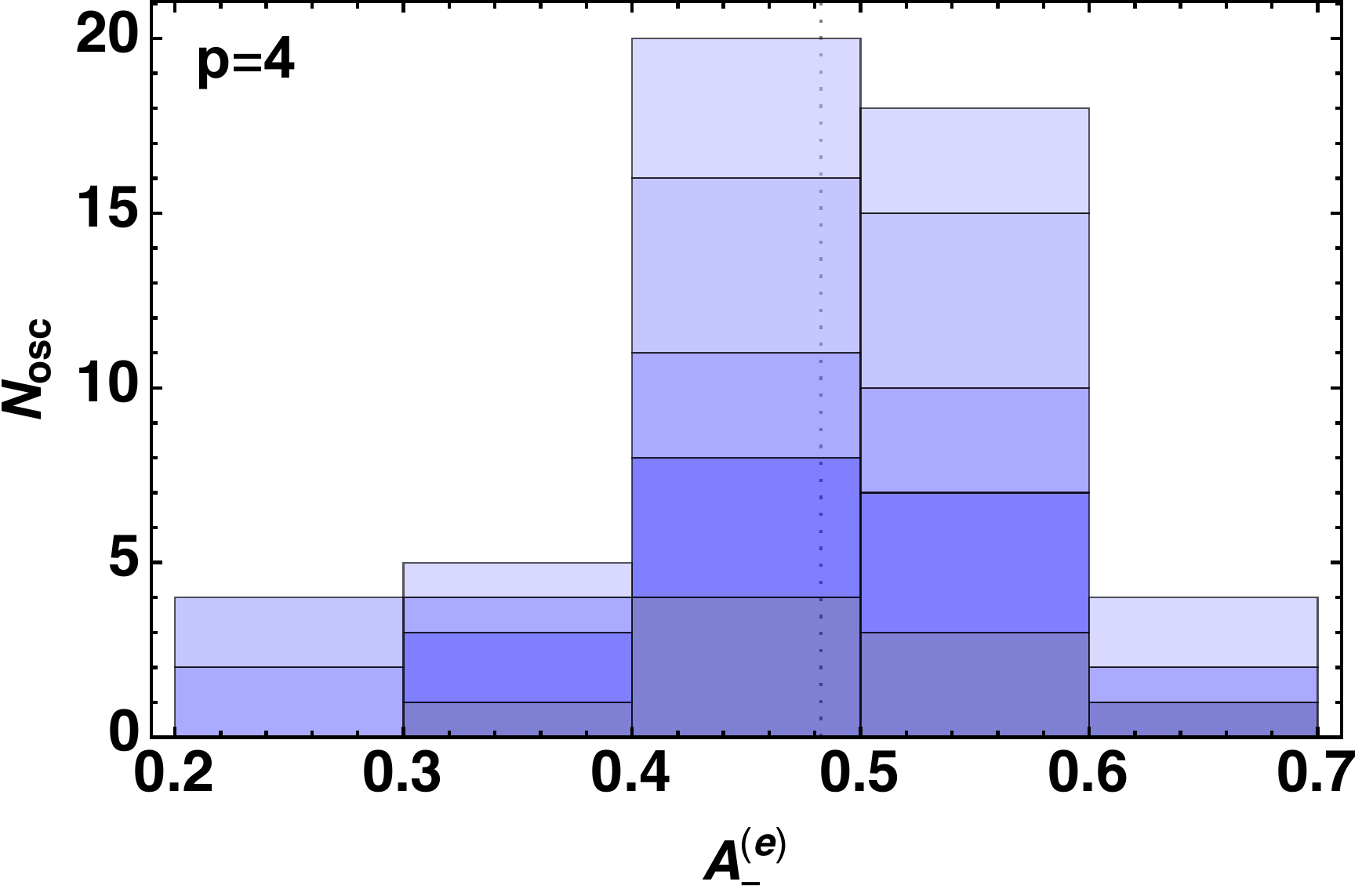} \hspace{0.15cm}
\includegraphics[width=0.3\textwidth]{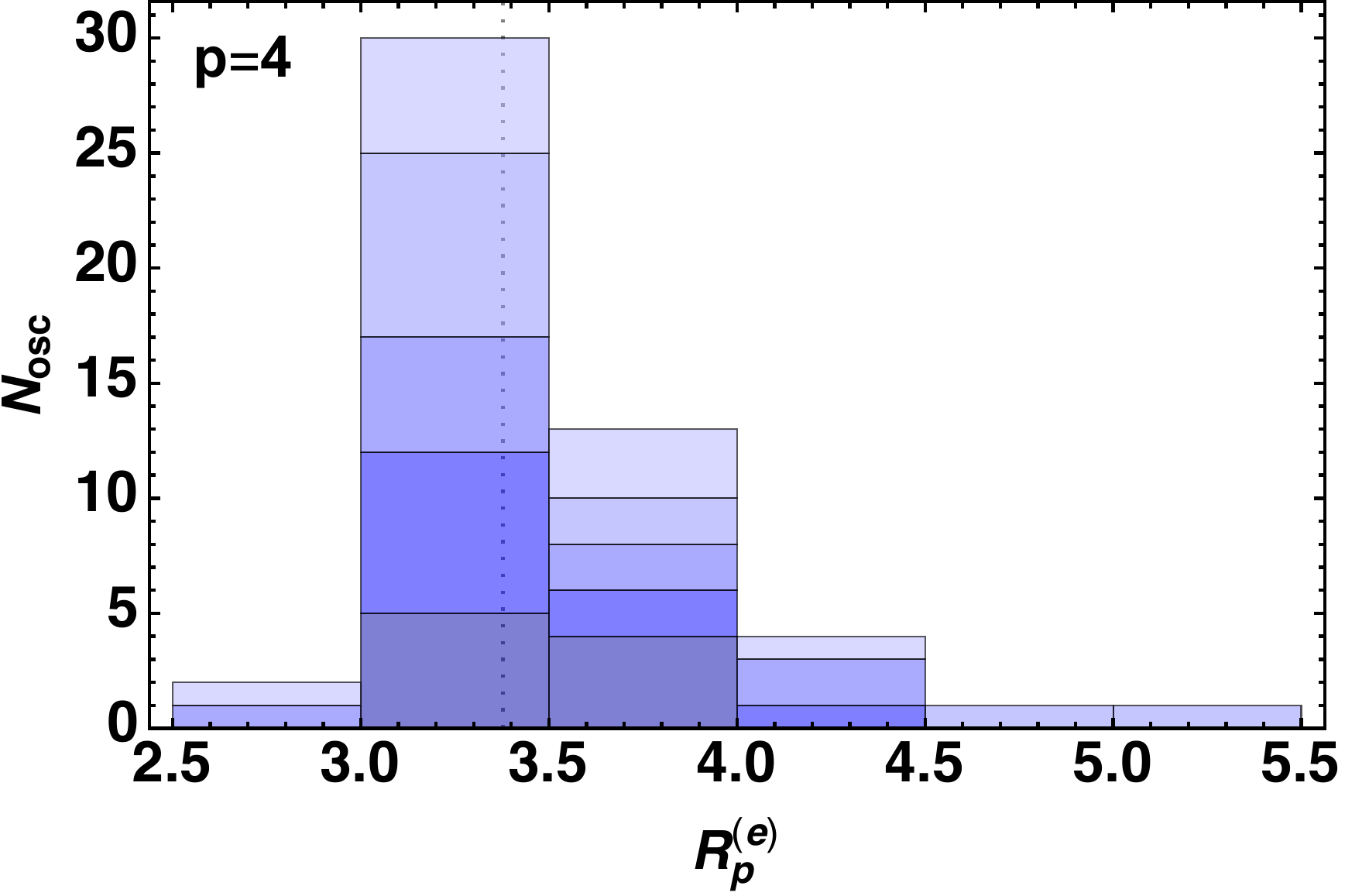}  \hspace{0.15cm}
\includegraphics[width=0.3\textwidth]{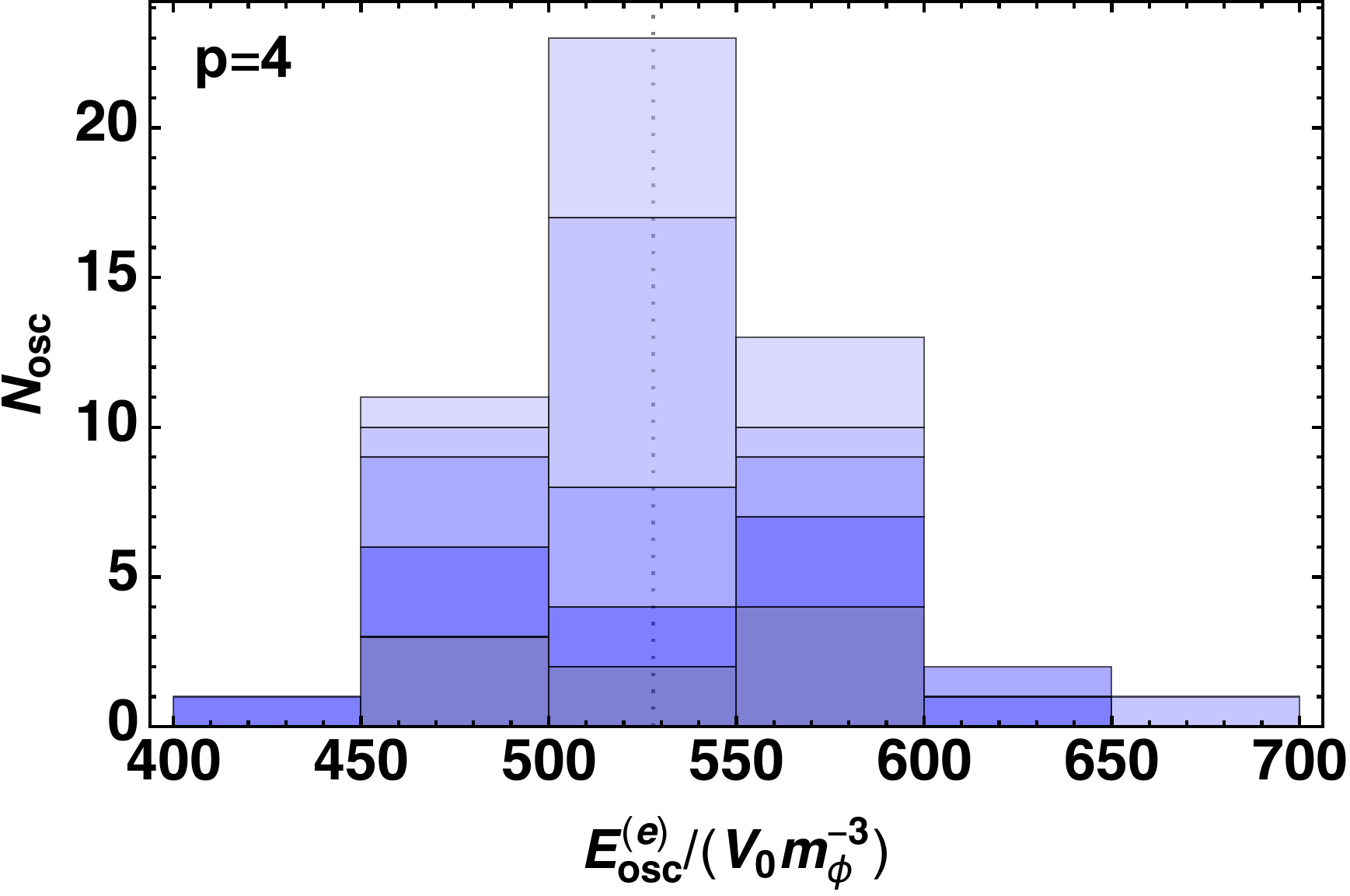} \\ \vspace{0.3cm}
\includegraphics[width=0.3\textwidth]{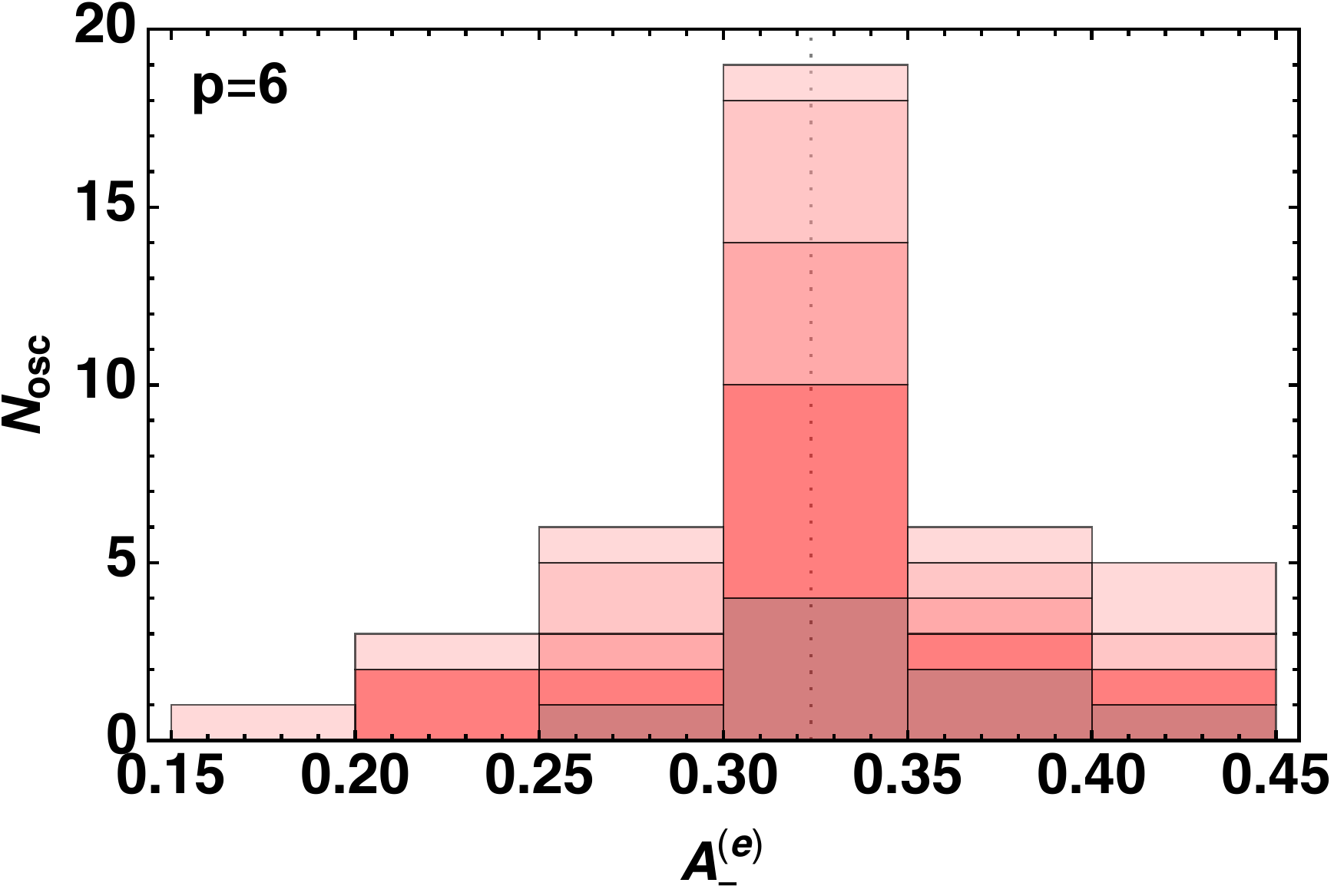} \hspace{0.15cm}
\includegraphics[width=0.3\textwidth]{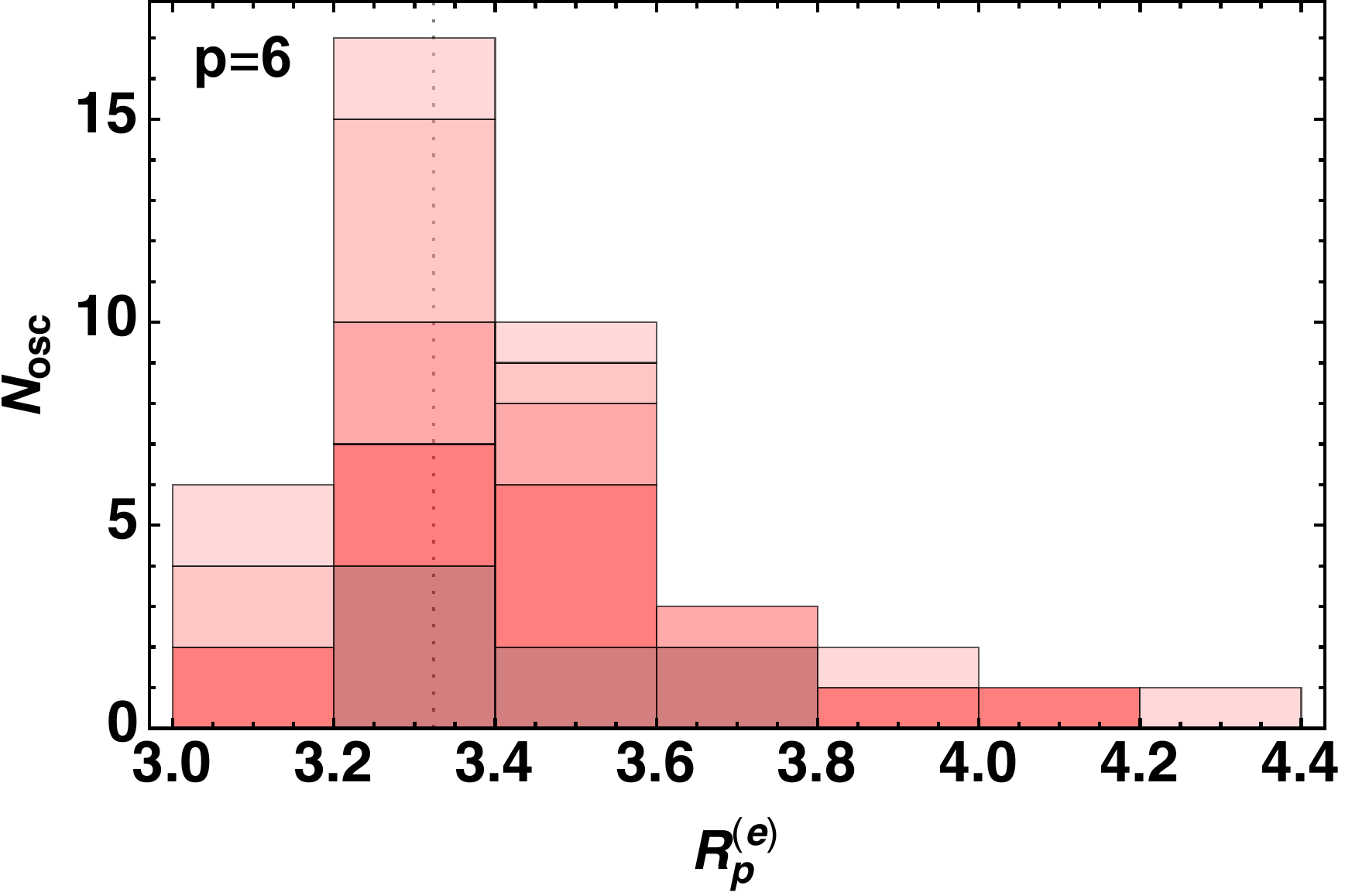}  \hspace{0.15cm}
\includegraphics[width=0.3\textwidth]{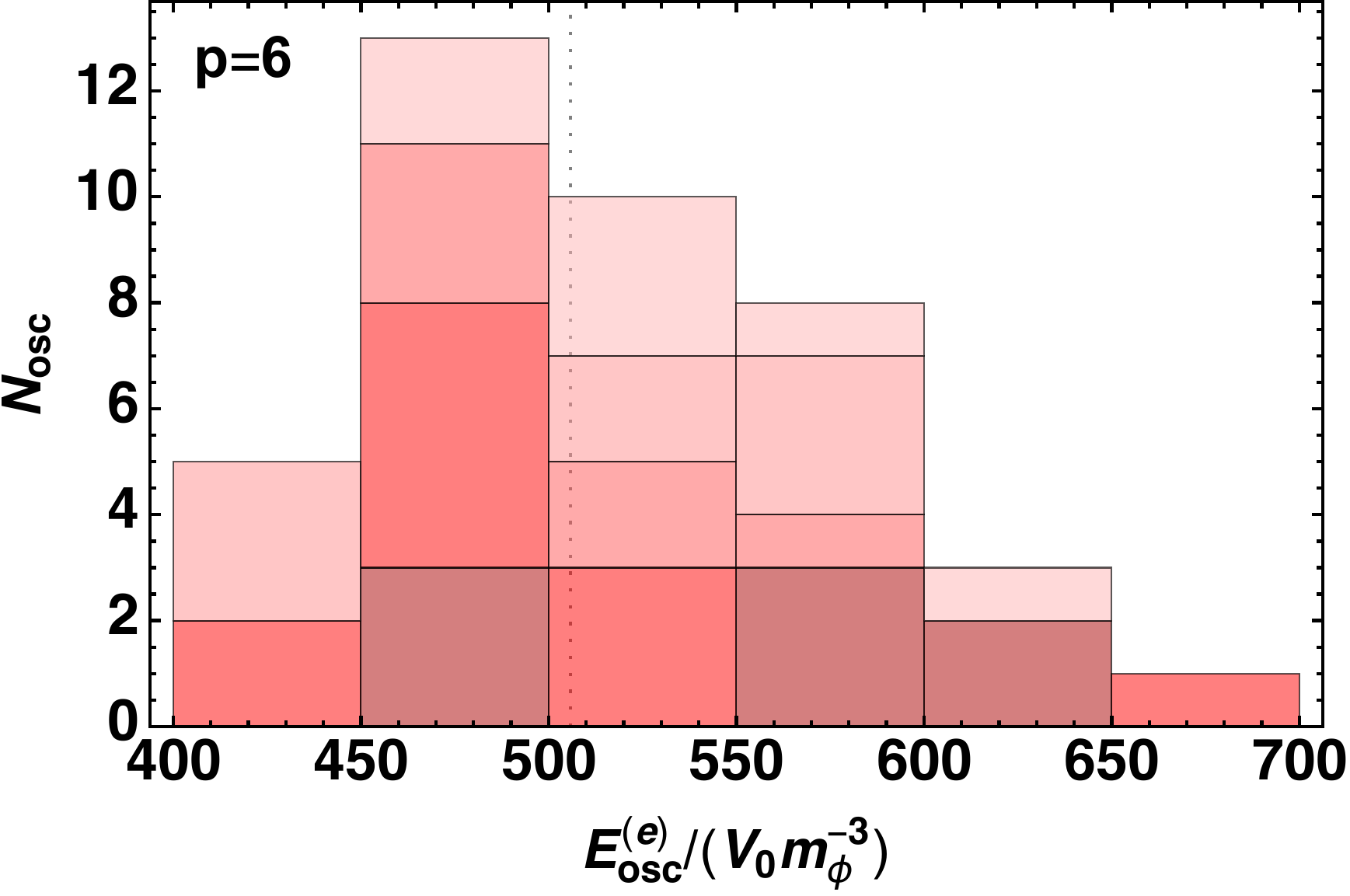} \\ \vspace{0.3cm}
\includegraphics[width=0.3\textwidth]{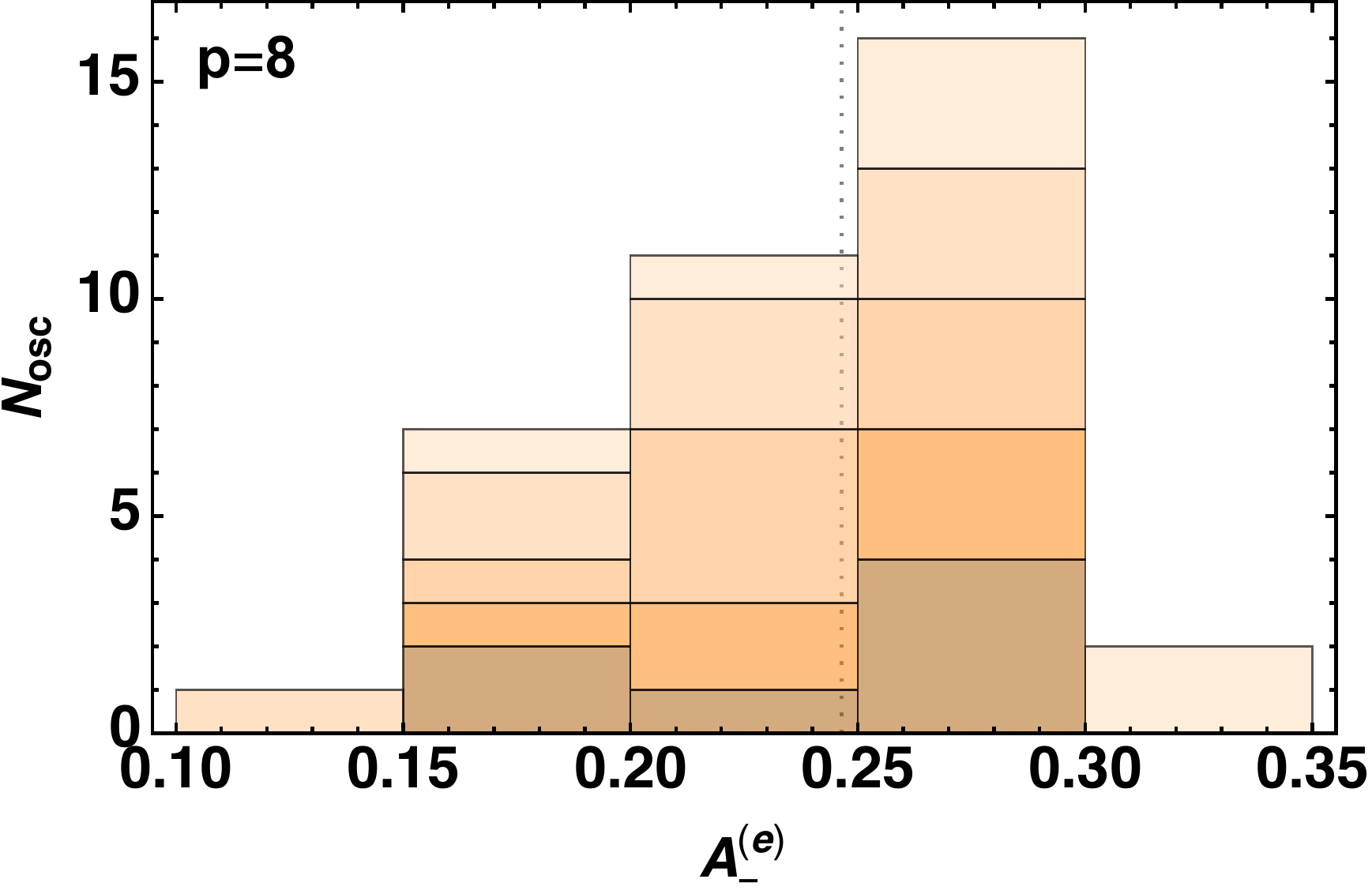} \hspace{0.15cm}
\includegraphics[width=0.3\textwidth]{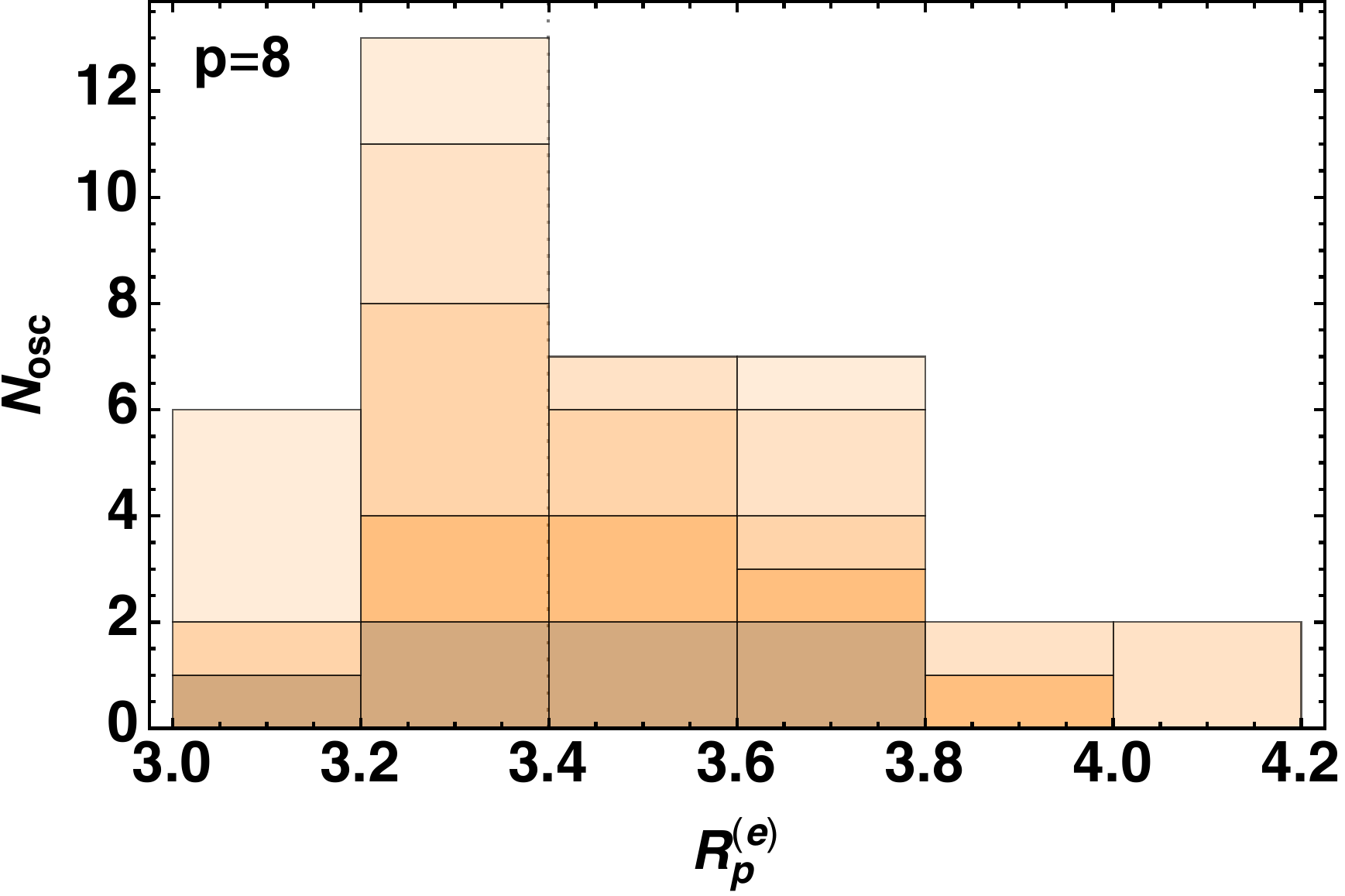}  \hspace{0.15cm}
\includegraphics[width=0.3\textwidth]{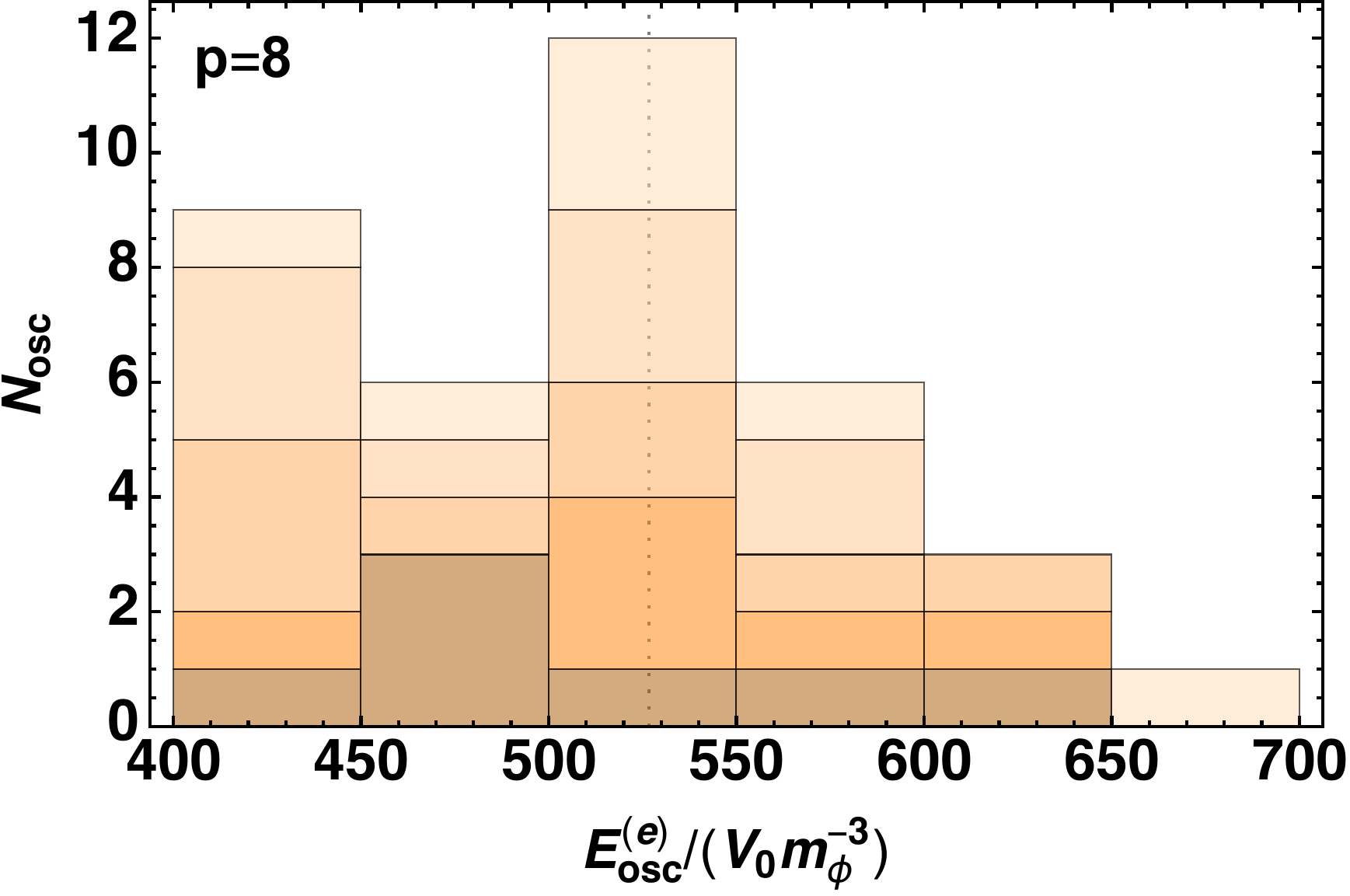} \\ \vspace{0.3cm}
\caption{These histograms show the distribution of oscillon amplitudes, radii, and energies, in hilltop potentials with power-law coefficients $p=4$ (top panels), $p=6$ (middle panels), and $p=8$ (bottom panels). Data has been extracted from the (3+1)-dimensional lattice simulations when the scale factor is $a \simeq a_e \equiv 5$. In order to have enough statistics, we have done five different lattice simulations for each power-law coefficient, with different random initial conditions (see bulk text). Each colour in each panel corresponds to a different lattice simulation. The dashed vertical lines show the median of each distribution.}\vspace{1cm}
\label{fig:oscillonshapes}

\includegraphics[height=3.2cm]{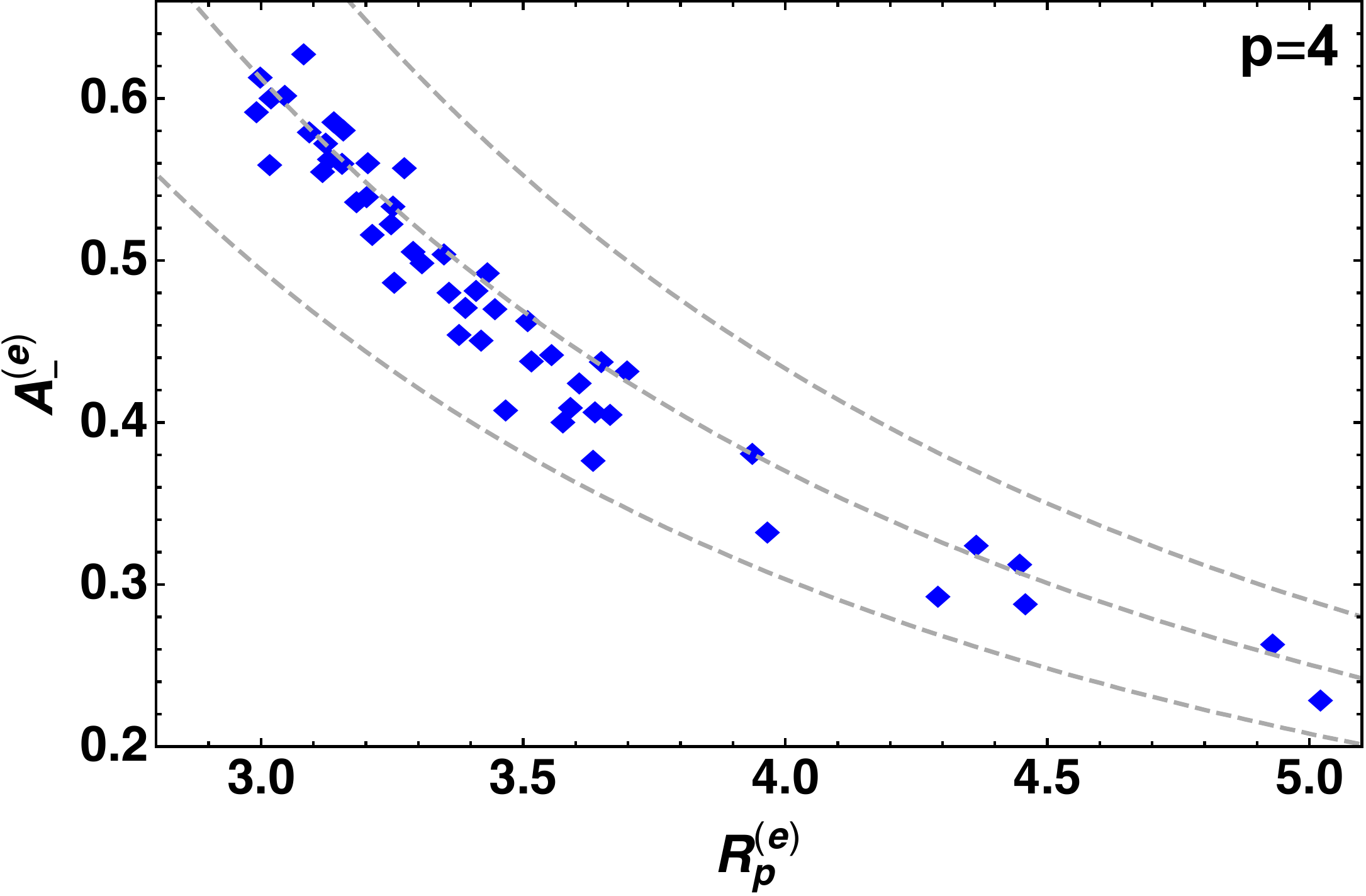} \hspace{0.1cm}
\includegraphics[height=3.2cm]{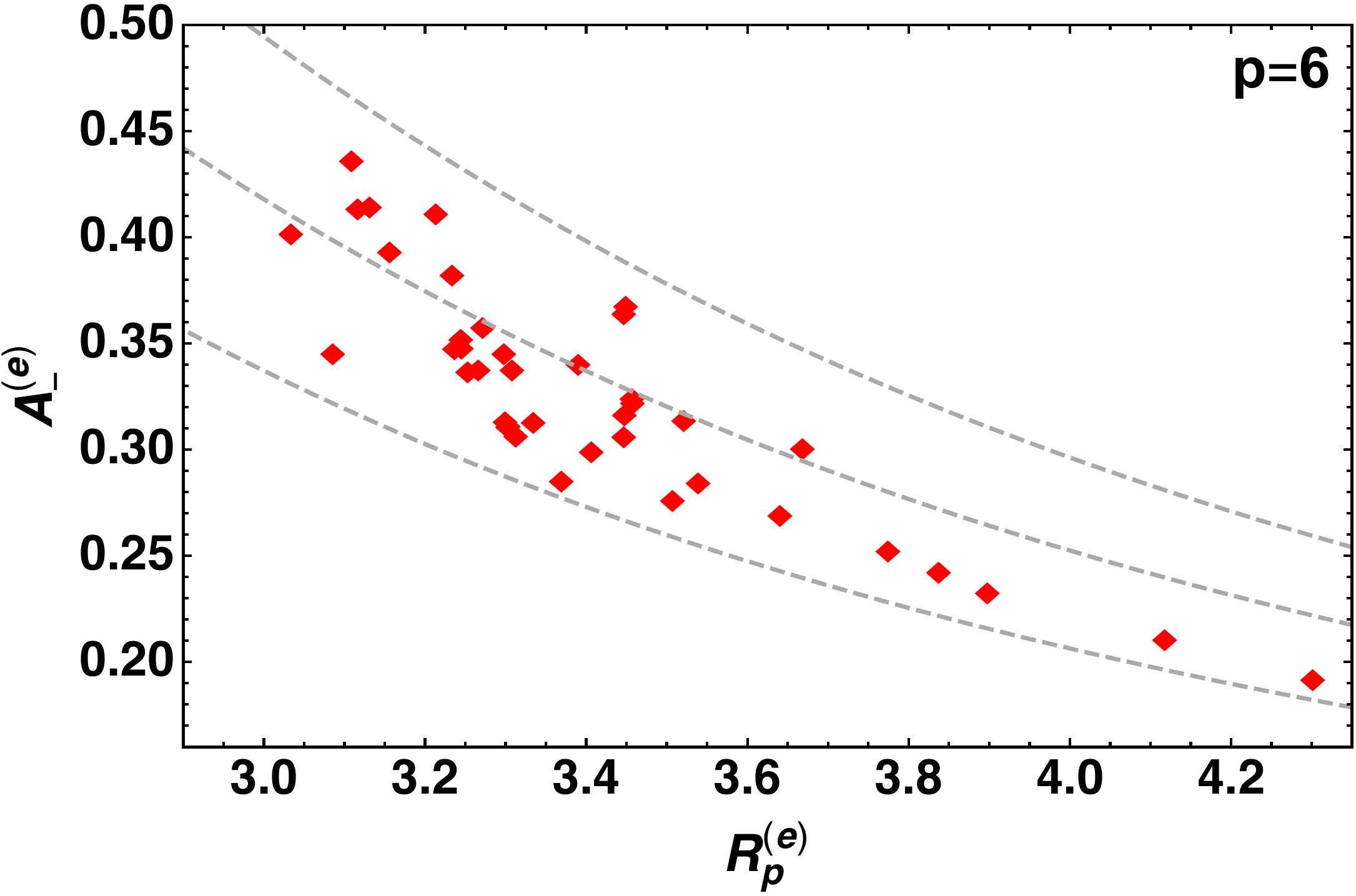} \hspace{0.1cm}
\includegraphics[height=3.2cm]{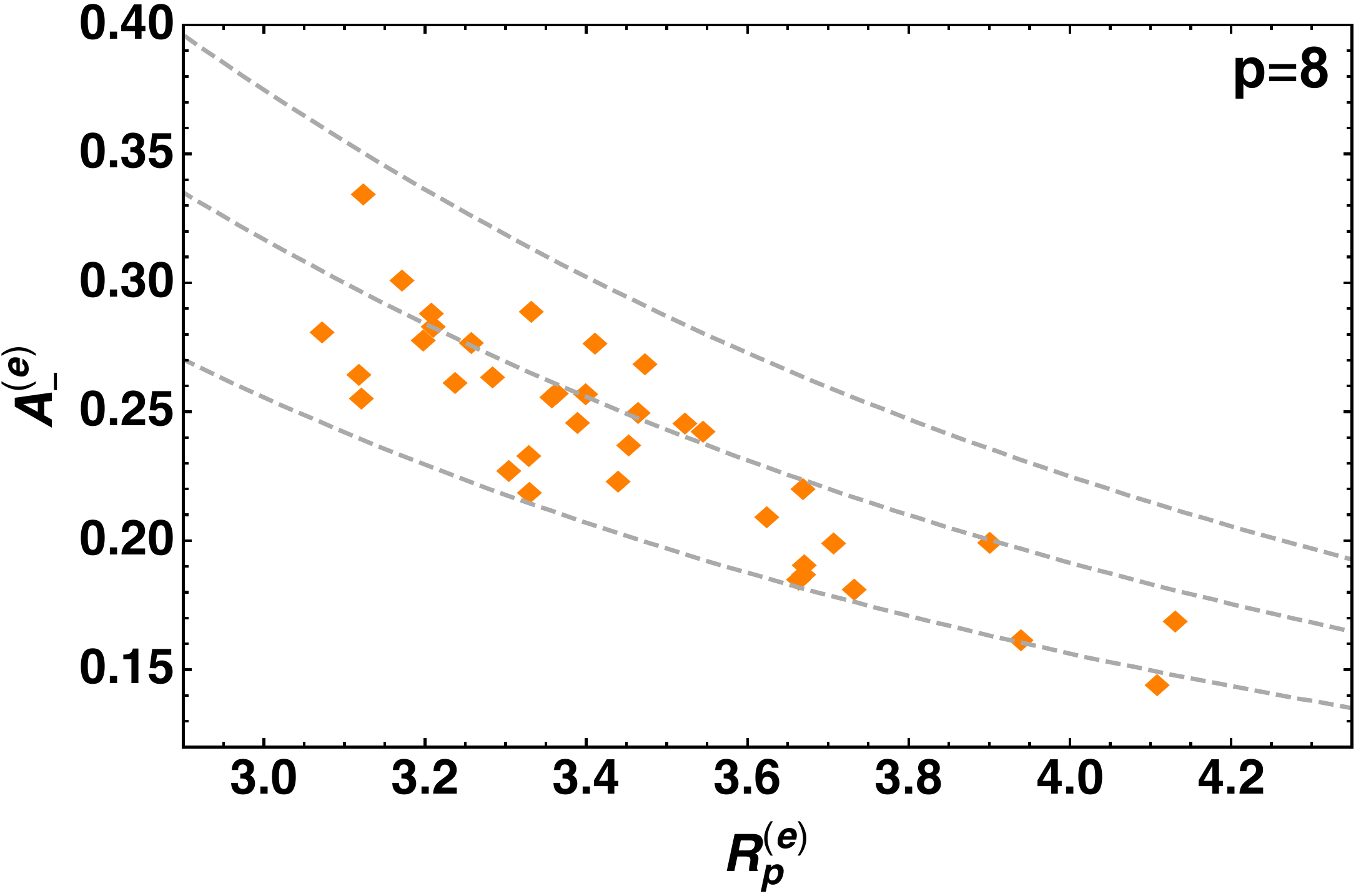}
\caption{We show the correlation between the amplitudes and radii of the oscillons extracted from the (3+1)-dimensional lattice simulations, for $p=4$ (left), $p=6$ (middle), and $p=8$ (right). The data is the same as in Fig.~\ref{fig:oscillonshapes}. The dashed gray lines indicate the combinations of amplitude and radii that give rise to oscillon energies $E_{\rm osc}^{\rm (e)} /(V_0 m_{\phi}^{-3}) = 375$ (lower line), $513$ (middle line), and $650$ (upper line), defined in Eq.~(\ref{eq:oscenergy_lat}).}
\label{fig:ampradius}
\end{figure}

Our results for the oscillon shapes are summarized in Fig.~\ref{fig:oscillonshapes}. There we plot a set of nine histograms, which show the distribution of oscillon amplitudes $A_-^{\rm (e)}$, physical radii $R_p^{\rm (e)}$, and energies $E_{\rm osc}^{\rm (e)}$ [defined below in Eq.~(\ref{eq:oscenergy_lat})], for the three power-law coefficients $p=4$ (top panels), $6$ (middle panels), and $8$ (bottom panels). The distributions show the \textit{accumulated} results, where each color in each panel indicates one of the five different lattice simulations. We also indicate the median of each distribution with a vertical dashed line. From these distributions, we can compute that 80\% of the measured amplitudes and radii are within the following intervals,
\be A_{-}^{\rm (e)}  \in \left\{ \begin{array}{ll}
        \, [0.32,0.59] \ , \,\, \hspace{1cm} & \text{if}  \,\,\, p = 4 \, , \vspace*{2mm}\\ 
        \, [0.24,0.40] \ , \,\, \hspace{1cm} & \text{if}  \,\,\, p = 6 \, , \vspace*{2mm}\\ 
        \, [0.18,0.29] \ , \,\, \hspace{1cm} & \text{if}  \,\,\, p = 8 \, , \\ 
        \end{array} \right. \hspace{1cm}
        R_{p}^{\rm (e)}  \in \left\{ \begin{array}{ll}
        \, [3.1,4.3] \ , \,\, \hspace{1cm} & \text{if}  \,\,\, p = 4 \, , \vspace*{2mm}\\ 
        \, [3.1,3.8] \ , \,\, \hspace{1cm} & \text{if}  \,\,\, p = 6 \, , \vspace*{2mm}\\ 
        \, [3.1,3.9] \ , \,\, \hspace{1cm} & \text{if}  \,\,\, p = 8 \, , \label{eq:shape-results}\\ 
        \end{array} \right.
\ee
where the values of $R_p^{\rm (e)}$ are in units of $m_{\phi}^{-1}$. We observe that as we increase $p$, the amplitude of the typical oscillons gets smaller. This is due to the shape of potential: as seen in Fig.~\ref{fig:hilltop-potential}, for larger $p$ the potential around the minimum is steeper, meaning that the field value which separates positive-curvature and negative-curvature regions of the potential gets closer to the minimum at $\Phi \approx 1$. On the other hand, the typical extracted values for the physical radius $R_p$ (middle panels) are quite independent of $p$, and in fact, the most observed outcome is approximately $R_p \sim 3.2$ for the three power-law coefficients. However, let us remark the range of extracted radii is larger for $p=4$ than for $p=6,8$. On the other hand, we show in Fig.~\ref{fig:ampradius} the correlation between the different amplitudes and radii of the oscillons, for $p=4,6,8$. We observe that these quantities are strongly anti-correlated: smaller oscillon radii imply larger oscillon amplitudes.

Let us define the \textit{oscillon energy} $E_{\rm osc}$ as the integral of the local energy density (\ref{eq:local-energy}) over the oscillon volume. Assuming spherical symmetry, it can be written as
\be E_{\rm osc}^{\rm (e)} \equiv \int_{V_{\rm osc}} d V \rho \simeq 4 \pi a^3  V_0 m_{\phi}^{-3}  \int_{\bar{r}=0}^{\bar{r} = 5 R^{(e) \rm }} d \bar{r} \bar{r}^2  \left( p^2 \Phi'^2 + \frac{p^2}{a^2} ( \partial_{\bar{r}} \Phi)^2 + (1 - \Phi^p)^2 \right) \ , \label{eq:oscenergy_lat} \ee
where $d V \equiv 4 \pi a^3 m_\phi^{-3} \bar{r}^2 d \bar{r}$ is the volume differential, and the quantity has been evaluated at time $t=t_{\rm e}$. For the Gaussian solution Eq.~(\ref{eq:Gaussian-fit}), the energy density goes to $\rho \rightarrow 0$ as $\bar{r} \rightarrow \infty$, so the upper limit of the integral $\bar{r} = 5 R^{(e) \rm }$ captures all the relevant energetic contributions. In the lattice, the oscillon is computed by summing the local energies at all lattice points within radial distance $\bar{r} < 5R^{\rm (e)}$ with respect to the center of the oscillon.

We show in the right panels of Fig.~\ref{fig:oscillonshapes} the distribution of oscillon energies, extracted from the (3+1)-dimensional lattice simulations. Remarkably, we observe that the distribution of energies [or more specifically, of the ratio $E_{\rm osc}^{\rm (e)} /(V_0 m_{\phi}^{-3})$] is quite similar for the three power-law coefficients $p=4,6,8$. In particular, we get that 80\% of the oscillons have energies within the ranges
\be \frac{E_{\rm osc}^{\rm (e)}}{V_0 m_{\phi}^{-3}}  \in \left\{ \begin{array}{ll}
        \, [470,580] \ , \,\, \hspace{1cm} & \text{if}  \,\,\, p = 4 \, , \vspace*{2mm}\\ 
        \, [440,600] \ , \,\, \hspace{1cm} & \text{if}  \,\,\, p = 6 \, , \vspace*{2mm}\\ 
        \, [420,610] \ , \,\, \hspace{1cm} & \text{if}  \,\,\, p = 8 \, , \\ 
        \end{array} \vspace*{2mm}\right. \label{eq:oscillon-energies} \ee
and that \textit{all} oscillons have energies between $E_{\rm osc}^{\rm (e)} /(V_0 m_{\phi}^{-3}) \in [400,700]$. Let us recall that the value of the potential energy at the plateau $V_0$ and the effective mass $m_{\phi}$ are different for each value of $p$ (see Table \ref{table:hilltop-parameters}), so the oscillon energies do depend implicitly on these parameters though. In Fig.~\ref{fig:ampradius} we have also depicted, with dashed lines, the combinations of amplitude and radius that correspond to oscillon energies $E_{\rm osc}^{\rm (e)} /(V_0 m_{\phi}^{-3}) = 375$, $513$, and $650$. We observe that remarkably, all extracted oscillons have relatively similar energies, and that the distribution of observed oscillons arranges along these lines of equal energy\footnote{According to Fig.~\ref{fig:oscillonshapes}, there is a small number of oscillons that have energies between $650 < E_{\rm osc}^{\rm (e)}/(V_0 m_{\phi}^{-3}) < 700$, but in Fig.~\ref{fig:ampradius} all oscillons have energies less than $E_{\rm osc}^{\rm (e)}/(V_0 m_{\phi}^{-3}) < 650$. The reason of this apparent contradiction is that the energies in Fig.~\ref{fig:ampradius} are computed directly from the lattice (by summing the local energies of all lattice points within the oscillon), where oscillons are slightly asymmetric, while the dashed lines in Fig.~\ref{fig:oscillonshapes} are computed with Eq.~(\ref{eq:oscenergy_lat}), which assumes spherical symmetry.}.

\begin{figure}
\centering
\includegraphics[width=9cm]{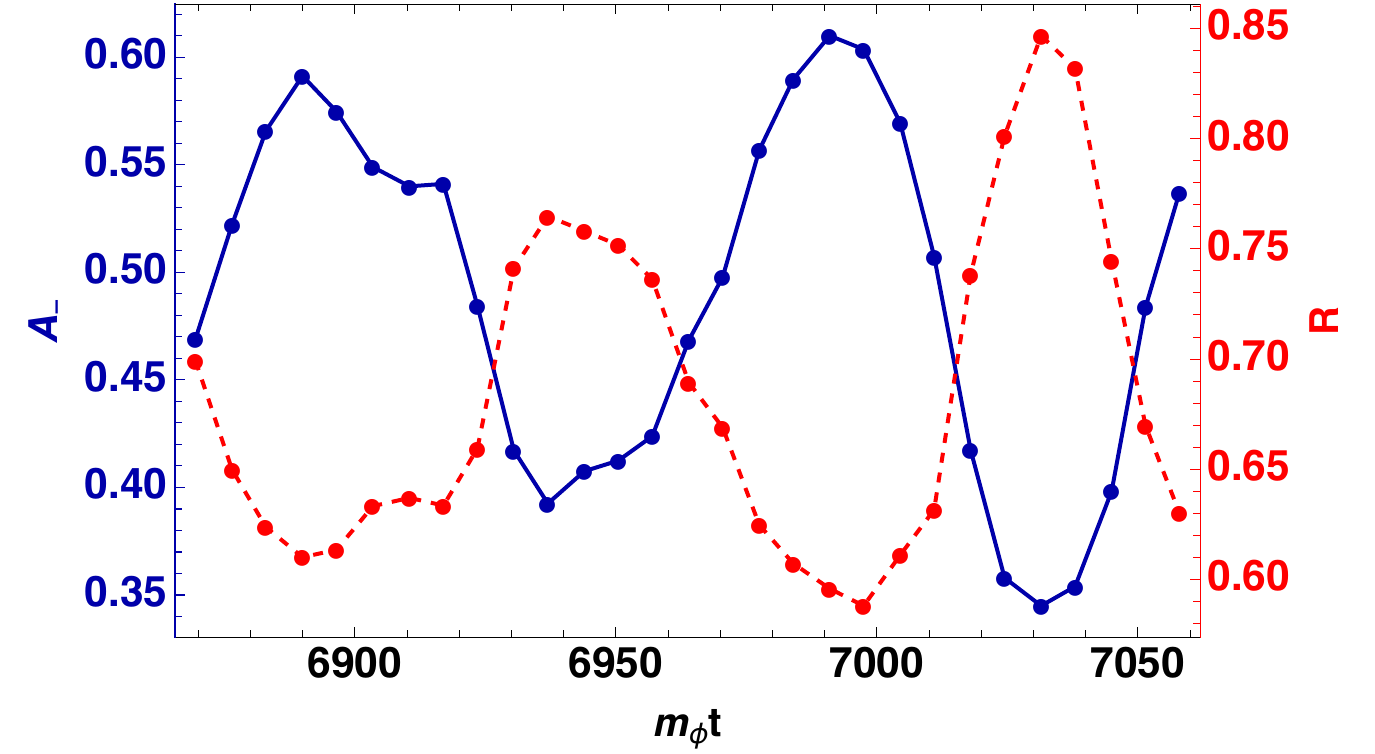}
\caption{We show the breathing mode for the same oscillon depicted in Fig.~\ref{fig:single-oscillon}. Data has been extracted from a lattice simulation in 3+1 dimensions with $p=4$. The blue continuous line shows the oscillon amplitude $A_{-} (t)$ as a function of time, defined in Eq.~(\ref{eq:amplitude_def}). The red dashed line shows the oscillon radius $R(t)$, measured by fitting the field solution to Eq.~(\ref{eq:Gaussian-fit}). We observe that both quantities oscillate in an opposite manner: the oscillon contracts and expands alternately, and correspondingly, the oscillon amplitude increases and decreases. This breathing mode is generically observed in other oscillons, but the amplitude and frequency of the breathing changes.}
\label{fig:breathing}
\end{figure}

The results presented in Figs.~\ref{fig:oscillonshapes} and \ref{fig:ampradius} allow to qualitatively characterize some of the most important properties of oscillons in hilltop potentials (amplitudes, radii, and energies). We have measured such properties well after oscillons have stabilized, at $a \simeq a_{\rm e} \equiv 5$. However, let us remark that such properties do not remain stable, but fluctuate in time. In particular, we have observed that both the amplitude and the radius of the oscillons oscillate with time in an opposite manner, which indicates the existence of a \textit{breathing mode}. Let us  show an example of this in Fig.~\ref{fig:breathing}. There we compare, for a single oscillon, the evolution of the oscillon amplitude $A_- (t) $ and radius $R(t)$ as a function of time, for some tens of oscillations of the inflaton homogeneous mode. We observe that during this time, the oscillon amplitude changes within the interval $0.34 < A_- (t) < 0.61$, while the radius changes between the range $0.59 < R (t) < 0.84$. The key aspect is that both quantities oscillate in an opposite manner: the larger the amplitude is, the smaller the radius becomes, and the other way around. In other words, the oscillons expand and contract, with an oscillation period $T_{\rm bth}$, such that $T_{\rm bth} > T_{\rm osc} \gtrsim 2 \pi$ in natural units (in the case of the figure, $T_{\rm bth} \approx 15 T_{\rm osc}$). Despite this breathing mode, the oscillon energy remains approximately constant, as we shall see in Section \ref{sec:spherical-symmetries}.

We have observed a similar breathing mode in other oscillons, both in the (3+1)-dimensional lattice simulations, as well as in the spherically symmetric simulations that we present in Section \ref{sec:spherical-symmetries}. However, the amplitude of the breathing is very dependent on the particular oscillon shape: for some the breathing is rather large, while for others it is almost absent. Let us also emphasize that, due to this breathing mode, the oscillons shapes extracted at $a \simeq a_{\rm e} \equiv 5$, and presented in Figs.~\ref{fig:oscillonshapes} and \ref{fig:ampradius} do not remain constant, but change in time. Hence, if we had extracted the shapes at a later time, these Figs.~would definitely be different. However, as the data corresponds to a relatively large number of oscillons, we expect to have captured qualitatively well the typical oscillon shapes.

\section{Oscillon lifetimes from lattice simulations with spherical symmetry}\label{sec:spherical-symmetries}
The lifetime of oscillons may depend on diverse factors, like the form of the scalar potential or the coupling of the field to secondary species. In this section, we study the lifetime of oscillons in inflationary hilltop potentials, assuming that the inflaton is decoupled from (or weakly coupled to) other species (see e.g.\ \cite{Antusch:2015ziz}). Moreover, we will focus only on the classical dynamics of the inflaton field. Quantum effects can  have a relevant impact on the longevity of oscillons \cite{Hertzberg:2010yz,Saffin:2014yka,Olle:2019skb}, but we ignore such effects in this work.

In this section we solve numerically the inflaton equation of motion for a \textit{single oscillon}. Oscillons evolve towards a more and more spherically symmetric shape as time goes on. Hence, here we will assume that oscillons are \textit{exactly symmetric} at time $t=t_{\rm e}$, and ignore possible contributions of the asymmetry to the oscillon lifetime. Moreover, we will approximate the oscillons at this time by a Gaussian function with different amplitudes and radii. In hilltop models, oscillons are generically found to be orders of magnitude smaller than the size of the horizon, so we can neglect the friction term in Eq.~(\ref{eq:eom3}). We can then write the field \textit{eom} (\ref{eq:eom3}) as
\be \frac{\partial^2\Phi}{\partial\overline{t}^2}   - \frac{1}{a^2} \left( \frac{\partial^2\Phi}{\partial \bar{r}^2} + \frac{2}{\bar{r}}\frac{\partial\Phi}{\partial \bar{r}} \right) + \frac{1}{p} \Phi^{p-1} (\Phi^p - 1) = 0 \ , \label{eq:rad-eom1} \ee
where $\Phi \equiv \Phi (\bar{r})$. As we shall see below, the physical radius of the oscillon $R_{p} \equiv a R$ does not change much during the time evolution of the system. Hence, when solving numerically the field equation, it is very useful to write it in terms of a new radial coordinate $\bar{r}_p \equiv a \bar{r} = a m_{\phi} r$, so that we ensure that the relevant length scales are always captured in the lattice. The field equation is then written as
\be \frac{\partial^2\Phi}{\partial\overline{t}^2}   -  \frac{\partial^2\Phi}{\partial \bar{r}_p^2} - \frac{2}{\bar{r}_p}\frac{\partial\Phi}{\partial \bar{r}_p}  + \frac{1}{p} \Phi^{p-1} (\Phi^p - 1) = 0 \ . \label{eq:rad-eom2} \ee
Note also that, in these variables, the scale factor does not appear explicitly in the field equation of motion\footnote{The potential can be expanded around the minimum as $V(\phi) \sim m_{\phi}^2 \phi^2 /2$, so the Universe expands approximately as matter-dominated after the initial tachyonic oscillatory regime. From the (3+1)-dimensional lattice simulations, we have been able to fit the scale factor for $v=0.01 m_p$ to the function
\be a(t) \simeq (1 + a_1  m_{\phi} t)^{2/3 - \delta } \ , \label{eq:scf-1}  \ee
where the numerical values of coefficients $a_1$ and $\delta$($\ll 2/3$)  are obtained from the fits. For $p=(4,6,8)$, we get ${a_1=(1.78 \cdot 10^{-3},1.19 \cdot 10^{-3},0.89 \cdot 10^{-3})}$ and $\delta=(0.043,0.048,0.052)$ respectively. These simulations end when the scale factor is $a \simeq a_{\rm e} \equiv 5$, but we can reasonably extrapolate our fit to later times.}.

In this section we are interested in setting the initial conditions of the field at the \textit{extraction} time $t = t_{\rm e}$, which was when we extracted the oscillon shapes from the (3+1)-dimensional lattice simulations. At this time, we initiate the field and its derivative as
\bea \Phi(\overline{t}=\overline{t}_{\rm e},\overline{r}_p)\,&=&\,1-A_{-}^{\rm (e)} \,e^{-\frac{1}{2}\left(\overline{r}_p /R_p^{\rm (e)}\right)^2}\,, \label{eq:initcond-1} \\
\Phi'(\overline{t}=\overline{t}_{\rm e},\overline{r}_p)\,&=&\,0 \ , \label{eq:initcond-2} \eea
where the values of $A_{-}^{\rm (e)}$ and $R_p^{\rm (e)}$ are typically within the intervals given in Eq.~(\ref{eq:shape-results}). We have set the initial field derivative to zero, which is a reasonable assumption when the oscillon is in a local minimum of its oscillations. 

Here we solve the equation of motion (\ref{eq:rad-eom2}) using (1+1)-dimensional lattice simulations, with initial conditions given by Eqs.~(\ref{eq:initcond-1})-(\ref{eq:initcond-2}). We will refer to these as \textit{radial} or  \textit{spherically symmetric} simulations. We solve the equation of motion for power-law coefficients $p=4,6,8$ and different initial oscillon shapes, and let the system evolve until the oscillon decays. In Section \ref{sec:numerics-radial} we explain the technical details of our numerical simulations, putting special emphasis on how we deal with boundary conditions. We will then present our results on the oscillon lifetimes in Section \ref{sec:lifetime-results}.

\subsection{Numerical technique for radial simulations}
\label{sec:numerics-radial}

Let us explain now the details of our numerical technique. We have performed simulations for a single spherically symmetric Gaussian oscillon centered at the origin ($\overline{r}_p=0$), with initial field distribution given by Eqs.~(\ref{eq:initcond-1})-(\ref{eq:initcond-2}). In particular, we solve a discrete version of the field \textit{eom} (\ref{eq:rad-eom2}) in a (1+1)-dimensional lattice of finite size $L$ and number of points $N$. All the results we present in Section \ref{sec:lifetime-results} are based on lattice simulations with 
\be L\,=\,100R_p^{\rm (e)} \,,\quad N\,=\,5000\,,\quad{\rm and}\quad \Delta {\rm x}\,\equiv\,\frac{L}{N}=0.02R_p^{\rm (e)} \label{eq:lat-par1}\, ,\ee
where $\Delta {\rm x}$ is the length step. This way, we adapt the size of the lattice to the initial physical radius of the oscillon, such that we have always the same number of nodes capturing the dynamics at scales within the oscillon. The oscillon \textit{eom} is solved using a staggered Leapfrog algorithm that is second-order accurate in time. In our simulations, the time step is chosen to be $\Delta {\rm t} = 0.5 \Delta {\rm x}$.  On the other hand, in order to discretize the spatial derivatives, we have decided to follow the same approach as in the (3+1)-dimensional simulations: there we approximated the continuous Laplacian by the central-scheme (\ref{eq:discrete-laplacian}), which is fourth-order accurate in space. Similarly, we substitute here the continuous spatial derivatives in the field \textit{eom}, by the following fourth-order accurate expressions \cite{DiscreteDerivatives},
\bea
\frac{\partial \Phi}{\partial \overline{r}_p} \,&\simeq\, &\frac{\Phi(\overline{r}_p-2\Delta {\rm x} ) - 8\Phi (\overline{r}_p-\Delta {\rm x} ) +8\Phi(\overline{r}_p+\Delta {\rm x} )  -\Phi (\overline{r}_p+2\Delta {\rm x} )}{12\,\Delta {\rm x}}  , \\[7pt]
\frac{\partial^2\Phi}{\partial\overline{r}_p^2}\,&\simeq \,&\frac{-\Phi (\overline{r}_p+2\Delta {\rm x} ) + 16\Phi (\overline{r}_p+\Delta {\rm x} )  -30\Phi (\overline{r}_p ) +16\Phi (\overline{r}_p-\Delta {\rm x} ) -\Phi (\overline{r}_p-2\Delta {\rm x} )}{12\,\Delta {\rm x}^2} \ .  \,\,\,\,\,\,\,\,
\eea 

Let us explain now how we deal with boundary conditions. According to Eq.~(\ref{eq:initcond-1}), the initial field distribution at time $t = t_{\rm e}$ decays exponentially as $\bar{r}_p \rightarrow \infty$. Due to this, and because of machine precision, the initial field amplitude in our simulations is \textit{exactly zero} for enough large radial distances $\bar{r} \gg R_p^{\rm (e)}$. As we have chosen the length of the box to be much larger than the oscillon radius, $L \gg R_p^{\rm (e)}$, the field amplitude is zero close to the boundary, and there are not unphysical interference effects initially, despite the existence of periodic boundary conditions in the lattice. However, oscillons continuously lose energy through the emission of small-amplitude scalar waves, which radially propagate away from the oscillon towards $\overline{r}_p\rightarrow\infty$. As we perform simulations in a finite box, these waves eventually hit the boundary at $\bar{r}_p = L$, thereby giving rise to unphysical reflections and interferences. In order to avoid such effects, our code smoothly truncates the field distribution whenever scalar waves come close to the boundary. 

More specifically, our method is based on the definition of a \textit{truncation function} $ \mathcal{G}(\overline{r}_p;\overline{r}_{_{\hspace{-0.06cm} \Delta}})$ as
\be \mathcal{G}(\overline{r}_p;\rDelta)\,\equiv\,\frac{e^{-\gamma(\overline{r}_p-\rDelta)}}{1+e^{-\gamma(\overline{r}_p-\rDelta)}}\ , \ee
where $\gamma$ is a free parameter, and $\rDelta$ is a length scale within the interval $0 < \rDelta  < L$. This function is antisymmetric around $\overline{r}_p=\rDelta$, and goes as $\mathcal{G}(\bar{r}_p;\rDelta) \rightarrow (1,1/2,0)$ in the limit $(\bar{r}_p - \rDelta) \rightarrow (-\infty, 0,\infty)$. 
Each time the wave becomes closer to the boundary, the program stops the computation, and reinitializes the field and its derivative according to

\be
\Phi(\overline{t},\overline{r}_p) \,=\, \begin{cases}
\Phi(\overline{t},\overline{r}_p)&\quad{\rm for}\quad \overline{r}_p<\rDelta- \Delta_R \,,\\
 1 + \Big[\Phi(\overline{t},\overline{r}_p) - 1\Big]\,\mathcal{G}(\overline{r}_p;\rDelta)&\quad{\rm for}\quad \rDelta-\Delta_R \le\overline{r}_p\le\rDelta+\Delta_R \,,\\
1&\quad{\rm for}\quad \overline{r}_p>\rDelta+\Delta_R \,,
\end{cases}
\ee
and
\be
\Phi'(\overline{t},\overline{r}_p) \,=\, \begin{cases}
\Phi'(\overline{t},\overline{r}_p)&\quad{\rm for}\quad \overline{r}_p<\rDelta-\Delta_R \,,\\
\Phi'(\overline{t},\overline{r}_p)\mathcal{G}(\overline{r}_p;\rDelta)&\quad{\rm for}\quad \rDelta-\Delta_R \le\overline{r}_p\le\rDelta+\Delta_R \,,\\
0&\quad{\rm for}\quad \overline{r}_p>\rDelta+\Delta_R  \ ,
\end{cases}
\ee
where $\Delta_R$ is also a free length scale. The truncation function smoothly pushes the amplitude of the field towards zero at $\bar{r} >  \rDelta - \Delta_R $, while keeping the field distribution intact for $\bar{r}  <   \rDelta - \Delta_R$.  Roughly speaking, parameter $\gamma$ controls how fast the transition is. For our simulations with lattice parameters (\ref{eq:lat-par1}), we have checked that good choices are
\be \Delta_R = 10 R_p^{\rm (e)} \,,\quad \gamma = 10\,,\quad{\rm and}\quad\rDelta=60R_p^{\rm (e)}\,. \label{eq:lat-par2}\ee

\begin{figure}
\centering
\includegraphics[height=4.7cm]{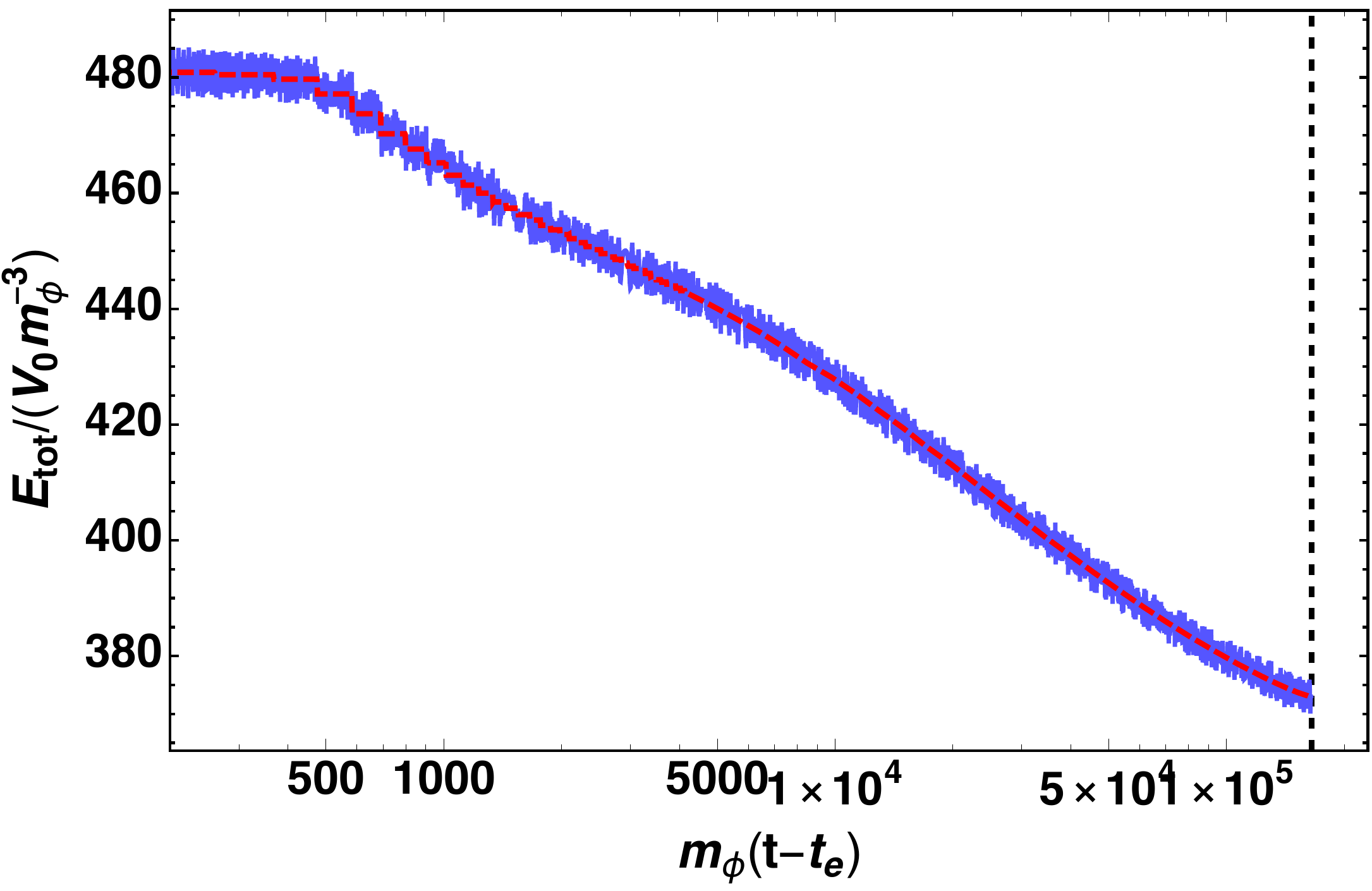} \hspace{0.2cm}
\includegraphics[height=4.7cm]{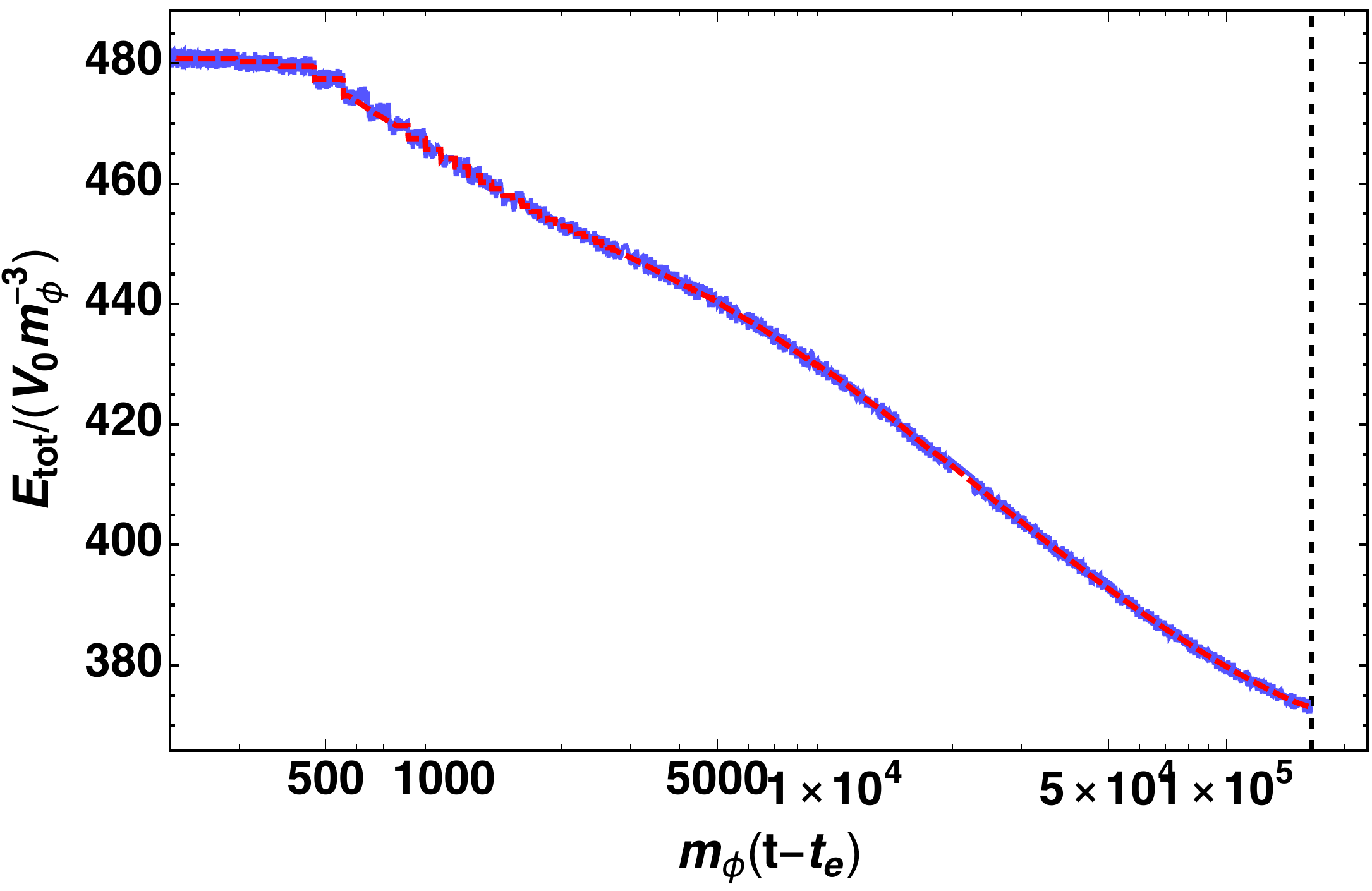} 
\caption{We track energy conservation for a radial simulations with $p=4$, initial amplitude $A_{-}^{\rm (e)} = 0.62$, and radius $R_p^{\rm (e)}=2.90$. The time step in the left panel is $\Delta {\rm t} = 0.5 \Delta {\rm x}$, while in the right panel is $\Delta {\rm t} = 0.2 \Delta {\rm x}$, with $ \Delta {\rm x} = 0.02 R_p^{\rm (e)}$. The blue lines show the total energy in the lattice as a function of time, which decreases due to the continuous truncation of scalar waves at large radial distances. The red lines show at first the initial energy in the box $E_{\rm tot}^{\rm (e)} /(V_0 m_{\phi}^{-3})\simeq 480$, but as time goes on, we subtract to it the lost energy due to the truncation technique. We can check energy conservation by comparing both quantities. As expected, energy is better preserved in the right panel, because the time step is smaller. The dashed vertical lines indicate the time when the oscillon decays, which is almost identical in both panels.}
\label{fig:energy-conservation}
\end{figure}

In order to check the robustness of our numerical method, we have performed additional numerical simulations with different lattices, time steps, and truncation parameters.  Typically, if the box is smaller, the scalar waves need less time to arrive at the boundary, so the field is reinitialized many more times. However, we have checked that for the oscillon dynamics and lifetimes, this is not relevant. As an additional test of our method, we have also reproduced the results presented in Figure~3a of Ref.~\cite{Copeland:1995fq}. Note that there exist alternative techniques to avoid the boundary condition problem. For example, in Ref.~\cite{Guzman:2004wj} an imaginary term is introduced in the potential, such that the field is truncated at large radial distances. Another possibility would be the implementation of absorbing boundary conditions (as in \cite{Salmi:2012ta,Andersen:2012wg}), or the addition of an artificial $\overline{r}$-dependent damping term, as proposed in~\cite{Gleiser:1999tj}. With our procedure, we have been able to accurately track  how much energy is lost due to the truncation of waves at large $\overline{r}_p$, which in turn allows to accurately check energy conservation within the box during the simulation. Due to the symplectic nature of the staggered leapfrog algorithm, energy is approximately conserved during the whole time evolution. For $\Delta {\rm t} = 0.5 \Delta {\rm x}$, energy is conserved up to a factor $\sim 10^{-2}$, but smaller time steps improve energy conservation, as seen in Fig.~\ref{fig:energy-conservation}. We have done different lattice simulations for different  time steps, but the estimated values for the oscillon lifetimes remain unchanged.

\subsection{Results for oscillon lifetimes}\label{sec:lifetime-results}

\begin{figure}
\centering
\includegraphics[width=0.3\textwidth]{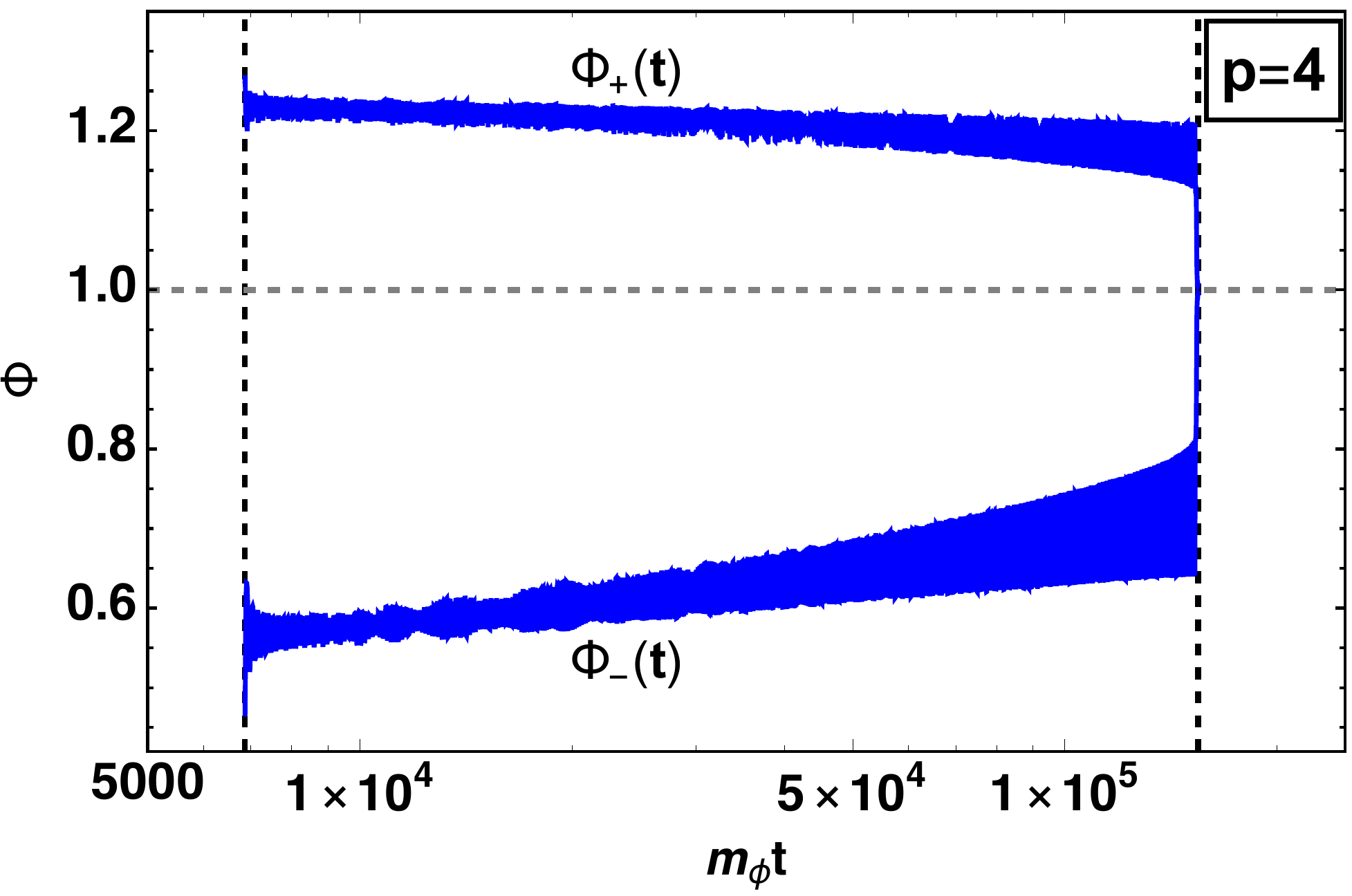} \hspace{0.1cm}
\includegraphics[width=0.3\textwidth]{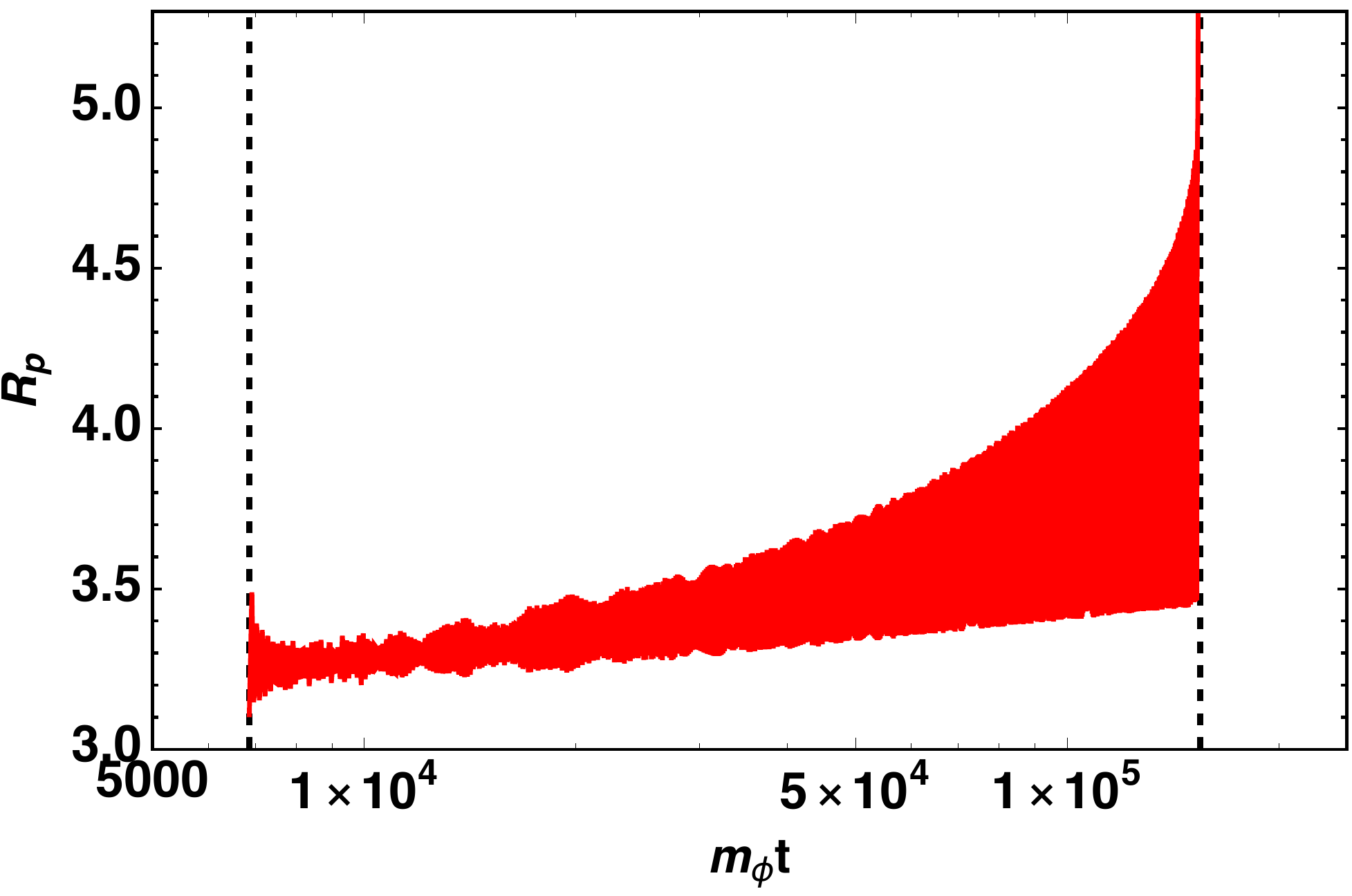} \hspace{0.1cm}
\includegraphics[width=0.3\textwidth]{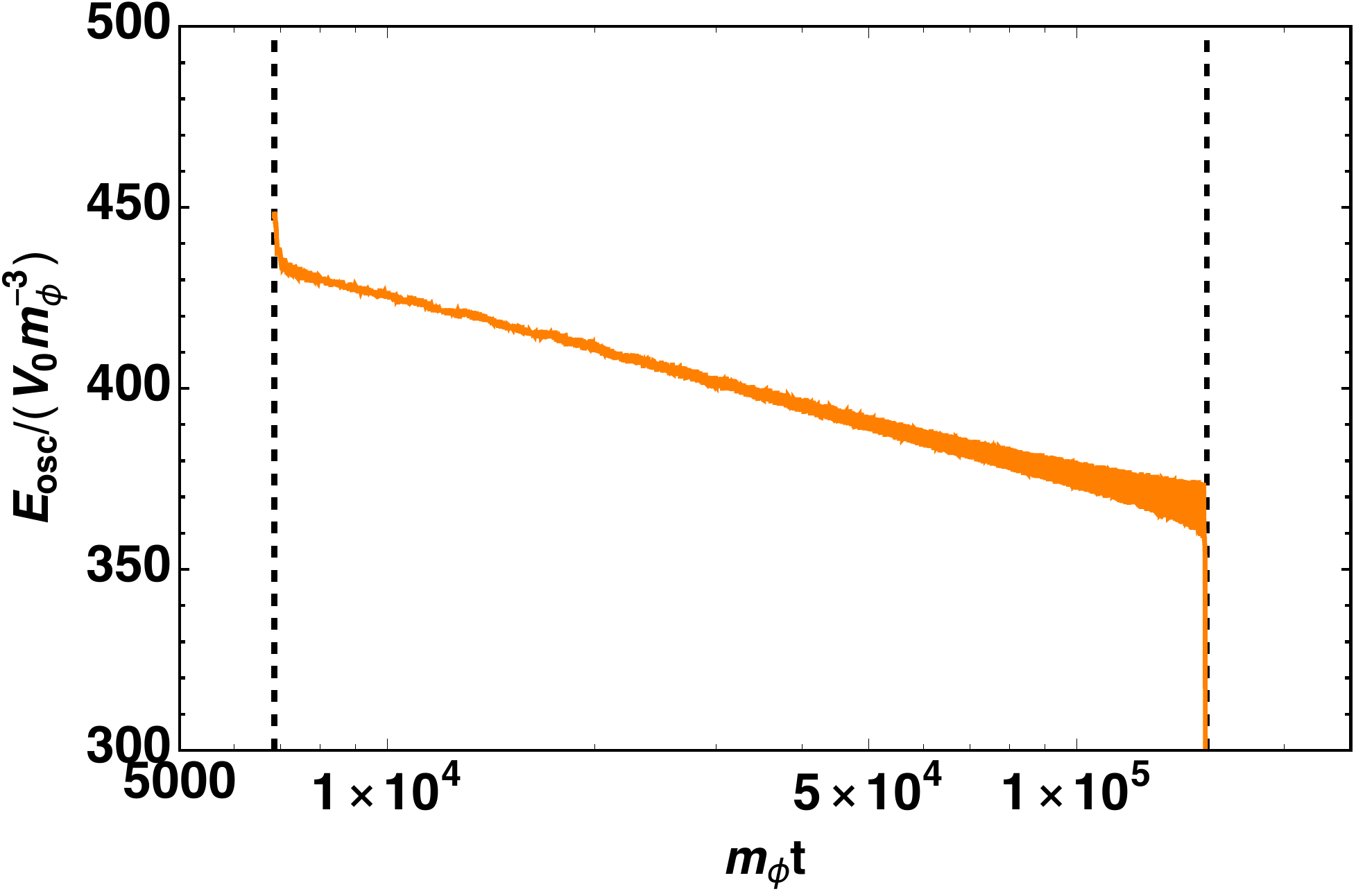} \vspace{0.2cm}\\
\includegraphics[width=0.3\textwidth]{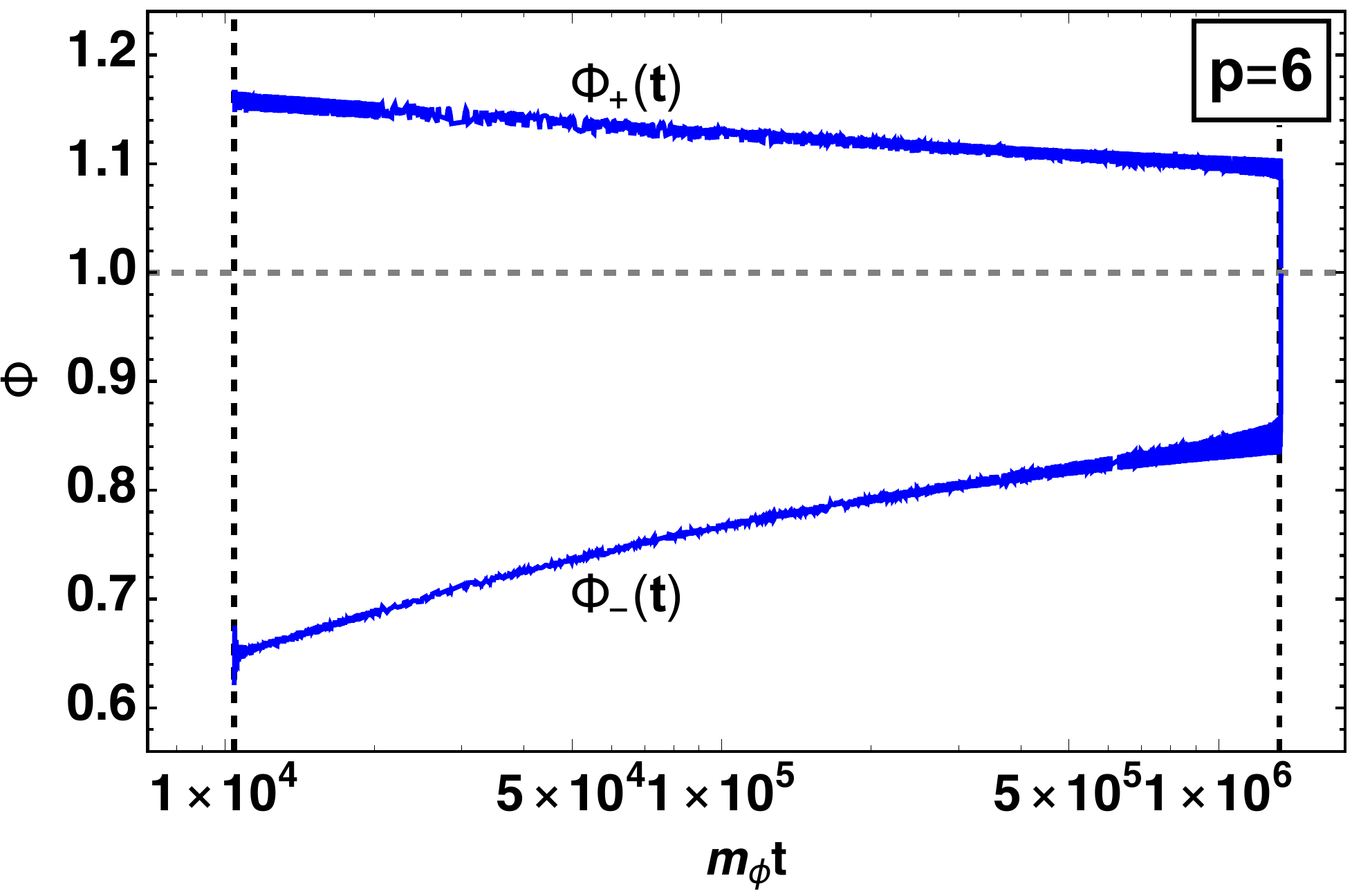} \hspace{0.1cm}
\includegraphics[width=0.3\textwidth]{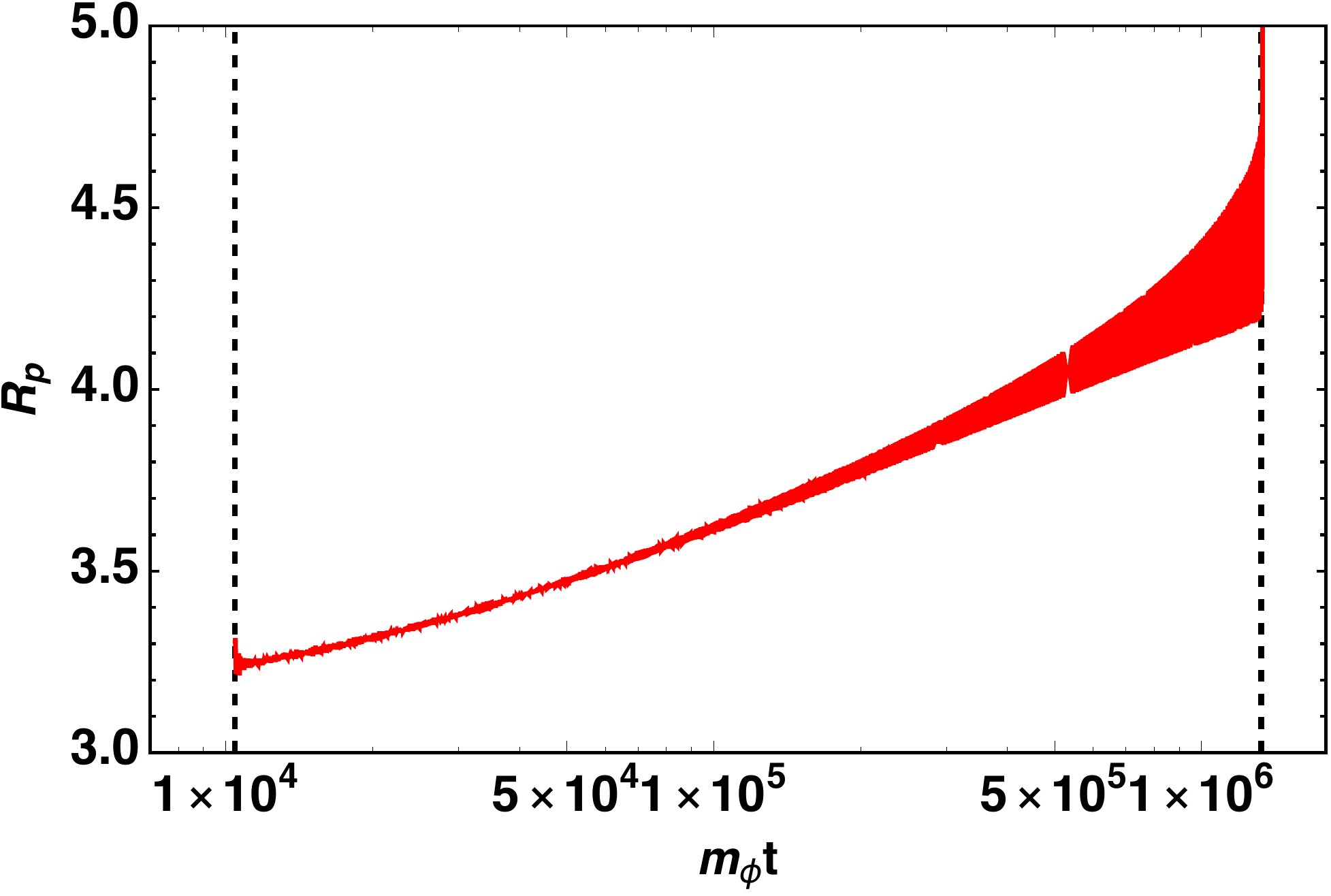} \hspace{0.1cm}
\includegraphics[width=0.3\textwidth]{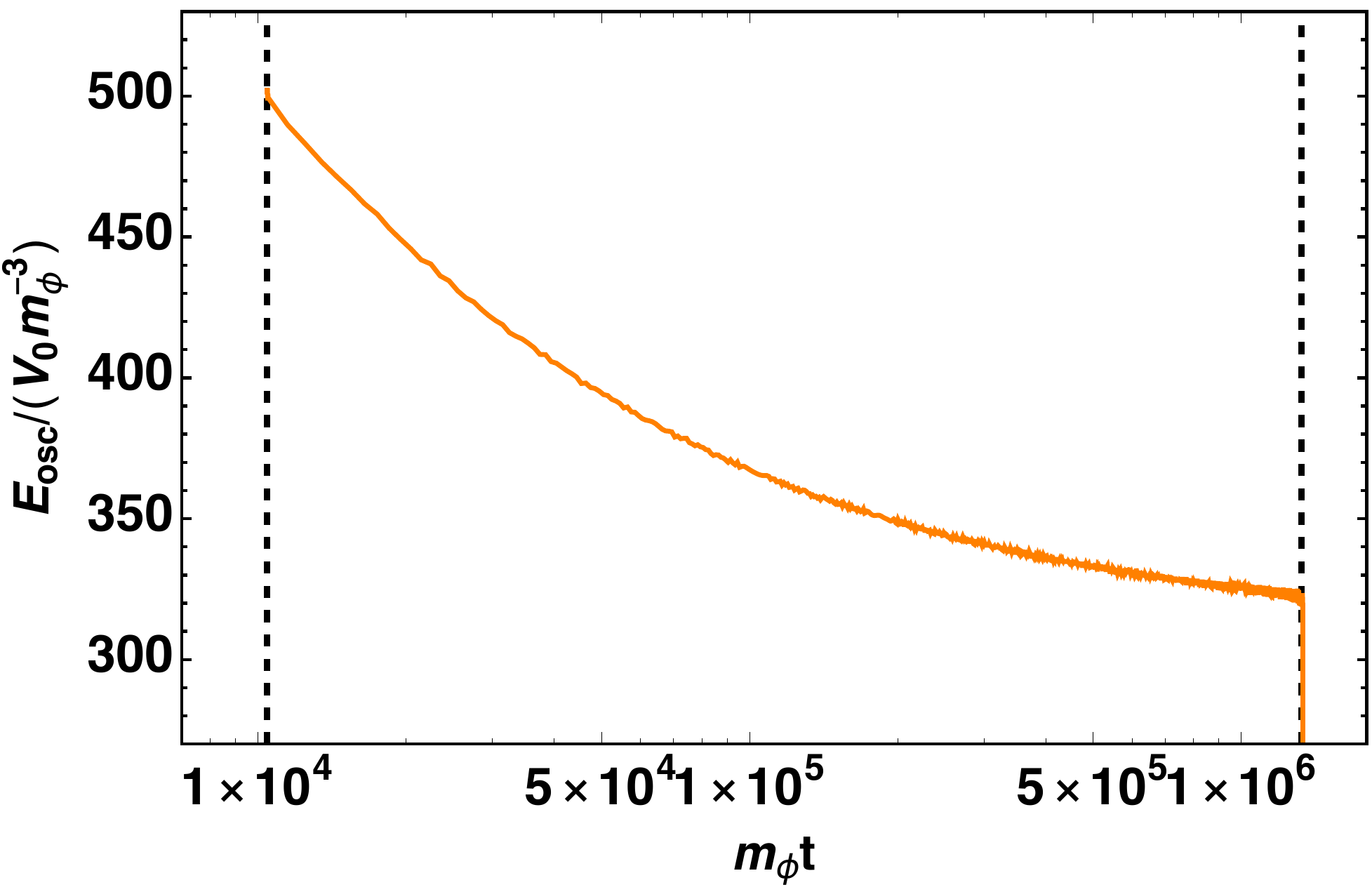}  \vspace{0.2cm}\\
\includegraphics[width=0.3\textwidth]{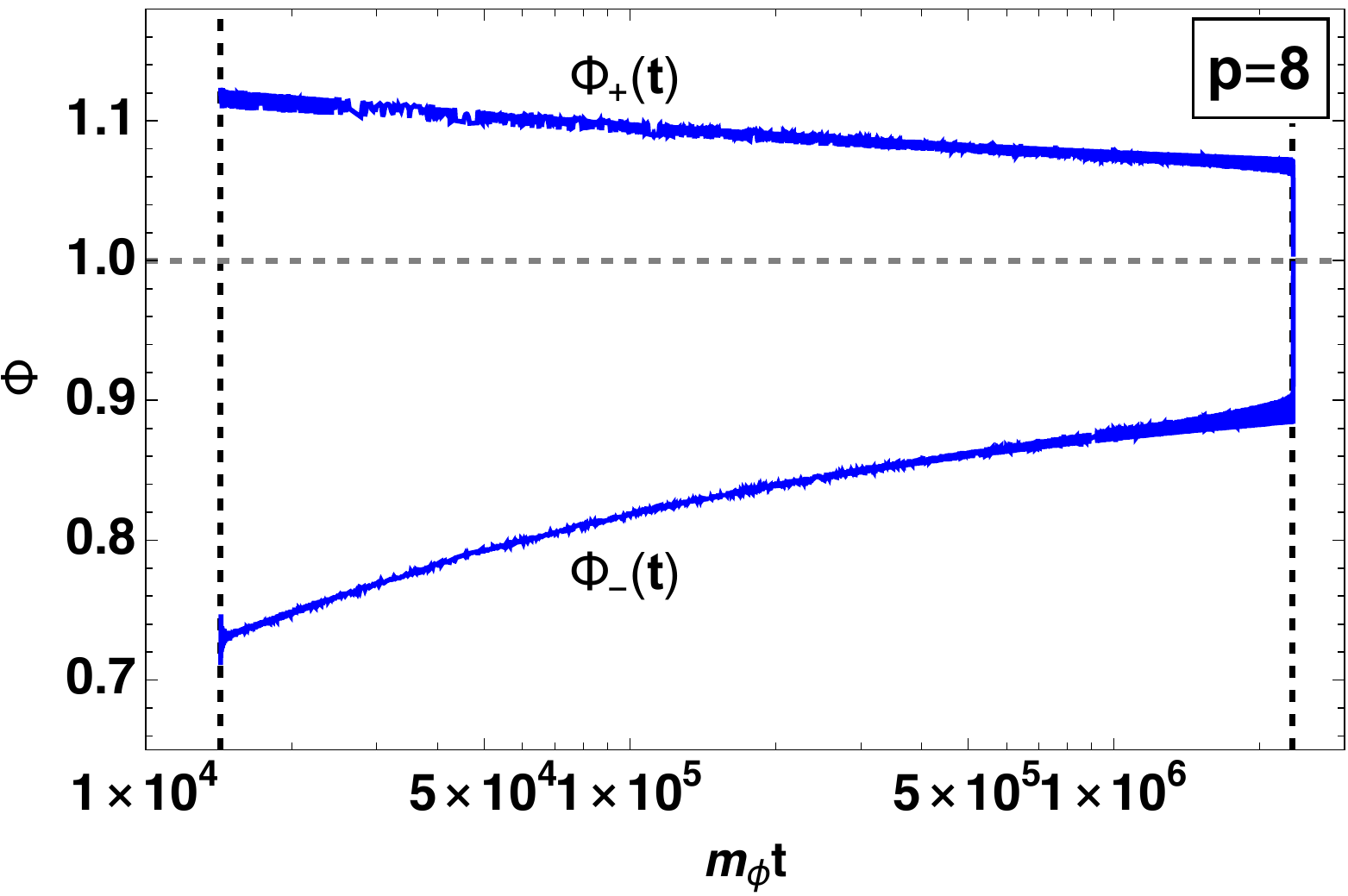} \hspace{0.1cm}
\includegraphics[width=0.3\textwidth]{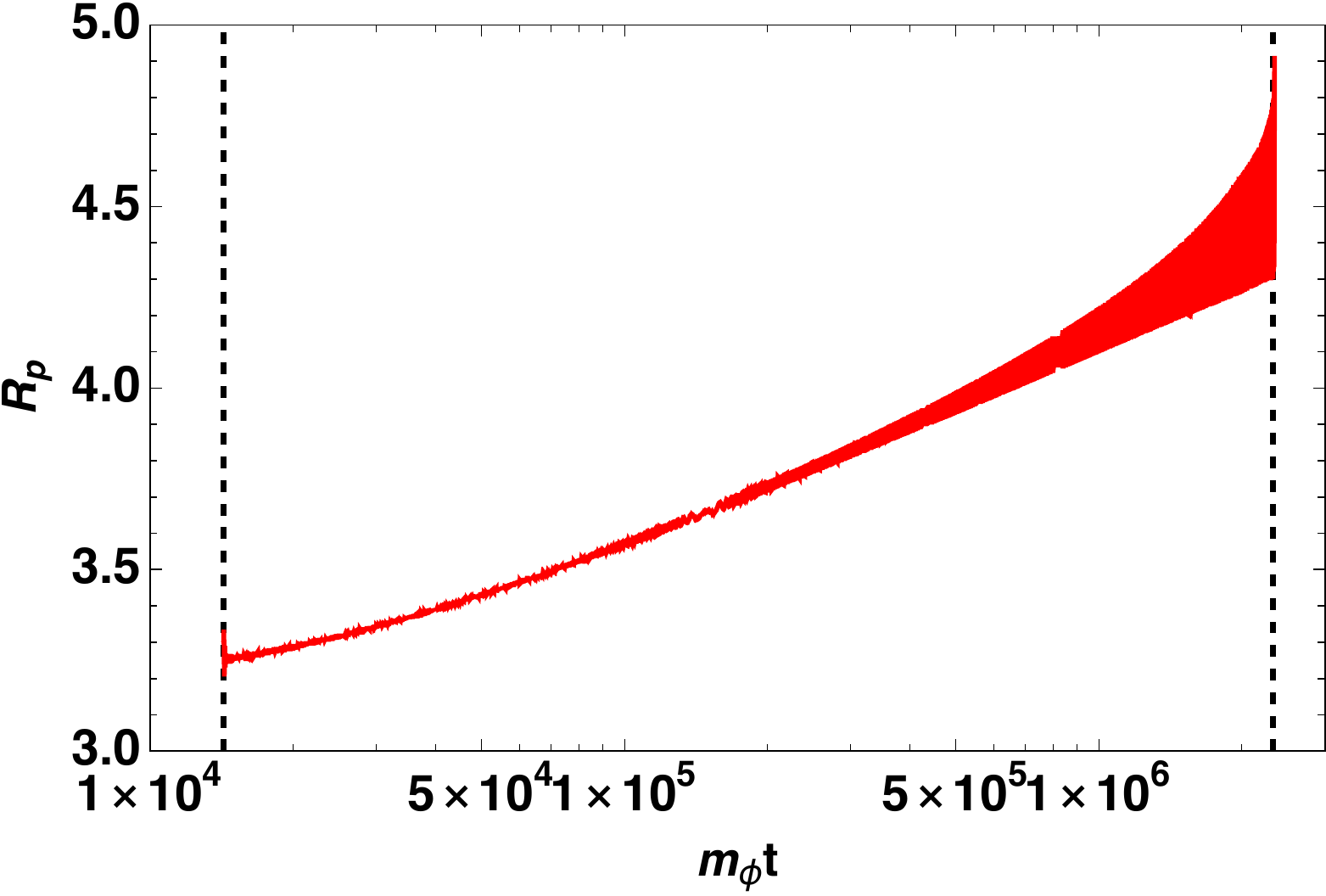} \hspace{0.1cm}
\includegraphics[width=0.3\textwidth]{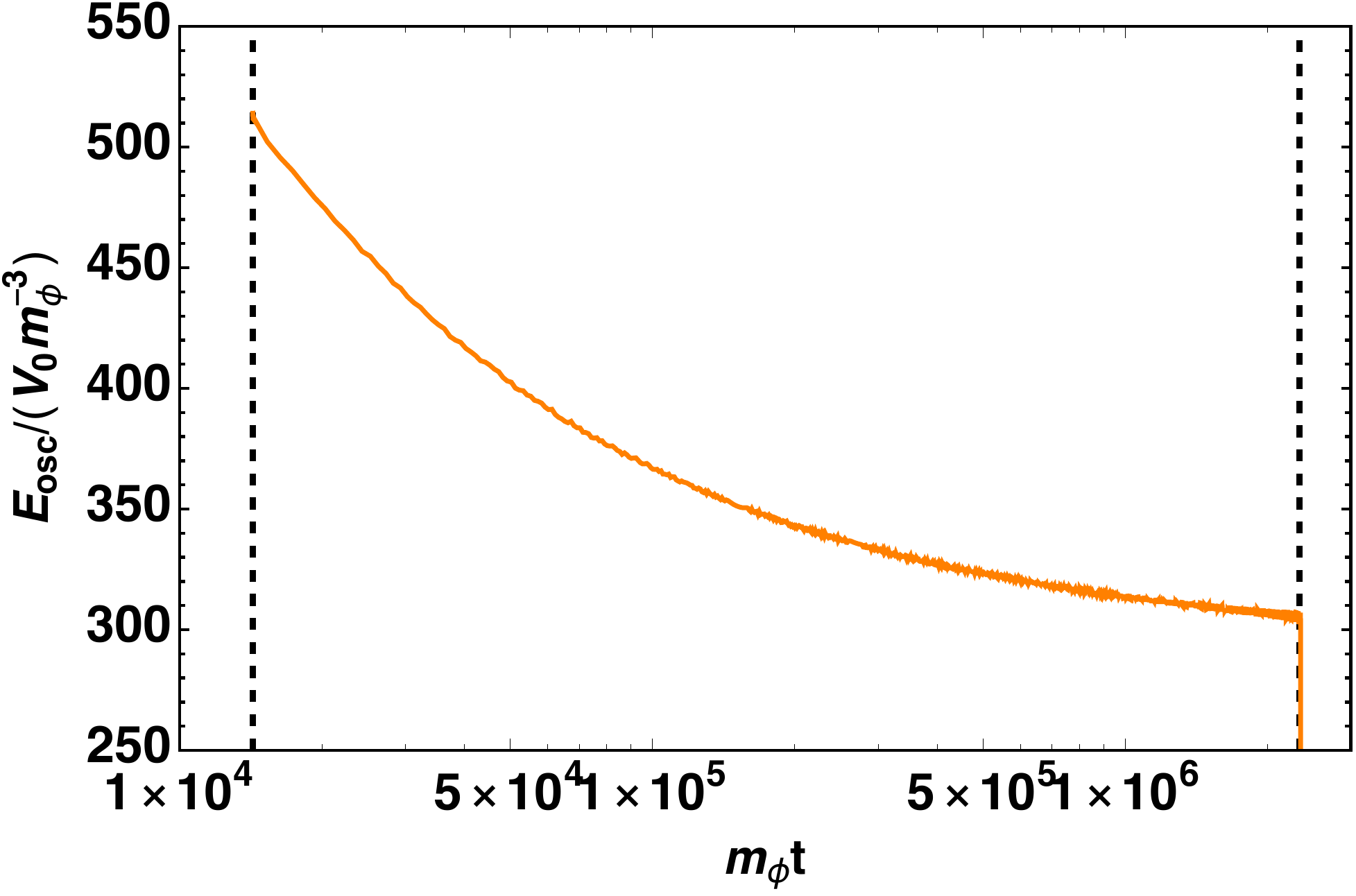}
\caption{We show the time evolution of three typical oscillons, obtained from radial simulations with $p=4$ (top panels), $p=6$ (middle panels), and $p=8$ (bottom panels). The initial amplitude and radius of the oscillons are given in Table \ref{tab:lifetimes}. The left panels show, for each simulation, the upper and lower envelops of the field oscillations $\Phi_{+}$ and $\Phi_{-}$, as measured at the center of the oscillon. The oscillon amplitude is defined as $A_{-} (t)= 1 - \Phi_- (t)$. The middle panels show the radius of the oscillon $R$, obtained by fitting the field solution to the Gaussian function (\ref{eq:Gaussian-fit}).  The right panels show the evolution of the oscillon energy, defined in Eq.~(\ref{eq:oscillon_energy_density}). In each panel we indicate, with vertical dashed lines, the \textit{extraction} time $m_{\phi} t_{\rm e}$ (when the simulations start), and the oscillon decay time $m_{\phi} t_{\rm dec}$.}
\label{fig:lifetime_oscillon}
\end{figure}

In this section we present our results from the spherically symmetric simulations, as well as give estimates for the typical oscillon lifetimes in hilltop models. To begin with, let us consider Fig.~\ref{fig:lifetime_oscillon}, where we show three different examples of simulations with power-law coefficients $p=4$ (top panels), $p=6$ (middle panels), and $p=8$ (bottom panels).  The initial shapes of the oscillons (i.e.\ the values of $A_{-}^{\rm (e)}$ and $R_p^{\rm (e)}$) are chosen, in each case, to represent some of the typical shapes extracted in Section \ref{sec:shapes-results}. The left dashed vertical line in each panel shows the initial time of the simulations, which correspond to the extraction times $m_{\phi} t_{\rm e} \approx 6870, 10450,14290$, for $p=4,6,8$, respectively.

Let us focus first on the left panels of Fig.~\ref{fig:lifetime_oscillon}. There we show the time evolution of the lower and upper envelopes of the oscillations, i.e. $\Phi_{+} (t)$ and $\Phi_{-} (t)$, measured at the center of the oscillon at $\bar{r}_p = 0$. Remember that according to our definition in Section \ref{sec:fitting-proc}, the oscillon amplitude is defined as $A_{-} (t) = 1 - \Phi_{-} (t)$. First we observe that both envelopes oscillate due to the existence of a breathing mode, in agreement with what we observed in the (3+1)-dimensional lattice simulations. The breathing amplitude is significantly larger for $p=4$ than for $p=6,8$. As the oscillons continuously lose energy due to the emission of small-amplitude scalar waves, the envelopes of the oscillons become increasingly closer to $\Phi = 1$. Due to this, the field amplitudes within the oscillon are, as time goes on, in the non-shallow region of the inflationary potential for increasingly longer periods of time. Eventually, there is a time when the oscillon is no longer stable and decays: the field amplitude becomes $\Phi \approx 1$ everywhere, and the oscillon amplitude becomes $A_{-} \approx 0$. This process takes only $m_{\phi} \Delta  t \sim 10^3$ units of time, which is very fast in comparison with the oscillon lifetime. Therefore, we can easily define a new quantity to signal this moment: the oscillon \textit{decay time} $t_{\rm dec}$.

Let us focus now on the middle panels of Fig.~\ref{fig:lifetime_oscillon}, which show, for the three simulations, the oscillon physical radius $R_p(t)$ as a function of time. The radius is obtained by fitting the field distribution at each time to the Gaussian function (\ref{eq:Gaussian-fit}). We observe that in the three cases, the physical radius slightly increases with time, albeit not much. For $p=4$, it grows from $R_p^{\rm (e)} = 3.2$ at $t=t_{\rm e}$, to $R_p (t \lesssim t_{\rm dec}) \sim 4.0$ just before the oscillon decays. Similarly, for $p=6$ it goes from $R_p^{\rm (e)} = 3.3$  to $R_p (t \lesssim t_{\rm dec})\sim 4.4$, and for $p=8$ it goes from  $R_p^{\rm (e)} = 3.3$  to $R_p (t \lesssim t_{\rm dec})\sim 4.5$. Let us also note that the radius also shows a breathing mode, whose amplitude increases as time goes on. The breathing mode is opposite to the one observed for the amplitude, in compatibility with the results presented in Section \ref{sec:oscillon-shapes}.

Finally, let us consider the right panels of Fig.~\ref{fig:lifetime_oscillon}, where we show the energy of the oscillons as a function of time, defined as
\be
 E_{\rm osc}(t)\,=\,4\pi V_0 m_{\phi}^{-3} \int_{0}^{5R_p}d\overline{r}_p\,\overline{r}_p^2\left[\frac{1}{2}\left(\frac{\partial\Phi}{\partial\overline{t}}\right)^2 + \frac{1}{2}\left(\frac{\partial\Phi}{\partial\overline{r}_p}\right)^2+V(\Phi) \right] \ ,
\label{eq:oscillon_energy_density}\ee
[recall Eq.~(\ref{eq:oscenergy_lat})].  The oscillon energy changes with time due to two reasons. First, the field amplitudes within the oscillon become closer to the vacuum of the potential, so the integrand in Eq.~(\ref{eq:oscillon_energy_density}) is smaller. And second, the (average of the) physical radius grows with time, so the integral in Eq.~(\ref{eq:oscillon_energy_density}) increasingly captures a larger volume. In Fig.~\ref{fig:lifetime_oscillon} we observe that the oscillon energies slowly decrease with time, so the first effect dominates over the second one: just before their decay, the oscillons have lost $\sim 15\%$, $\sim 35\%$ and $\sim 40\%$ of their initial energy, for $p=4,6,8$, respectively.

In Fig.~\ref{fig:lifetime_oscillon} we clearly see that the oscillon lifetimes depend very strongly on the power-law coefficient $p$.  We have displayed the numerical values for the observed decay times $m_{\phi} t_{\rm dec}$ in Table \ref{tab:lifetimes}. We get that, approximately,
\be (m_{\phi} t_{\rm dec} )^{[p=8]} \approx  2 (m_{\phi} t_{\rm dec} )^{[p=6]}  \approx  20 (m_{\phi} t_{\rm dec} )^{[p=4]} \ , \ee
so the larger $p$ is, the larger the lifetime of the oscillons in natural units becomes. However, let us remark that the inflaton effective mass $m_{\phi}$ [defined in Eq.~(\ref{eq:inf-mass})] depends explicitly on $p$, as well as on $V_0$ (which is also different for each value of $p$). Therefore, we have decided to translate these times to number of e-folds of expansion after inflation. Using the extrapolation for the scale factor given in Eq.~(\ref{eq:scf-1}), we get that typical oscillons in hilltop models  live $N \sim 3.5,4.6,4.7$ e-folds for power-law coefficients $p=4,6,8$ respectively.

\begin{figure}
\centering
\includegraphics[height=6cm]{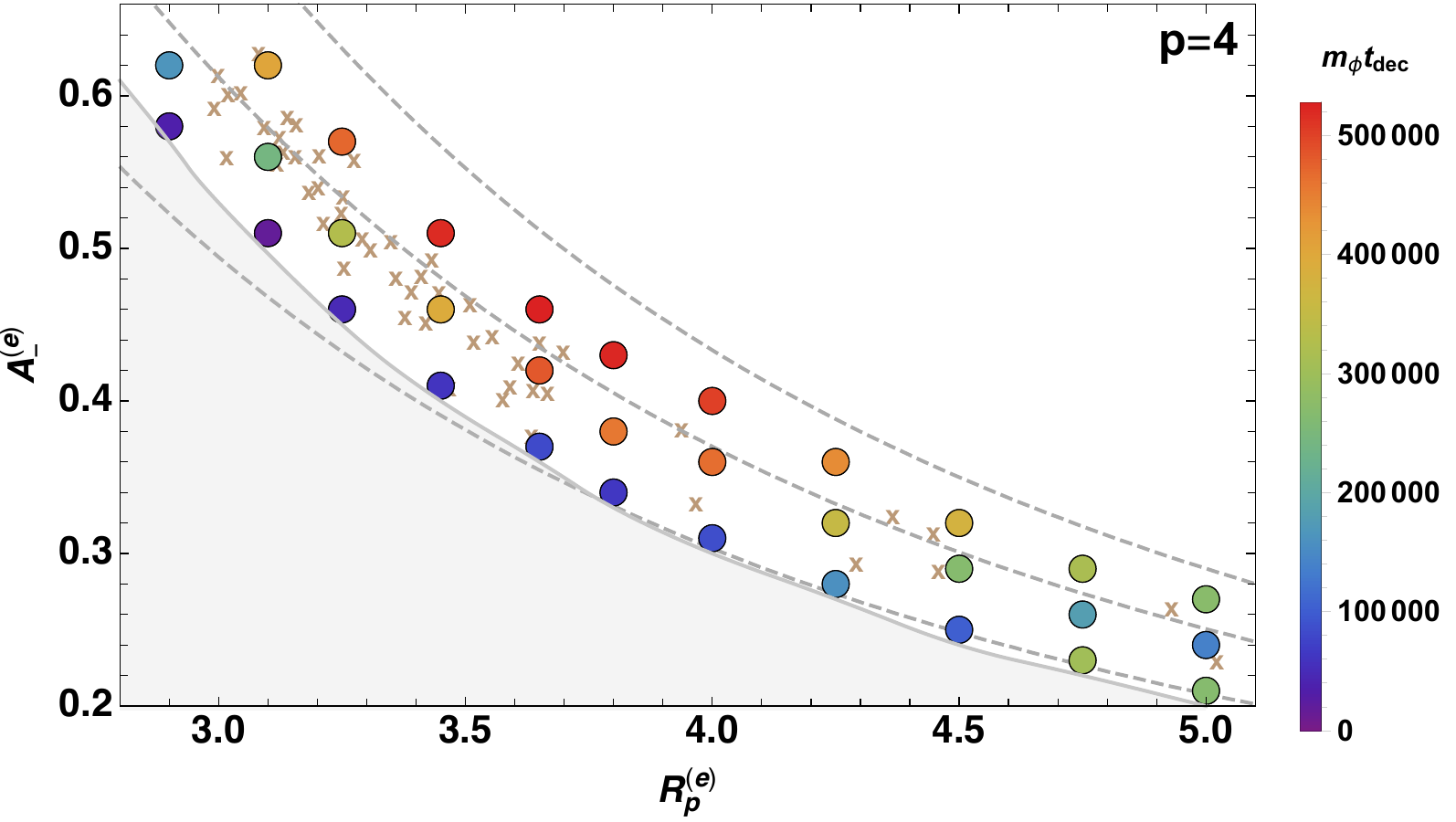} \\ \vspace{0.2cm}
\includegraphics[height=6cm]{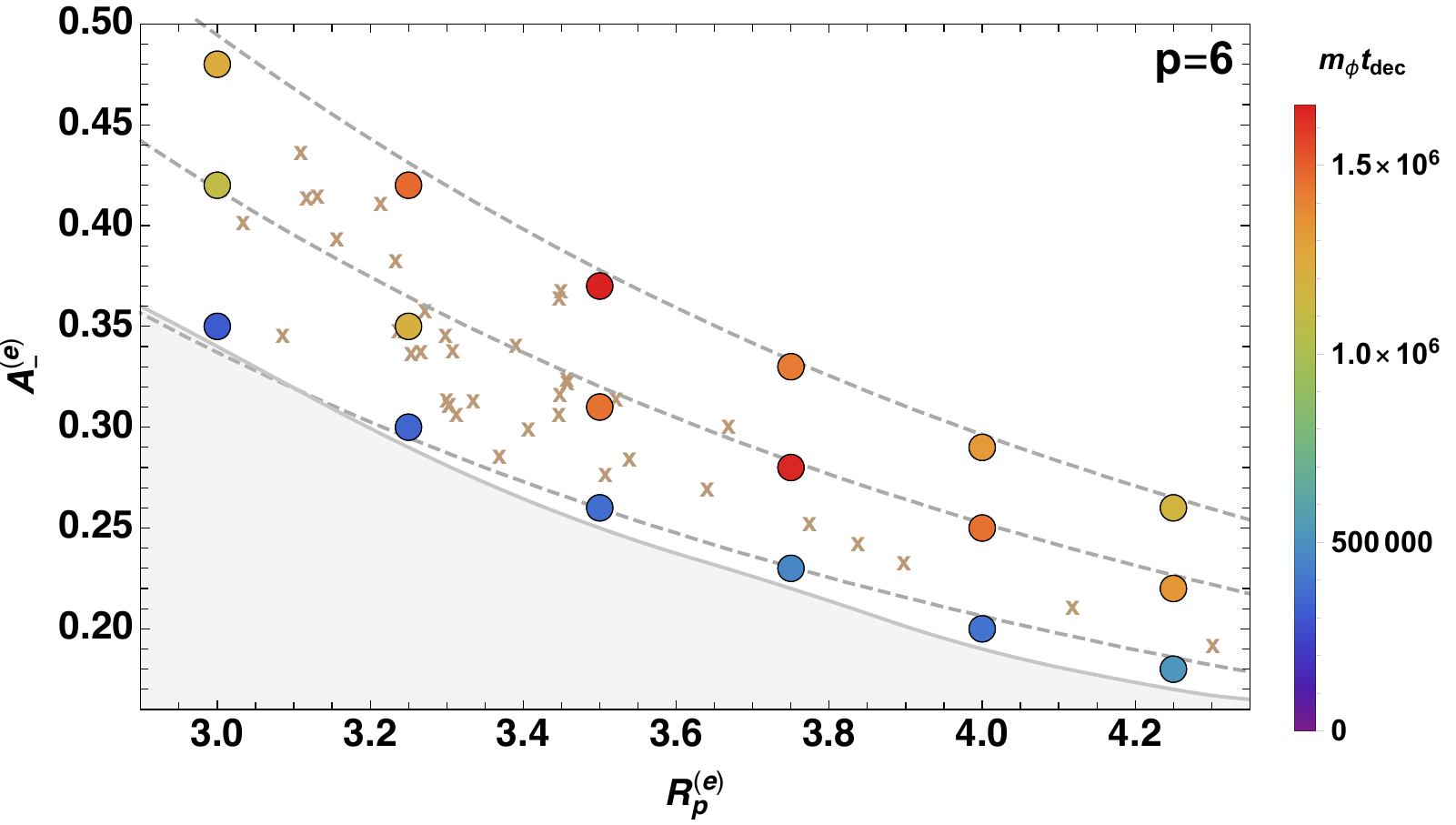} \\ \vspace{0.2cm}
\includegraphics[height=6cm]{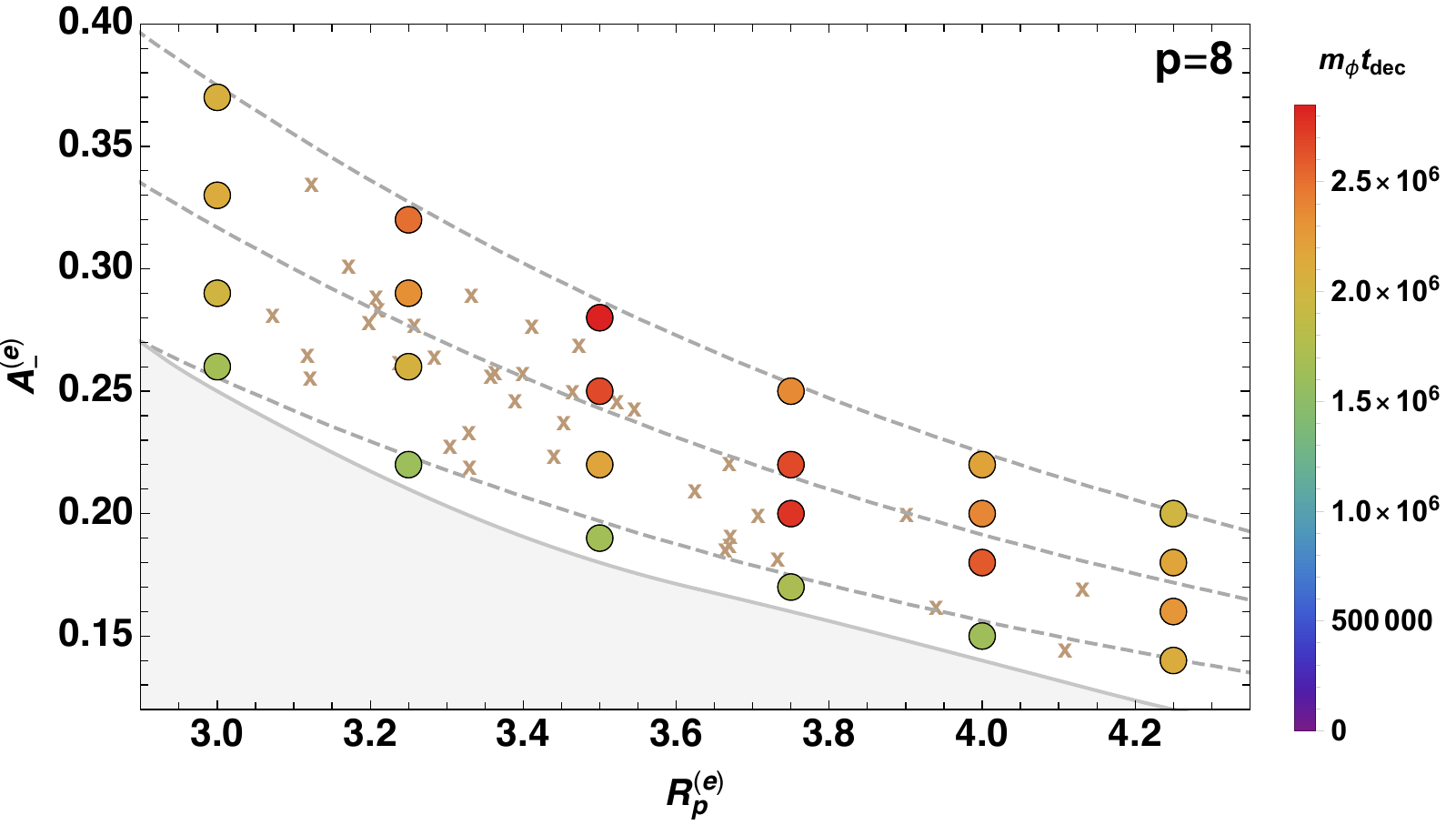} \\ \vspace{0.2cm}
\caption{We show the decay time of the oscillons for $p=4$ (top), $p=6$ (middle), and $p=8$ (bottom), obtained from the spherically symmetric simulations. Each circle corresponds to a single simulation, where the initial shape of the oscillon is parametrized by the physical radius $R_p^{\rm (e)}$ (horizontal axis) and amplitude $A_{-}^{\rm (e)}$ (vertical axis). The color of each circle indicates the measured decay time, going from purple (shorter times) to red (larger times). The gray areas in each panel show the regions in the ($A_{-}^{\rm (e)}$,$R_p^{\rm (e)}$) parameter space where the solution is unstable, i.e. when the initial Gaussian oscillon in the spherically symmetric simulation immediately decays. The brown `x' show the amplitudes and radii of the oscillons observed in the (3+1)-dimensional simulations and extracted at $a \simeq a_e \equiv 5$, see Fig.~\ref{fig:ampradius}. Finally, the dashed lines indicate the values of $A_{-}^{\rm (e)}$ and $R^{\rm (e)}$ that correspond to oscillon energies $E_{\rm osc}^{\rm (e)} /(V_0 m_{\phi}^{-3}) = 375$ (bottom line), 513 (middle line) and 650 (top line), see Eq.~(\ref{eq:oscillon_energy_density}). }
\label{fig:decaytime}
\end{figure}

Of course, the results we have just presented correspond to a particular choice for the initial oscillon shape. However, as observed in Section \ref{sec:shapes-results}, oscillons are seen to have a large variety of shapes, which translate into a larger spectrum of possible $A_{-}^{\rm (e)}$ and $R_p^{\rm (e)}$. We have observed that, indeed, the oscillon decay time is very dependent on the initial oscillon shape. Therefore, we have decided to study how this quantity changes for different reasonable initial configurations. Our findings on this issue are summarized in Fig.~\ref{fig:decaytime}, where we plot the decay times extracted from our spherically symmetric simulations for $p=4,6,8$, and for different choices of $A_{-}^{\rm (e)}$ and $R_p^{\rm (e)}$. We have chosen values that are in approximate agreement with the ones extracted from the (3+1)-lattice simulations in Section \ref{sec:oscillon-shapes}. For each of the three power-law coefficients $p=4,6,8$, we can divide the $A_{-}^{\rm (e)}$-$R_p^{\rm (e)}$ parameter space in two clear regions: the \textit{instability region}, depicted in gray, and the \textit{stability region}, depicted in white. For oscillon configurations within the instability region, the oscillon decay is seen to be immediate, so $m_{\phi} t_{\rm dec} \approx 0 $.  On the other hand, for oscillons with initial shapes within the stability region, the solution for the field distribution obtained from the radial simulations correspond to the one of an oscillon. In these cases, the oscillon dynamics is qualitatively similar to the one described above, and we can identify a non-zero decay time $m_{\phi} t_{\rm dec} > 0$. We can parametrize the line dividing the stability and instability regions for each of the power-law coefficients as
\bea
A_{-}^{\rm (st)} &\simeq & 0.53 (R_p^{\rm (st)}/3)^{-2} \ , \hspace{0.6cm} \text{for  $p=4$ }\ , \label{eq:AmpRad-st1} \\
A_{-}^{\rm (st)} &\simeq & 0.34 (R_p^{\rm (st)}/3)^{-2} \ , \hspace{0.6cm} \text{for  $p=6$ } \ , \label{eq:AmpRad-st2}\\
A_{-}^{\rm (st)} &\simeq & 0.25 (R_p^{\rm (st)}/3)^{-2} \ , \hspace{0.6cm} \text{for  $p=8$ } \ . \label{eq:AmpRad-st3}
\eea
We observe that the smaller the radius, the larger the amplitude must be for the oscillon to be stable. Note also that in Fig.~\ref{fig:decaytime}, we have also depicted the values for the oscillon shapes extracted from the (3+1)-dimensional lattice simulations: as expected, all the extracted shapes are within the stability region, which is a good consistency check.

\begin{table}
\centering
\begin{tabular}{| c | c | c || c | c | c | }
 \hline
 $p$ & $A_{-}^{\rm (e)}$ &  $R_p^{\rm (e)}$  & $m_{\phi} t_{\rm dec}$ & $N_{\rm osc}$  & $N$ \\ \hline 
 4 & 0.50 & 3.2 &  $1.47 \cdot 10^5$  & $2.4 \times 10^4$ & $3.5$   \\ \hline
 6 & 0.35 & 3.3 &  $1.32 \cdot 10^6$  & $2.1 \times 10^5$ & $4.6$   \\ \hline
 8 & 0.27 & 3.3 &  $2.34 \cdot 10^6$  & $3.7 \times 10^5$ & $4.7$   \\ \hline
\end{tabular}
\caption{We show the lifetimes of three typical oscillons for $p=4,6,8$, obtained from the spherically symmetric simulations. The dynamics of these oscillons have been shown in Fig.~\ref{fig:lifetime_oscillon}. The number of oscillations is $N_{\rm osc} \approx m_{\phi} t_{\rm dec} / (2 \pi)$, and the number of efolds $N$ is computed by using Eq.~(\ref{eq:scf-1}).} \label{tab:lifetimes}
\end{table}

Let us focus on the distribution of decay times observed in the stability regions of Fig.~\ref{fig:decaytime}. As expected, oscillons with initial shapes closer to the instability region have shorter lifetimes, while if the initial energy of the oscillon is large enough, the oscillons lifetime can be significantly larger. The maximum lifetime observed in the spherically symmetric simulations is,
\begin{eqnarray}
 \text{for $p=4$:} & \hspace{0.4cm} & m_{\phi} t_{\rm dec}^{\rm (max)} \simeq 5.3 \cdot 10^5 \hspace{0.3cm} \rightarrow \hspace{0.3cm}  N_{\rm osc}^{\rm (max)} \approx 8.4 \cdot 10^4  \ ,  \hspace{0.3cm}   N^{\rm (max)} \approx 4.3 \ , \nonumber \\  
  \text{for $p=6$:} & \hspace{0.4cm} & m_{\phi} t_{\rm dec}^{\rm (max)} \simeq 1.7 \cdot 10^6 \hspace{0.3cm} \rightarrow \hspace{0.3cm} N_{\rm osc}^{\rm (max)} \approx 2.7 \cdot 10^5  \ ,  \hspace{0.3cm}  N^{\rm (max)} \approx 4.7  \ ,\nonumber \\  
   \text{for $p=8$:} & \hspace{0.4cm} & m_{\phi} t_{\rm dec}^{\rm (max)} \simeq 2.9 \cdot 10^6 \hspace{0.3cm} \rightarrow \hspace{0.3cm} N_{\rm osc}^{\rm (max)} \approx 4.6 \cdot 10^5   \ ,  \hspace{0.3cm}  N^{\rm (max)} \approx 4.8  \ ,\nonumber  
\end{eqnarray}
where we have quoted the oscillon lifetime in number of oscillations $N_{\rm osc} \simeq m_{\phi} t_{\rm dec} / (2 \pi)$ and in efolds of expansion $N$, using Eq~(\ref{eq:scf-1}).  We observe that for $v=0.01 m_p$, oscillons do not live for more that 5 e-folds, for any of the considered power-law coefficients and for reasonable oscillon shapes\footnote{For $p=4$, our (3+1)-dimensional lattice simulations run until time $m_{\phi} t_{\rm e} \simeq 6870$, while the typical lifetimes observed in the spherically symmetric simulations are $m_{\phi} t_{\rm dec} \simeq \mathcal{O} (10^5)$. However, this does not imply that, by simply increasing the simulated time in the (3+1)-dimensional simulation by a factor $\sim 10$, we would be able to observe the oscillon decay there.  The reason is that, as time goes on, the volume of the oscillon decreases in the lattice, so we would lack enough spatial resolution to resolve it accurately at late times. }.

\begin{figure}
\centering
\includegraphics[height=3.2cm]{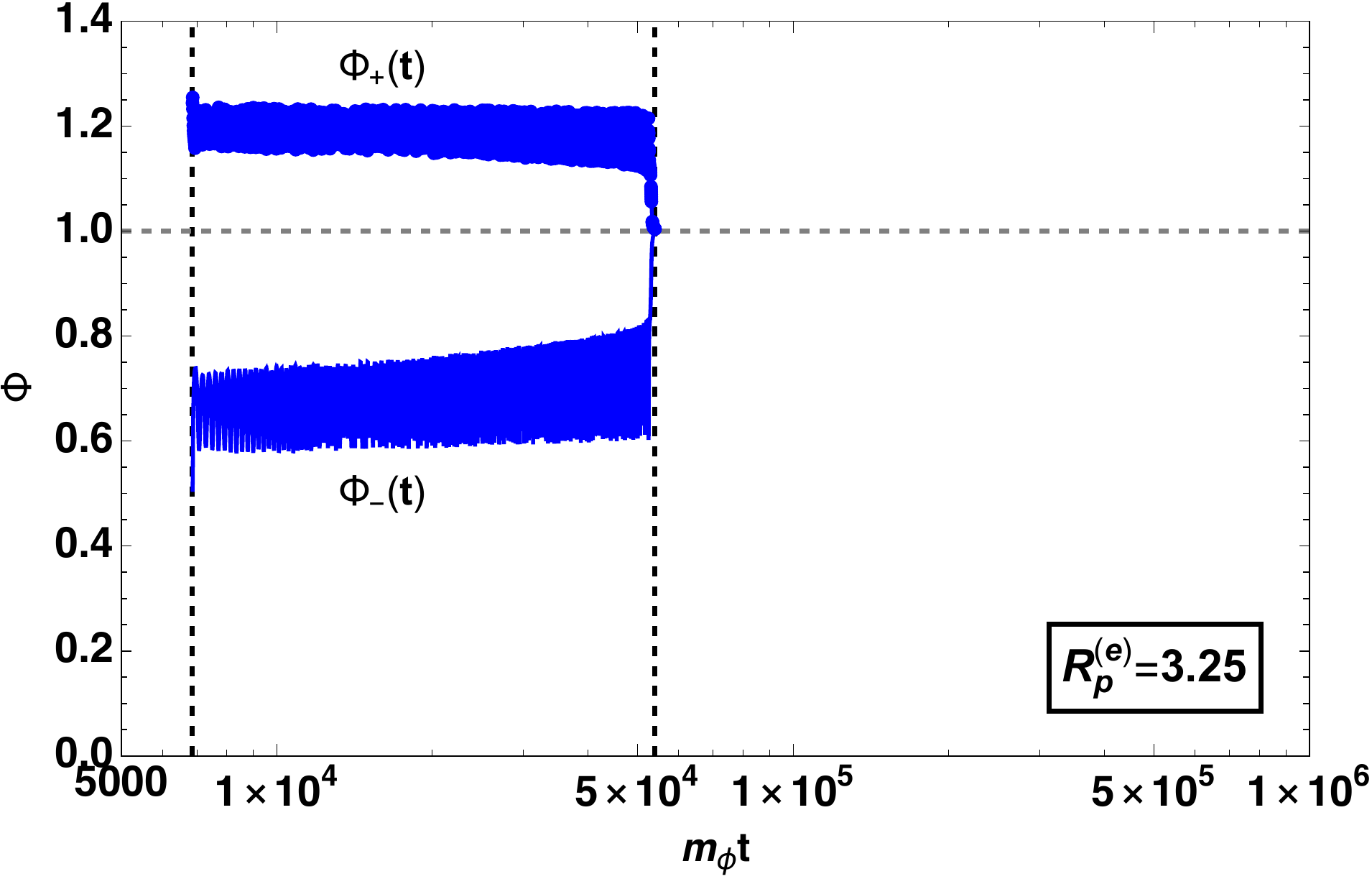}
\includegraphics[height=3.2cm]{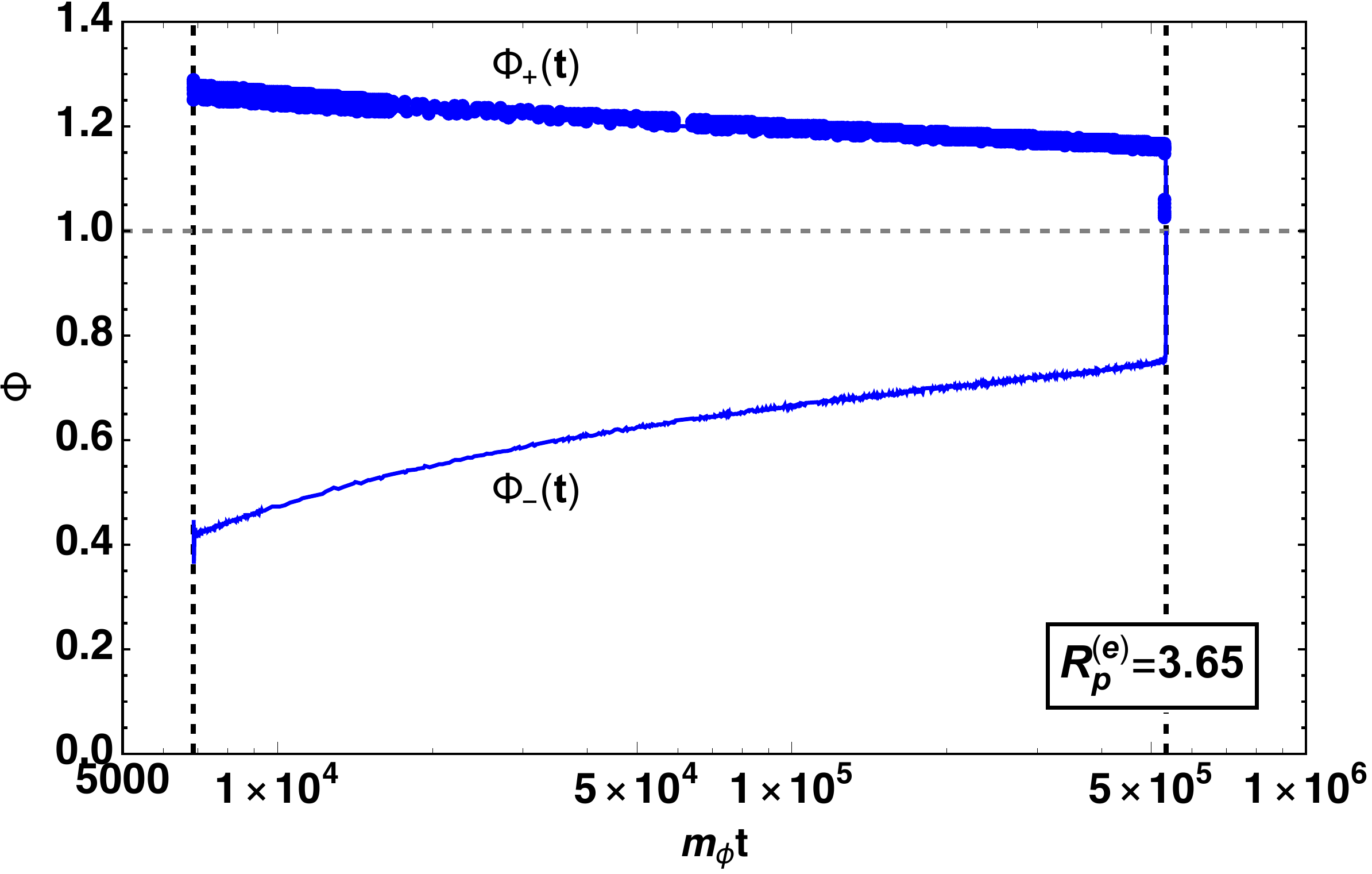}
\includegraphics[height=3.2cm]{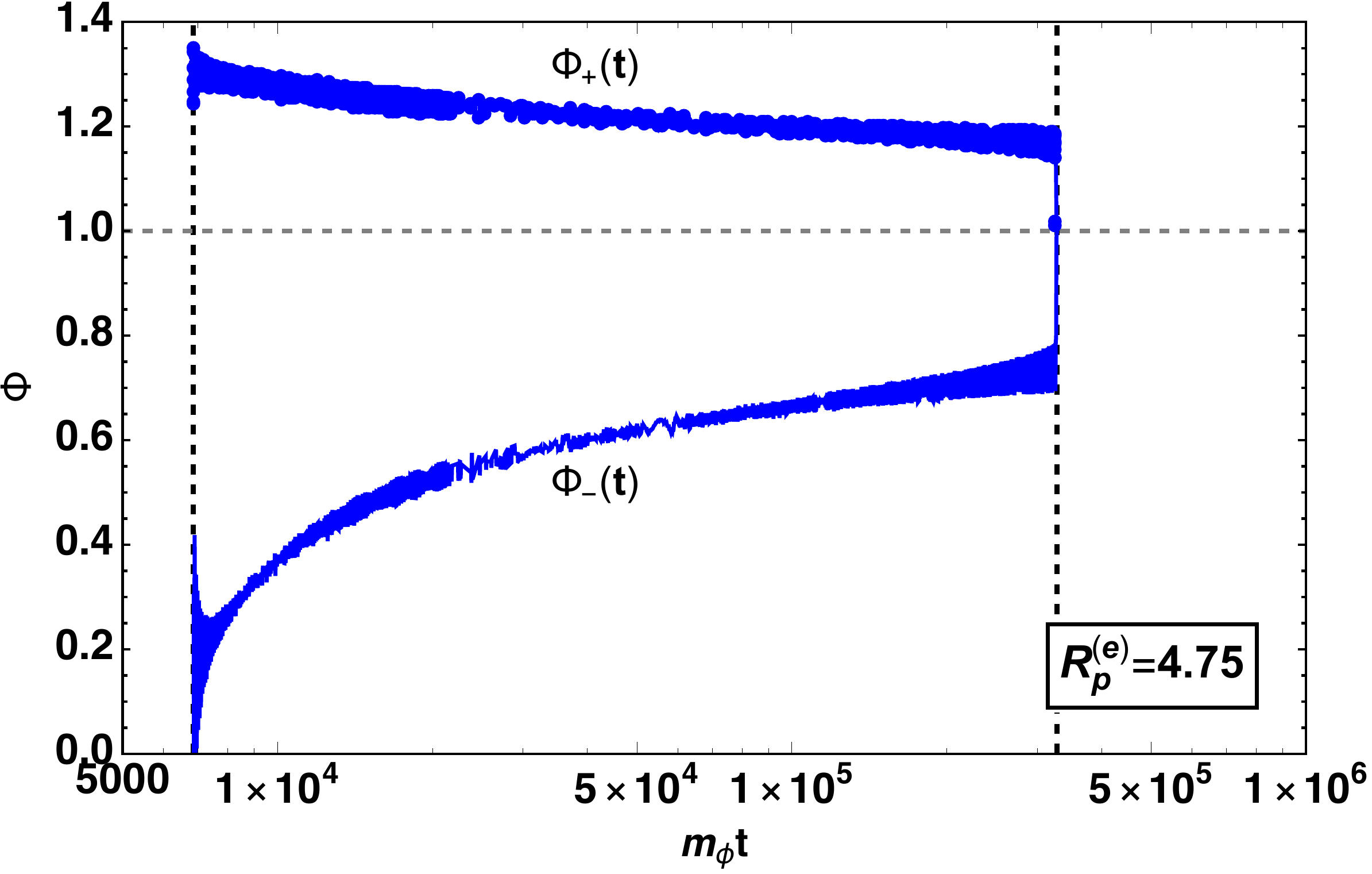}
\caption{We show, for three different oscillons, the upper and lower envelopes $\Phi_{+,-} (t)$ of the oscillations as a function of time. The power-law coefficient is $p=4$. The initial oscillon amplitude is the same in the three panels, $A_{-}^{\rm (e)}=0.46$, but the radius is different: $R_p^{\rm (e)}=3.25$ (left), 3.65 (middle), 4.75 (right). We indicate, with vertical dashed lines, the \textit{extraction} time $m_{\phi} t_{\rm e}$ (i.e. when the spherically symmetric simulations begin) and the oscillon decay time $m_{\phi} t_{\rm dec}$. Depending on the initial radius of the oscillon, both the lifetime and the breathing mode are different.}
\label{fig:breathing-radialsims}
\end{figure}

As a final remark, let us comment that we have observed a strong correlation between the amplitude of the breathing mode, and the oscillon decay time. To see this, let us focus on  Fig.~\ref{fig:breathing-radialsims}, where we show the upper and lower envelopes of the field oscillations as a function of time, for three different oscillons. The power-law coefficient is $p=4$. The initial amplitude of the oscillon is the same in the three cases ($A_{-}^{\rm (e)}=0.46$), but the physical radius differs. The oscillon on the left panel has a initial radius $R_p^{\rm (e)} = 3.25$, which puts it initially slightly above the line  separating the stability and instability regions of the amplitude-radius parameter space, $R_p^{\rm (e)} \gtrsim R_p^{\rm (st)} \simeq 3.22$ [see Eq.~(\ref{eq:AmpRad-st1})]. Due to this, the lifetime of this oscillon is remarkably short: only $m_{\phi} t_{\rm dec} \approx 5 \cdot 10^4$. On the other hand, as observed in Fig.~\ref{fig:breathing-radialsims}, the amplitude of the breathing mode is very large. Let us write the oscillon amplitude as
\be A_{-}(t) \simeq \langle A_{-} (t)\rangle_{\rm osc} \times (1 + \Delta A_{-}(t) ) \ , \ee
where $\langle \dots \rangle_{\rm osc}$ means average over many oscillations. For this oscillon, we get approximately $\Delta A_{-} (t) \sim 0.1$. Let us now focus on the oscillon in the middle panel, which has an initial radius slightly larger: $R_p^{\rm (e)} = 3.65$. Due to this, the oscillon lifetime is significantly larger, $m_{\phi} t_{\rm dec} \simeq 5.2 \cdot 10^5$. On the other hand, it is clearly seen that the amplitude of the breathing mode is much smaller than in the former case: $\Delta A_{-} (t) \sim 0.01$. Finally, in the right panel we show an oscillon with an exceptionally large initial radius $R_p^{\rm (e)} = 4.75$. Note that this oscillon configuration was not observed in the equivalent (3+1)-dimensional lattice simulations, but we show it here for completeness. In this case, the oscillon lifetime is slightly shorter than in the second case, $m_{\phi} t_{\rm dec} \simeq 3.1 \cdot 10^5$, but the amplitude of the breathing mode is larger, $\Delta A_{-} (t) \sim 0.01$. Hence, we observe than oscillons which show a stronger breathing typically have shorter lifetimes. This behaviour has also been observed for other amplitudes, as well as for other power-law coefficients. A more detailed study of the breathing mode and its effect on the lifetime of the oscillons is beyond the scope of this work, but it would be worth studying it elsewhere.

\section{Summary and Discussion}\label{sec:discussion}
\label{sec:summary_and_conclusions}

Oscillons have attracted large interest in the cosmology community  (see e.g.~\cite{Bogolyubsky:1976nx,Segur:1987mg,Gleiser:1993pt,Copeland:1995fq,Honda:2001xg,Copeland:2002ku,Adib:2002ff,Broadhead:2005hn,Farhi:2005rz,Fodor:2006zs,Graham:2006vy,Farhi:2007wj,Gleiser:2007te,Fodor:2008es,Amin:2010jq,Amin:2010dc,Gleiser:2011xj,Amin:2011hj,Amin:2013ika,Achilleos:2013zpa,Antusch:2015nla,Bond:2015zfa,Antusch:2015ziz,Lozanov:2017hjm,Hong:2017ooe,Cotner:2018vug,Graham:2006xs,Hindmarsh:2006ur,Saffin:2006yk,Gleiser:2008ty,Gleiser:2009ys,Hertzberg:2010yz,Salmi:2012ta,Andersen:2012wg,Saffin:2014yka,Gleiser:2014ipa,Mukaida:2016hwd,Ikeda:2017qev,Gleiser:2018kbq,Ibe:2019vyo,Gleiser:2019rvw,Olle:2019kbo,Muia:2019coe}). They constitute a fascinating example of a non-perturbative out-of-equilibrium phenomenon in the early Universe, where they can form e.g.\ after inflation when the potential sourcing the expansion of the Universe is shallower than quadratic away from the minimum, at least for some field amplitudes. It has been observed that oscillons can live for very long times, but the lifetime depends strongly on several factors, such as the shape of the potential, the initial shape of the oscillons, or the coupling of the scalar field to secondary species.

In this work, we have carried out an extensive analysis of the properties of oscillons in hilltop-type potentials, focusing on the hilltop inflation models specified in Eq.~(\ref{eq:scalar_potential}). The potentials feature a \textit{plateau} at $\phi \approx 0$, where inflaton happens, but have their minimum at non-zero vacuum expectation value $\phi = \pm v \,\neq 0$. They are also characterised by the power-law coefficient $p$, and we considered for our analysis the cases $p=4,6,8$. Larger $p$ (for fixed $v$) implies a  \textit{flatter} plateau, and a \textit{steeper} decline towards the minimum of the potential. As seen in \cite{Antusch:2015nla,Antusch:2015ziz,Antusch:2016con,Antusch:2017flz}, the post-inflationary tachyonic oscillations of the inflaton field in these models can efficiently induce the creation of oscillons, which may be very long-lived.

In Section \ref{sec:oscillon-shapes} we have studied and parametrized the oscillon shapes and energies. Our analysis has been based on a set of classical lattice simulations in (3+1) dimensions, which capture the first $\mathcal{O} (10^3)$ oscillations of the inflaton.  In hilltop models, oscillons tend to become approximately spherically symmetric, and their profile can be approximated by a Gaussian function. By using an appropriate algorithm to track the individual oscillons and monitor their evolution, we have been able to extract the oscillon shapes (which are parametrized in terms of the fitted amplitudes and radii) and their energies. Our results are summarized in Figs.~\ref{fig:oscillonshapes} and \ref{fig:ampradius}, as well as in Eqs.~(\ref{eq:shape-results}) and (\ref{eq:oscillon-energies}). We have observed that in hilltop models, the oscillon amplitudes and radii are strongly anti-correlated, such that their combination give rise to oscillons of similar energies. We have also observed that oscillons \textit{breathe}: their radii contract and expand as time goes on, while the maximal oscillon amplitude simultaneously increases and decreases. The oscillation period of the breathing mode is significantly larger than the oscillation period of the oscillons themselves.
 
In Section \ref{sec:spherical-symmetries} we have studied and parametrized the oscillon lifetimes, using spherically symmetric simulations for single oscillons. A truncation technique has been introduced, in order to avoid unphysical boundary effects on the oscillon dynamics. This way, we have been able to check energy conservation in our simulations. We have then simulated the oscillon dynamics until their decay, for different initial oscillon configurations and different power-law coefficients of the hilltop potential. We have observed that typical oscillons in these potentials live up to $10^4$ - $10^5$ field oscillations, and that their lifetime tends to increase with the value of $p$ (cf.\ Fig.~\ref{fig:decaytime} and Table \ref{tab:lifetimes}). For $v=0.01 m_p$, this corresponds to approximately 4-5 e-folds of expansion.

We would like to emphasize that our results are only a first step towards fully understanding and quantifying the properties of oscillons in hilltop potentials. Various aspects could affect the oscillon shapes and lifetimes, that we have not taken into account or only treated in some approximation. To start with, we have simulated the oscillons in an expanding flat FLRW background, but neglected gravitational interactions as well as quantum effects. Also, our extraction procedure of the oscillon shapes from the (3+1)-dimensional simulations assumes that oscillons are perfectly spherically symmetric. However, even at the comparatively late extraction time when the scale factor is $a = 5$, the oscillons were still somewhat asymmetric, which is not captured in our subsequent spherically symmetric simulations. Furthermore, although isolated oscillons are expected to become more and more spherical with time, it is conceivable that they could develop an asymmetry again when the oscillons become unstable and decay. Another effect that could affect the lifetime is the fact that the spherically symmetric simulations do not allow to take the effects of neighbouring oscillons into account, which can e.g.\ distort the oscillon shape via oscillon-oscillon interactions.  

Understanding better the lifetime of oscillons is important for various reasons. Generally speaking, since a significant part of the energy density after inflation can be concentrated in oscillons, they can have a strong impact on the (p)reheating phase. The longer the lifetime of the oscillons, the more pronounced the consequences can be. For example, they could have an important effect on the post-inflationary equation of state (e.g.~\cite{Gleiser:2011xj,Gleiser:2014ipa,Lozanov:2017hjm}) which could affect the inflationary observables. Furthermore, oscillons not only produce a stochastic gravitational wave background when they are produced and during their asymmetric oscillations \cite{Zhou:2013tsa,Antusch:2016con, Liu:2017hua,Antusch:2017flz,Antusch:2017vga,Amin:2018xfe,Kitajima:2018zco,Liu:2018rrt,Lozanov:2019ylm}, but potentially also when they decay. The resulting gravitational wave spectrum will depend on the amount of asymmetry the oscillons develop during their decay, as well as on the (distribution of) oscillon lifetimes.   

\vspace*{-0.15cm}

\section*{Acknowledgements}
This work has been supported by the Swiss National Science Foundation. F.T. thanks Mustafa Amin and Kaloian Lozanov for useful discussions.

\end{document}